\documentclass[]{spie}  
\usepackage{amsmath,amsfonts,amssymb}
\usepackage{graphicx}
\usepackage{setspace}
\usepackage{tocloft}
\usepackage{deluxetable}
\usepackage{threeparttablex}
\usepackage{booktabs}
\usepackage{hyperref}
\usepackage{chngcntr}
\usepackage{aas_macros}
\newcommand{\NA}{---}
\usepackage[symbol]{footmisc}
 
\usepackage{longtable}

\title{Comparing Non-Redundant Masking and Filled-Aperture Kernel Phase for Exoplanet Detection and Characterization}

\author[a,b,*]{Steph Sallum}
\author[a]{Andy Skemer}
\affil[a]{UC Santa Cruz, Department of Astronomy and Astrophysics, 1156 High St. Santa Cruz, CA, USA 95062}
\affil[b]{NSF Astronomy and Astrophysics Postdoctoral Fellow}

\cftpagenumbersoff{figure}
\cftpagenumbersoff{table} 
\begin{document} 
\maketitle

\begin{abstract}
The limitations of adaptive optics and coronagraph performance make exoplanet detection close to $\lambda/D$ extremely difficult with conventional imaging methods. 
The technique of non-redundant masking (NRM), which turns a filled aperture into an interferometric array, has pushed the planet detection parameter space to within $\lambda/D$. 
For high Strehl, the related filled-aperture kernel phase technique can achieve resolution comparable to NRM, without the associated dramatic decrease in throughput. 
We present non-redundant masking and kernel phase contrast curves generated for ground- and space-based instruments. 
We use both real and simulated observations to assess the performance of each technique, and discuss their capabilities for different exoplanet science goals such as broadband detection and spectral characterization.   
\end{abstract}


{\noindent \footnotesize\textbf{*}Steph Sallum,  ssallum@ucsc.edu}

\section{INTRODUCTION}
\label{sec:intro} 
Direct imaging has recently emerged as a viable planet detection and characterization method.
Near- to mid-infrared observations are particularly useful for discovering giant planets, since they have relatively low contrast at these wavelengths\cite{2014ApJ...792...17S}.
Several molecules that are expected to be in giant planet atmospheres have significant opacity in this wavelength range\cite{2008ApJ...683.1104F}, making infrared spectroscopy useful for constraining atmospheric composition.
Furthermore, infrared imaging of young planets can constrain evolutionary scenarios such as hot- versus cold-start models, and can distinguish between planetary atmosphere and circumplanetary accretion disk emission.\cite{2008ApJ...683.1104F,2015ApJ...799...16Z,2015ApJ...803L...4E}

Typical direct imaging planet searches are limited to angular separations of a few $\lambda / D$.
Point-spread function (PSF) deconvolution algorithms are less effective within these separations due to the small number of resolution elements available.\cite{2006ApJ...641..556M}
Phase leakage in even the highest performance coronagraphic observations prevents high contrast detections within $\sim1-2~\lambda/D$.\cite{2014ApJ...780..171G}
These limitations make semi-major axes less than $\sim10$ AU accessible for only the most nearby stars.\cite{2016PASP..128j2001B}
Expanding this detection parameter space to smaller semi-major axes and/or to more distant stars (including nearby star forming regions\cite{2007ApJ...671.1813T}) requires novel imaging techniques such as interferometry.

\subsection{Non-Redundant Masking}
Non-redundant masking\cite{2000PASP..112..555T} is well suited for detecting young, giant planets around more distant stars than those targeted by typical direct imaging surveys.
NRM uses a pupil-plane mask to turn a conventional telescope into an interferometric array.
The images are the interference fringes formed by the mask, which we Fourier transform to calculate complex visibilities.
Since the mask is non-redundant, no two baselines have the same orientation and separation; information from each baseline is encoded in a unique location in Fourier space.

Non-redundancy means that the instrumental component of each Fourier phase can be written as a linear combination of two pupil phases.
Calculating closure phases,\cite{1986Natur.320..595B} sums of phases around baselines forming a triangle, eliminates these instrumental phases to first order, leaving behind sums of phases intrinsic to the target.\footnote{We note that if the instrumental pupil-plane phases for each sub-aperture are treated as static pistons, then closure phases cancel instrumental phase exactly. However, spatial and temporal pupil-plane phase variations, as well as amplitude variations such as scintillation, lead to higher order closure phase errors that require calibration.\cite{2013MNRAS.433.1718I}}
Closure phases are particularly powerful for close-in companion detection because they are sensitive to asymmetries.
Since closure phases are correlated, we project them into linearly-independent combinations called kernel phases\cite{2010ApJ...724..464M,2013MNRAS.433.1718I,2015ApJ...801...85S}.
We also calculate squared visibilities, the powers on each mask baseline.

Despite the low throughput ($\sim10\%$), NRM's superior PSF characterization probes angular separations at and even within the diffraction limit.
It has led to detections of both stellar\cite{2012ApJ...753L..38B} and substellar companions\cite{2012ApJ...745....5K,2015Natur.527..342S}, as well as circumstellar disks\cite{2001ApJ...562..440D,2015ApJ...801...85S,2017ApJ...844...22S} at these small angular separations. 
This angular resolution means that spatial scales of $\sim10$ AU can be resolved for stars $\sim150$ pc away.
NRM led to the discovery of a promising system for planet formation studies, the protoplanetary candidates in LkCa 15.\cite{2012ApJ...753L..38B}
Dual-aperture LBTI masking observations recently resolved a solar-system sized disk around the star MWC 349A\cite{2017ApJ...844...22S}, at a distance of $\gtrsim 1.2$ kpc from Earth.
With a maximum baseline of 23 meters, this is a preview of NRM's potential on 30-meter class telescopes.
Non-redundant masking on current telescopes and on future facilities will expand the exoplanet detection parameter space.

\subsection{Filled-Aperture Kernel Phase}
Extreme adaptive optics systems have made filled-aperture kernel phase\cite{2010ApJ...724..464M} an interesting alternative to non-redundant masking.
Kernel phase involves treating a conventional telescope as if it were a redundant array.
Redundancy prevents Fourier phases from being written as a linear combination of pupil-plane phases in general.
However, in the high Strehl regime (where instrumental pupil plane phases are small), one can justify Taylor expanding the instrumental wavefront.
This means that the redundant instrumental Fourier phases can be approximated as a linear operator on the pupil plane phases.
Finding the nullspace of this linear operator yields kernel phases, linearly independent combinations of Fourier phases that are robust to instrumental phase errors to first order.
Like closure (and kernel phase) in non-redundant masking, filled-aperture kernel phases are sensitive to asymmetries and are thus powerful for close-in companion detection.

Kernel phase has been demonstrated on archival \emph{Hubble Space Telescope} observations of ultracool dwarfs\cite{2013ApJ...767..110P}.
In addition to confirming several known companions to these L dwarfs, it led to the detection of five new binary systems at angular separations of $\sim40-80$ mas at $\sim1.1-1.7~\mu$m.
It has also been applied to ground-based data from Keck\cite{2014IAUS..299..199I} and Palomar\cite{2016MNRAS.455.1647P}, where it led to the detection of stellar and substellar companions at ($\sim1-2~\lambda/D$).
The Keck observations were taken at $\mathrm{M_s}$ band, where the high sky background reduces the SNR of masking observations.
These kernel phase datasets provided comparable resolution to that expected for NRM, with a shorter integration time. 

\subsection{Outline of this Paper}
Here we compare non-redundant masking and filled-aperture kernel phase in a controlled way, with a specific focus on exoplanet science.
We generate simulated datasets for three imagers: NIRC2 on Keck, and NIRCam and NIRISS on \emph{James Webb Space Telescope (JWST)}.
We also generate observations for two integral field spectrographs (IFSs): OSIRIS on Keck, and NIRSpec on \emph{JWST}.
Integral field spectroscopy is a particularly interesting kernel phase application, since dispersing the light and applying a pupil plane mask would require long integration times for high signal to noise.
Furthermore, when broadband kernel phase observations of bright stars may be unfeasible due to detector saturation, IFS kernel phase observations may not saturate.

We describe the simulated datasets in Section \ref{sec:simdata}.
We use the simulated observations to generate contrast curves for both filled-aperture kernel phase and NRM, and to place detection limits on planetary atmosphere and circumplanetary accretion disk models.
We present these results in Section \ref{sec:results}.
In Sections \ref{sec:discussion} and \ref{sec:conclusions} we discuss the potential of each technique as a method for exoplanet detection and characterization.

\section{SIMULATED OBSERVATIONS}\label{sec:simdata}
Here we simulate non-redundant masking and filled-aperture kernel phase observations for three imagers - Keck NIRC2, \emph{JWST} NIRCam, and \emph{JWST} NIRISS - and kernel phase observations for two integral field spectrographs - Keck OSIRIS, and \emph{JWST} NIRSpec.
For the IFS simulations, we use the central wavelength bin to estimate the average performance for each instrument and mode.
For all instruments, we compare the two techniques by setting the total amount of observing time, including overheads, to be the same.
To ensure a systematic comparison, we use simulated observations so that we can control the various noise sources.
When possible, we anchor these simulations using real datasets.
We include random noise sources from the star, sky background, and detector.
We also simulate changing optical path difference maps for each instrument to account for quasi-static (``speckle") noise that can lead to kernel phase calibration errors.
To simulate these calibration errors, we divide the total observing time for all instruments into observations of a science target and of a PSF calibrator with the same brightness as the science target, but with different quasi-static errors.
We break the observations up into ``visits." each of which is a single datacube containing a number of frames on a single target (science or PSF calibrator). 

We calculate kernel phases using an updated version of the data reduction pipeline presented in Sallum \emph{et al.} (2017)\cite{2017ApJS..233....9S}.
For both Keck and \emph{JWST} filled-aperture observations, we assume one sub-aperture per mirror segment. 
This is an arbitrary choice, and while it prevents phase noise due to jumps between mirror segments, denser pupil plane sampling may lead to improved results. 
We use the ``Martinache" kernel phase projection\cite{2010ApJ...724..464M}, which makes orthonormal combinations of Fourier phases (see Appendix \ref{app:kpweights}).
We note that other phase combinations can provide higher contrast away from the PSF core; one example is the ``Ireland" projection\cite{2013MNRAS.433.1718I}, which makes statistically independent combinations of Fourier phases using the kernel phase covariance matrix. 
This has been shown to boost contrast away from the PSF core in the photon noise limit\cite{2013MNRAS.433.1718I}.
Since some of our simulated observations are in the calibration error limit, where the advantages of the ``Ireland" projection do not apply\cite{2013MNRAS.433.1718I}, we use the ``Martinache" projection.
The contrast curves shown in Section \ref{sec:results} could be improved by using different projections for probing different angular separations or for different noise regimes.

For both techniques, we apply a super-Gaussian window to create inter-pixel correlations in the Fourier plane before calculating kernel phases.
These inter-pixel correlations are helpful for reducing the random kernel phase scatter induced by noise sources such as sky background and detector noise. 
For NRM datasets we tune the super-Gaussian so that it has a value of 0.9 at the null in the PSF of an individual hole ($\lambda / d_h$).
For filled-aperture kernel phase the super-Gaussian has a value of 0.9 at $3~\lambda / D$.
To minimize the impact of sky background and detector noise beyond these regions - where the PSFs have low signal - we use an 8th order super-Gaussian.
We note that alternative techniques such as the ``Monnier" method\footnote{The ``Monnier" method involves averaging the bispectra for many pixel triangles when calculating the closure phase for a triangle of holes in a non-redundant mask.}\cite{1999PhDT........19M}, and image-plane fringe fitting\cite{2015ApJ...798...68G} can reduce random phase scatter without windowing; these have been demonstrated for NRM observations and similar techniques could be applied to filled-aperture kernel phase.

For all instruments and bandpasses, we simulate a grid of stellar apparent magnitudes between $\sim 6$ and 13. 
This corresponds to absolute magnitudes between $\sim 0$ and 7, at the distance of Taurus ($\sim 140$ pc\cite{2007ApJ...671.1813T}).
While this does not cover the bright end of expected young stars' absolute magnitudes\cite{2012MNRAS.427..127B}, for filled-aperture kernel phase observations, the apparent magnitude 6 stars are close to or beyond the saturation limit for NIRC2 and NIRCam. 
We thus focus on this range of stellar brightnesses to compare the two techniques on these instruments.
We generate contrast curves for each simulated dataset by comparing the $\chi^2$ of the null model ($\chi^2_{null}$; a single point source) to a grid of single companion models with different separations, contrasts, and position angles.
For each single companion separation, we calculate the average $\chi^2$ value for all sampled position angles ($\chi^2_{sep}$), and take the 5$\sigma$ contrast to be the contrast at which $\chi^2_{sep} - \chi^2_{null} = 25$.
The details of the individual simulations can be found in the following subsections and in Appendix \ref{app:obsplanning}.

\subsection{Keck NIRC2}
\label{sec:nirc2}

We simulate kernel phases for filled-aperture and masked Keck NIRC2 observations at Ks = 2.15 $\mu$m, L$'$ = 3.78 $\mu$m, and Ms = 4.67 $\mu$m.
The filled-aperture and NRM modes yield 45 and 28 kernel phases, respectively.
For each target observation, we simulate a cube of $n_{frames}$ frames, each of which is the sum of $n_{coadd}$ coadds (see Tables \ref{tab:ksnrmobs} - \ref{tab:mskpobs}).
We add sky background, dark current, and readout noise to the images according to the number of coadds and exposure times; Table \ref{tab:nirc2detpars} lists the relevant noise parameters and other detector information for each bandpass. 
For all observations, we background subtract each frame and apply any window functions. We then calculate complex quantities for each kernel phase that are analogous to bispectra for closure phases (see Appendix \ref{app:kpweights}).
Like averaging bispectra for calculating mean closure phases, we average these complex quantities over the cube of images before taking the phase as the mean kernel phase for a single visit.

We simulate optical path difference maps to account for quasi-static PSF aberrations.
Keck AO is known to be affected by low-order residual wavefront errors; modeling these as segment piston errors has been shown to produce realistic point spread functions \cite{2018SPIE10700E..1DR}.
To simulate point spread functions, we create optical path difference (OPD) maps that are a combination of low- and high-order residual wavefront errors.
We generate low-order errors over the entire Keck pupil and on each individual mirror segment.
For the entire pupil, we make combinations of the first 20 Zernike modes ($Z_n$; excluding tip and tilt), drawing each coefficient from a uniform distribution with a width ($2~f(n)$) that depends on the Zernike order (n):
\begin{equation}
OPD_{low} = \sum_{n=0}^{20} \mathrm{Unif}\left(-f(n),f(n)\right) Z_n = \sum_{n=0}^{20} C_{low} Z_n.
\end{equation}
The function $f(n)$ has the form $n^\alpha$, and we tune $\alpha$ to match real observations (see below).
For the individual mirror segments, we add tip, tilt, and piston errors, drawing each coefficient from a uniform distribution with a fixed width:
\begin{equation}
OPD_{seg} = \sum_{n=0}^{2} \mathrm{Unif}\left(-g,g\right) Z_n = \sum_{n=0}^{2} C_{seg} Z_n.
\end{equation}

\begin{deluxetable}{l cccccccc}
\tablecaption{NIRC2 Noise and Detector Parameters\label{tab:nirc2detpars}}
\tablecolumns{9}
\tablewidth{0pt}
\tablehead{\colhead{Band} & \colhead{$\lambda_c$\tablenotemark{a}} & \colhead{$\lambda/D$\tablenotemark{b}} & \colhead{S\tablenotemark{c}} & \colhead{ZP\tablenotemark{d}} & \colhead{DC\tablenotemark{e}} & \colhead{RN\tablenotemark{f}} & \colhead{WD\tablenotemark{g}} & \colhead{g\tablenotemark{h}}\\
\colhead{} & \colhead{($\mu$m)} & \colhead{(pix)} & \colhead{(mag/arcsec$^2$)} & \colhead{(mag)} & \colhead{(e$\mathrm{^-}$ / pix / s)} & \colhead{(e$\mathrm{^-}$ / pix)} & \colhead{(DN)} & \colhead{(e$\mathrm{^-}$/DN)}
}
\startdata
K$\mathrm{_s}$ & 2.146 & 4.45 & 12.66 & 24.84 & 0.1 & 15 & 18,000 & 4\\
L$'$ & 3.776 & 7.83 & 3.01 & 23.6 & 0.1 & 15 & 18,000 & 4\\
M$\mathrm{_s}$ & 4.670 & 9.69 & 0.23 & 21.42 & 0.1 & 15 & 18,000 & 4\\
\enddata
\tablenotetext{a}{Central wavelength in microns}
\tablenotetext{b}{Detector PSF sampling in pixels}
\tablenotetext{c}{Sky background in magnitudes per square arcsecond}
\tablenotetext{d}{Zero point in magnitudes for Strehl = 1.0}
\tablenotetext{e}{Dark current in electrons per pixel per second}
\tablenotetext{f}{Read noise in electrons per pixel}
\tablenotetext{g}{Well depth in counts at 5\% linearity}
\tablenotetext{h}{Gain in electrons per count}
\end{deluxetable}

\begin{deluxetable}{l c}
\tablecaption{NIRC2 OPD Parameters\label{tab:opdpars}}
\tablecolumns{2}
\tablewidth{0pt}
\tablehead{}
\startdata
Low-order RMS & 160 nm \\
Low-order order dependence ($\alpha$) & 0.2 \\
Low-order evolution parameter ($h$) & 0.3 \\
Segment RMS & 280 nm \\
Segment evolution parameter ($h$) & 0.3 
\enddata
\end{deluxetable}

\begin{figure}[h!]
\begin{center}
\begin{tabular}{c} 
\includegraphics[width=0.8\textwidth]{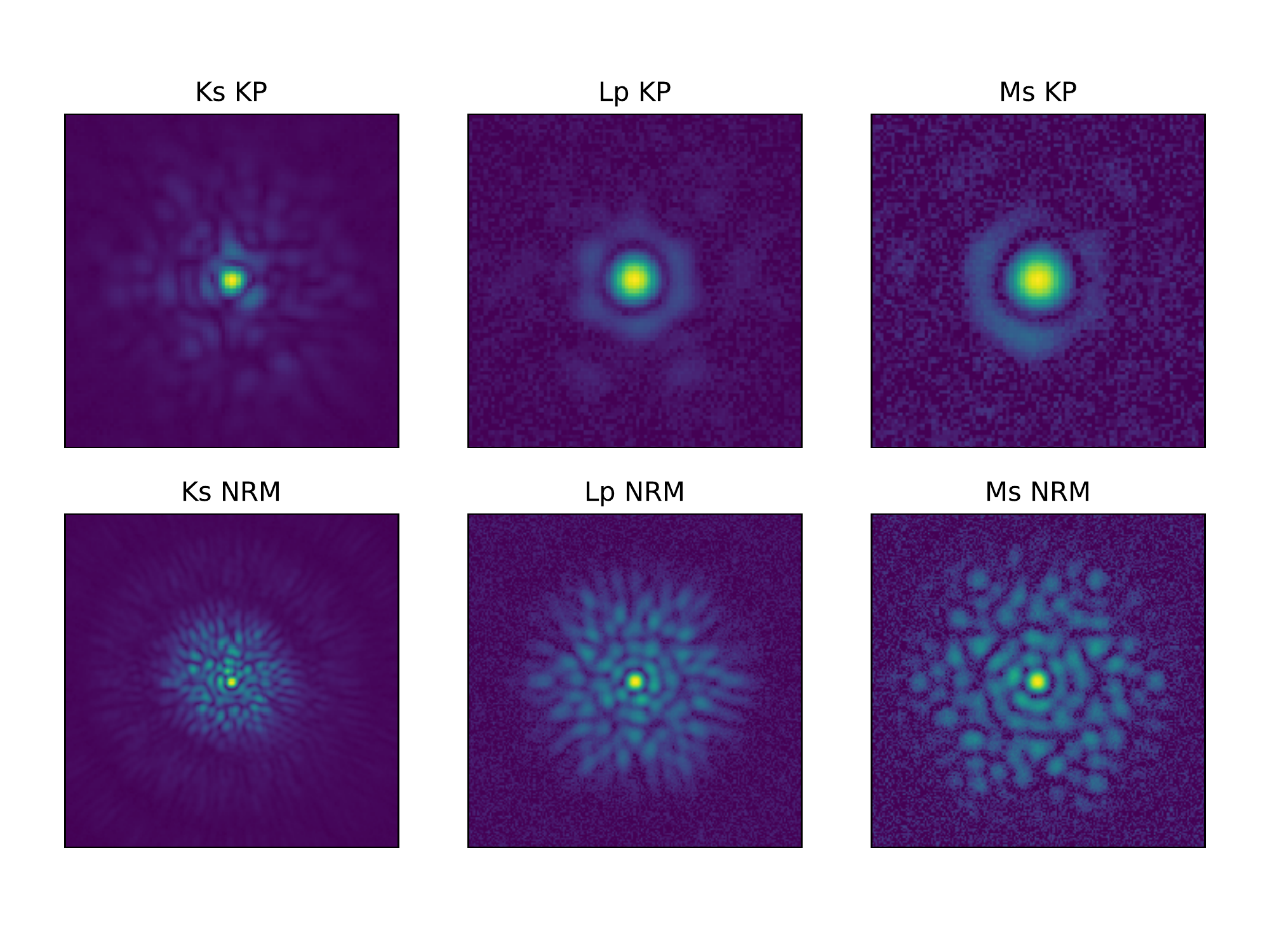}
\end{tabular}
\end{center}
\caption[Simulated NIRC2 Point Spread Functions] 
{ \label{fig:nirc2psfs}
Simulated individual K$\mathrm{_s}$ (left), L$'$ (center), and M$\mathrm{_s}$ (right) images for the observing parameters listed in Table \ref{tab:nirc2detpars}. 
The top row shows filled-aperture data, and the bottom row non-redundant masking data.
The images are displayed on a square root scale to highlight structure in the wings of the point spread function.
The fields of view are different for the two techniques: the kernel phase images are 45 mas on a side, and the NRM images are 100 mas on a side.}
\end{figure}

\begin{figure} [h!]
\begin{center}
\begin{tabular}{c} 
\includegraphics[width=0.8\textwidth]{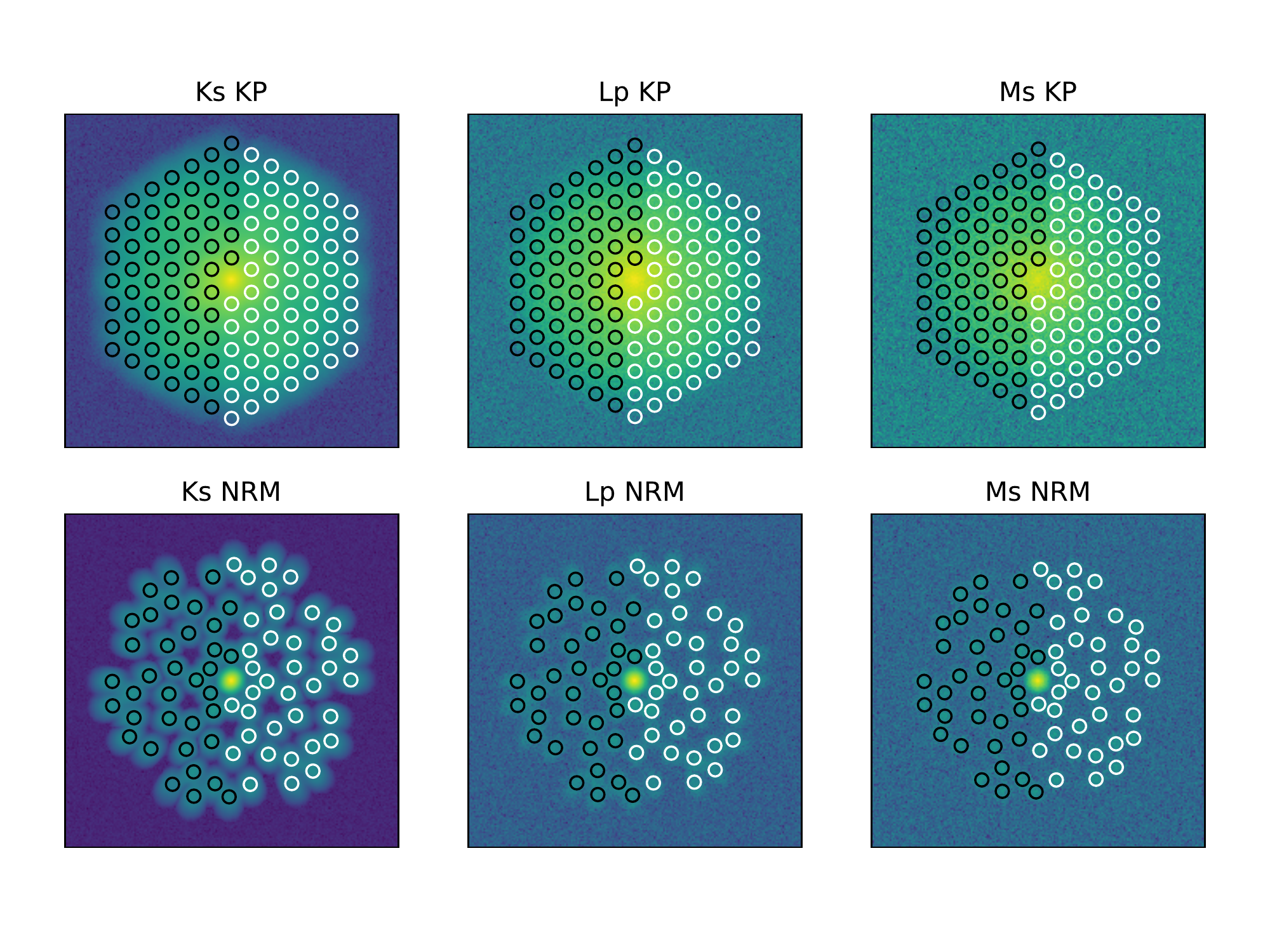}
\end{tabular}
\end{center}
\caption[Simulated NIRC2 Power Spectra] 
{\label{fig:nirc2pspecs} Simulated K$\mathrm{_s}$ (left), L$'$ (center), and M$\mathrm{_s}$ (right) power spectra for a single frame using the observing parameters listed in Table \ref{tab:nirc2detpars}. 
The top row shows filled-aperture kernel phase data, and the bottom row non-redundant masking data.
Random noise from the large sky background is apparent in the L$'$ and M$_\mathrm{s}$ power spectra; building signal to noise in the long baselines requires combining more frames.
The scattered points show the sampling locations for calculating kernel phases; they are shown in two colors since the Fourier transform is symmetric.
}
\end{figure}

When we apply this prescription, we use arbitrary normalizations for $f(n)$ and $g$, and then re-scale the final OPD maps to have an RMS tuned to match real observations.
We also tune the dependence of the low order Zernike coefficients on Zernike order ($\alpha$).
The best OPD parameters are listed in Table \ref{tab:opdpars}.
These lead to uncalibrated kernel phase scatters of $\sim 1.2^\circ$ for L$'$ NRM observations (compared to $\sim1.0-1.5 ^\circ$ for real observations), and Strehl values of 0.55, 0.85, and 0.9 at Ks, L$'$, and Ms (consistent with typical Strehls observed using NIRC2), respectively. 
They also match observed PSF radial profiles to within $\sim4 \%$ fractional error.
Figures \ref{fig:nirc2psfs} and \ref{fig:nirc2pspecs} show example images and power spectra, respectively, for NRM and kernel phase for each bandpass.

\begin{figure} [ht]
\begin{center}
\begin{tabular}{c} 
\includegraphics[width=\textwidth]{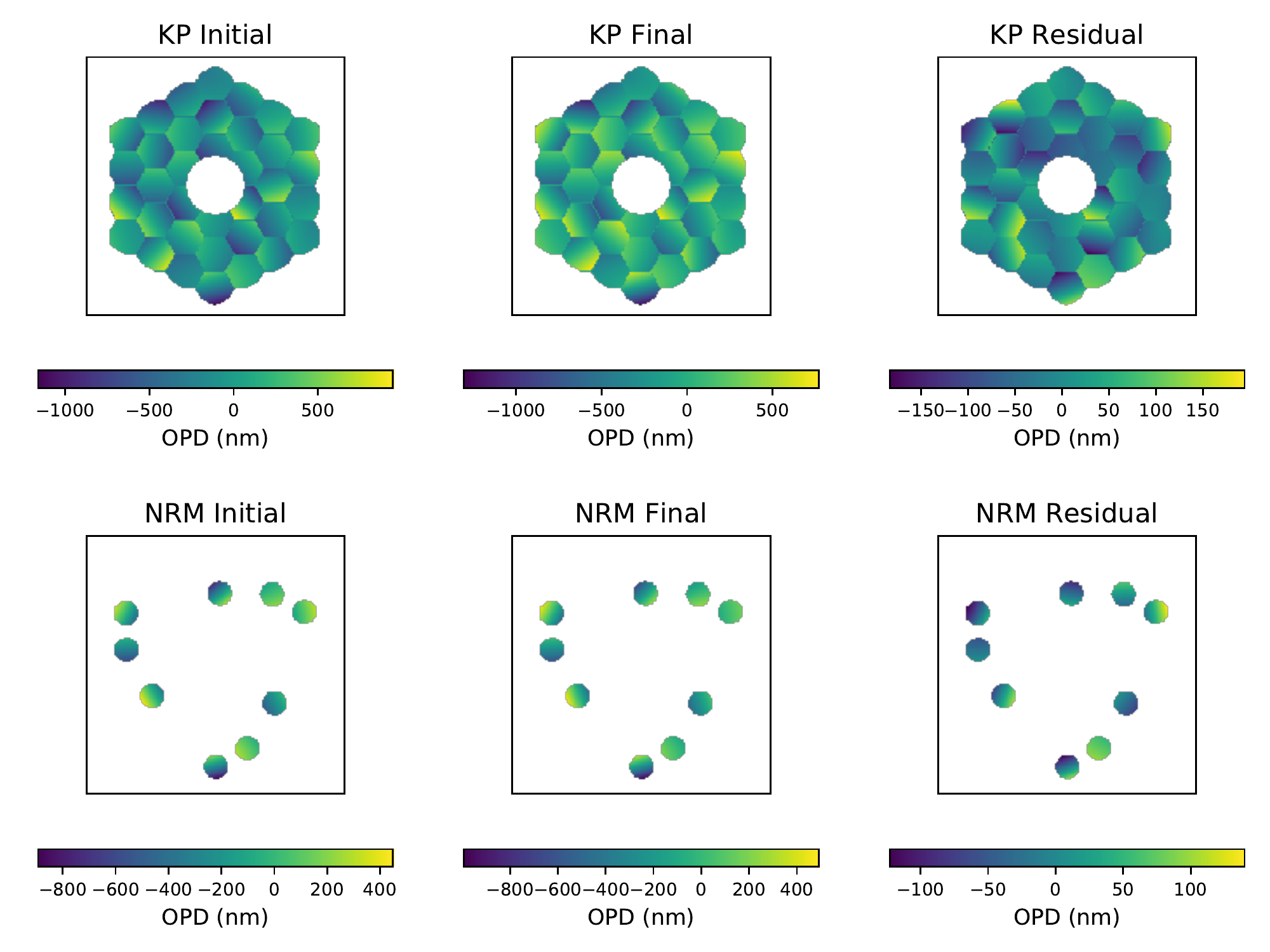}
\end{tabular}
\end{center}
\caption[NIRC2 OPD Evolution] 
{\label{fig:n2opd}
Example OPD evolution for NIRC2 filled-aperture kernel phase (top) and non-redundant masking (bottom). 
The left shows an initial OPD made up of both low- and high-order wavefront errors, and the center panel shows the result of evolving the OPD map.
The right side shows the difference between the two; it has an rms OPD of 60 nm.}
\end{figure}

Keck's low-order errors are quasi-static \cite{2014SPIE.9148E..5IR} and contribute to calibration errors in the final kernel phases.
We thus generate calibrator observations after evolving the low-order and segment OPD Zernike coefficients by a factor drawn from a one-mean uniform distribution with width $2h$.
Equations \ref{eq:opdev1} and \ref{eq:opdev2} describe this procedure: 
\begin{equation}
C_{n,low,cal} = \mathrm{Unif}\left(1-h,1+h\right) C_{n,low,targ}
\label{eq:opdev1}
\end{equation}
\begin{equation}
C_{n,seg,cal} = \mathrm{Unif}\left(1-h,1+h\right) C_{n,seg,targ}.
\label{eq:opdev2}
\end{equation}
We use the original and evolved optical path difference maps to generate point spread functions for each target and calibrator observation, respectively.
To calibrate, we subtract the simulated calibrator kernel phases from the target kernel phases.
We tune the evolution parameter ($h$) so that the calibration decreases the kernel phase scatter by the same amount as it does in typical NIRC2 NRM observations ($h = 0.3$).
This results in calibrated NRM kernel phase scatters of $\sim0.2-0.3 ^\circ$ for observations of a L$'$ = 5.8 star, consistent with the observed scatter in PSF calibrator observations from Keck at L$'$.
Figure \ref{fig:n2opd} shows an example of OPD evolution for both techniques; the residual OPD map has $\sim$60 nm rms.
We note that this prescription does not include residual AO errors such as fitting error, bandwidth error, or aliasing, which would change on timescales comparable to the AO frame rate (much shorter than the integrations presented here).
Adding these effects would make the OPD simulations more realistic, but since they would average over many AO loops and since the simplified prescription can match real observations, we do not include them.

We design a half night of observations for each target star magnitude and bandpass, accounting for overheads. 
We split the half night into a number of visits to target and PSF calibrator stars. 
The length of each visit depends on the number of photons collected from the star and sky; for a small number of source photons, kernel phase errors may be dominated by random noise.
As the number of source photons increases, the random kernel phase scatter decreases, and the visit approaches the contrast limit - where quasi-static errors dominate over random errors.
Longer integrations in this limit do not decrease the kernel phase scatter significantly.
Thus we design our visits to be either long enough to reach the contrast limit, or to take 3 hours (half the observing time), since the observing is split evenly between target and PSF calibrator observations.
Datasets on brighter stars will contain more visits than datasets on fainter stars.
When multiple visits can fit into a half night, we treat them as snapshots at different parallactic angles that change in $5^\circ$ increments and have independent calibration errors.

To account for NIRC2 overheads, we treat each visit as a collection of frames and each frame as the sum of a number of coadds.
We design each coadd to be long enough to reach 5000 - 10,000 maximum counts - this is to minimize overheads but remain in a linear regime on the detector. 
We assume a 512 by 512 pixel subframe, which results in a single coadd overhead of 0.05 seconds.
We enforce a minimum frame time of 20 seconds, which adds an additional overhead of 2.18 seconds every 20 seconds.
We assume that the observations are dithered every 200 seconds to enable sky subtraction, which adds an overhead of 6 seconds for each dither. 
Lastly, we assume target acquisition overheads of $\sim 90$ seconds per pointing\cite{2006PASP..118..297W}.
Tables \ref{tab:ksnrmobs} - \ref{tab:mskpobs} in Appendix \ref{app:obsplanning} provide frame, coadd, and dither details for each target star brightness.
Figures \ref{fig:n2histsbright} and \ref{fig:n2histsfaint} show histograms of the raw (uncalibrated) and calibrated kernel phases for each technique and bandpass for two different target star brightnesses.

\subsection{\emph{JWST} NIRCam and NIRISS}

We use the Pandeia engine \cite{2016SPIE.9910E..16P} and WebbPSF software \cite{2012SPIE.8442E..3DP} to simulate NIRCam kernel phase and NIRISS aperture masking (AMI) observations to compare the two techniques.
NIRCam imaging and NIRISS AMI yield 21 and 15 kernel phases, respectively.
Since exoplanets are relatively bright in the near- to mid-infrared, we simulate data for filters centered on wavelengths $\gtrsim 4~\mu$m that can be used with NIRISS AMI and NIRCam imaging (F430M and F480M).
Since slew and telescope roll overheads are large for \emph{JWST}, we do not allow the number of visits for the different target stars to vary like we do for NIRC2.
We rather assume the observations are made up of two pairs of target - PSF calibrator visits taken at different telescope roll angles 45$^\circ$ apart. 
We assume that the length of each visit is 1.5 hours, for a total observation time of 6 hours, excluding slew and roll angle overheads.
When designing the observations, we account for detector overheads within each 1.5 hour visit.

\emph{JWST} visits are split into sets of integrations, each of which is composed of a number of groups.
To calculate overheads, for each target magnitude we first find the maximum number of groups ($n_g$) that can be used in a single integration without saturation.
For NIRCam imaging, this means choosing the readout mode that provides the maximum integration time, but for NIRISS AMI only one readout mode is available.
We then assume that each visit is composed of the maximum number of $n_g$-group integrations ($n_i$) that can be acquired in 1.5 hours, given that each integration comes with a readout overhead (0.0494 seconds for NIRCam in a sub64P subarray, and 0.0745 seconds for NIRISS in a sub80 subarray).
When the remaining time after $n_i$ integrations allows for more than a single group, we add an additional integration containing ($n_{g,rem}$) groups. 
Tables \ref{tab:nircamf430} - \ref{tab:nirissf480} in Appendix \ref{app:obsplanning} list $n_g$, $n_i$, and $n_{g,rem}$ for each instrument.
We use Pandeia's ``Detector" output for the observational parameters in Tables \ref{tab:nircamf430} - \ref{tab:nirissf480}.
This generates a long-exposure image for the entire visit, which contains the Poisson noise from the target and \emph{JWST's} ``medium" thermal background, and read noise.\footnote{
The data product for each real JWST exposure will be a cube of integrations, allowing for coadding of smaller pieces of each exposure, or for averaging kernel phases over all short integrations. 
We do not explore different options for coadding or averaging of short-exposure kernel phases using Pandeia. 
However, we do test the effects of using a single summed image versus averaging many short-exposure images using our Keck simulations.
Contrast curves generated from these two extreme options are nearly identical in the bright limit, where OPD errors dominate. 
In the limit where random noise dominates, averaging short exposures performed better by $\sim$0.5 mag; averaging kernel phases generated from short exposures would thus improve the achievable contrast for the faintest simulated targets ($\gtrsim$ 11th mag). 
We note that with real \emph{JWST} observations, averaging the kernel phases for many short exposures will be more practical, since effects such as pointing jitter and noise sources such as cosmic rays will degrade long-exposure datasets.}
Tables \ref{tab:nircamobs} and \ref{tab:nirissobs} list the relevant noise and detector parameters for these observations.

\begin{deluxetable}{l cc}
\tablecaption{NIRCam Noise and Detector Parameters\label{tab:nircamobs}}
\tablecolumns{3}
\tablewidth{0pt}
\tablehead{}
\startdata
$\lambda_c$\tablenotemark{a} $~$ $\mu$m&\hspace{25pt}& 4.3, 4.8\\
$\lambda/D$\tablenotemark{b} $~$ (pix) &\hspace{20pt}&  2.10, 2.34 \\
DC\tablenotemark{c} $~$ (e$\mathrm{^-}$ / pix / s)&\hspace{50pt} & 0.027 \\
RN\tablenotemark{d} $~$ (e$\mathrm{^-}$)&\hspace{20pt} & 13.5 \\
WD\tablenotemark{e} $~$ (DN)&\hspace{20pt} & 83,300 \\
g\tablenotemark{f} $~$ (e$\mathrm{^-}$/DN)&\hspace{20pt} & 1.82
\enddata
\tablenotetext{a}{Central wavelength in microns}
\tablenotetext{b}{Detector PSF sampling in pixels}
\tablenotetext{c}{Dark current in electrons per pixel per second}
\tablenotetext{d}{Read noise in electrons per pixel}
\tablenotetext{e}{Well depth in counts}
\tablenotetext{f}{Gain in electrons per count}
\end{deluxetable}

\begin{deluxetable}{l cc}
\tablecaption{NIRISS Noise and Detector Parameters\label{tab:nirissobs}}
\tablecolumns{3}
\tablewidth{0pt}
\tablehead{}
\startdata
$\lambda_c$\tablenotemark{a} $~$ $\mu$m&\hspace{25pt}& 4.3, 4.8\\
$\lambda/D$\tablenotemark{b} $~$ (pix)&\hspace{25pt} &  2.10, 2.34 \\
DC\tablenotemark{c} $~$ (e$\mathrm{^-}$ / pix / s)&\hspace{50pt} & 0.0125 \\
RN\tablenotemark{d} $~$ (e$\mathrm{^-}$) &\hspace{25pt}& 18.32 \\
WD\tablenotemark{e} $~$ (DN) &\hspace{25pt}& 100,000 \\
g\tablenotemark{f} $~$ (e$\mathrm{^-}$/DN) &\hspace{25pt}& 1.61
\enddata
\tablenotetext{a}{Central wavelength in microns}
\tablenotetext{b}{Detector PSF sampling in pixels}
\tablenotetext{c}{Dark current in electrons per pixel per second}
\tablenotetext{d}{Read noise in electrons per pixel}
\tablenotetext{e}{Well depth in counts}
\tablenotetext{f}{Gain in electrons per count}
\end{deluxetable}

\begin{figure} [h!]
\begin{center}
\begin{tabular}{c} 
\includegraphics[width=\textwidth]{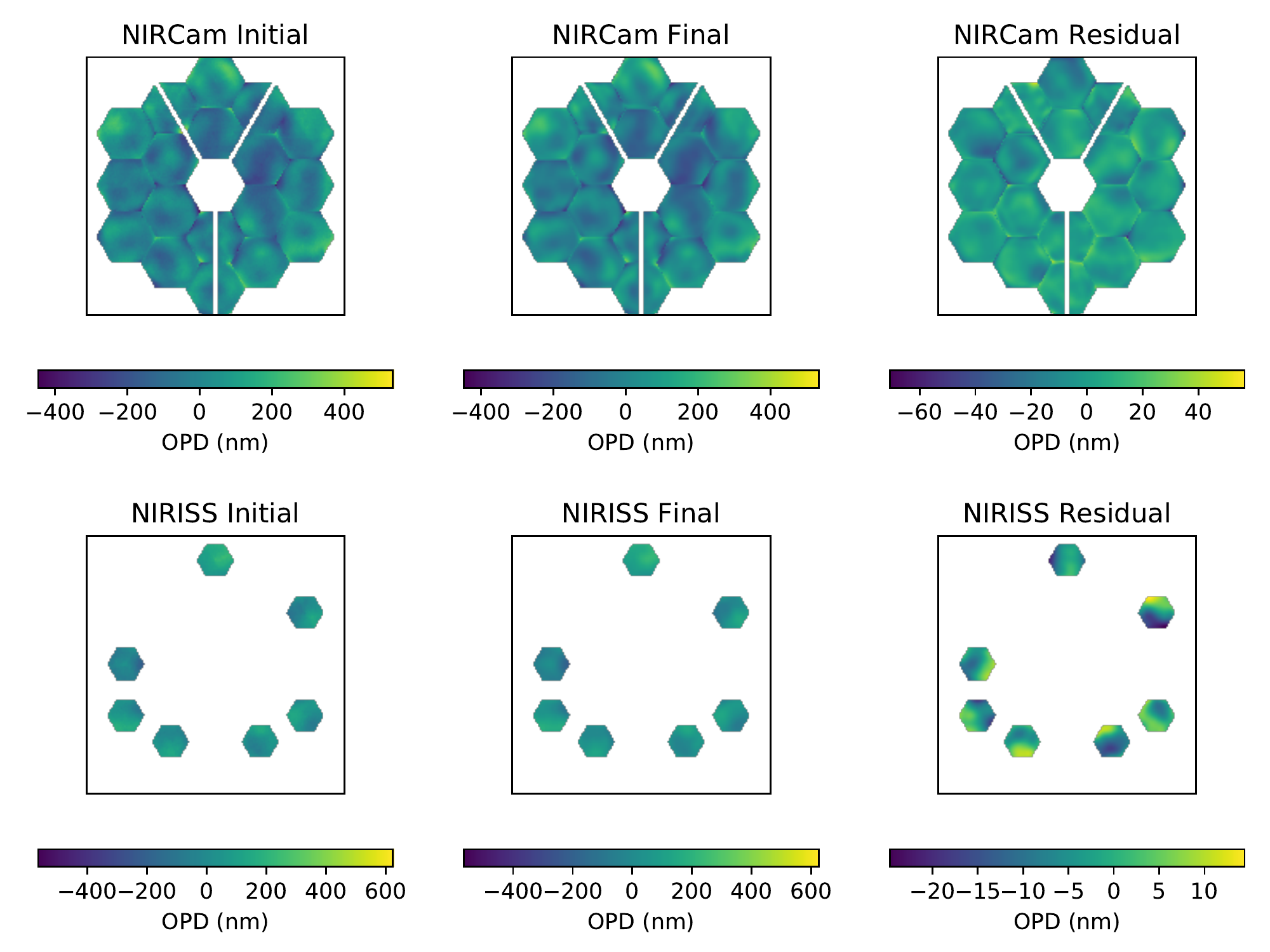}
\end{tabular}
\end{center}
\caption[NIRCam and NIRISS OPD Evolution] 
{ \label{fig:jwstopd} 
Example JWST optical path difference (OPD) evolution. 
The left panels show one of ten predicted NIRCam (top) and NIRISS (bottom) optical path difference files from WebbPSF. 
The center panels show the resulting OPDs after fitting hexikes to each JWST mirror segment and evolving their coefficients by a small fraction in random directions.
The right panels show the difference between the first two OPD maps, and have rms residual optical path differences of $\sim 10$ nm.
}
\end{figure}

\begin{figure} [h!]
\begin{center}
\begin{tabular}{c}  
\includegraphics[width=0.6\textwidth]{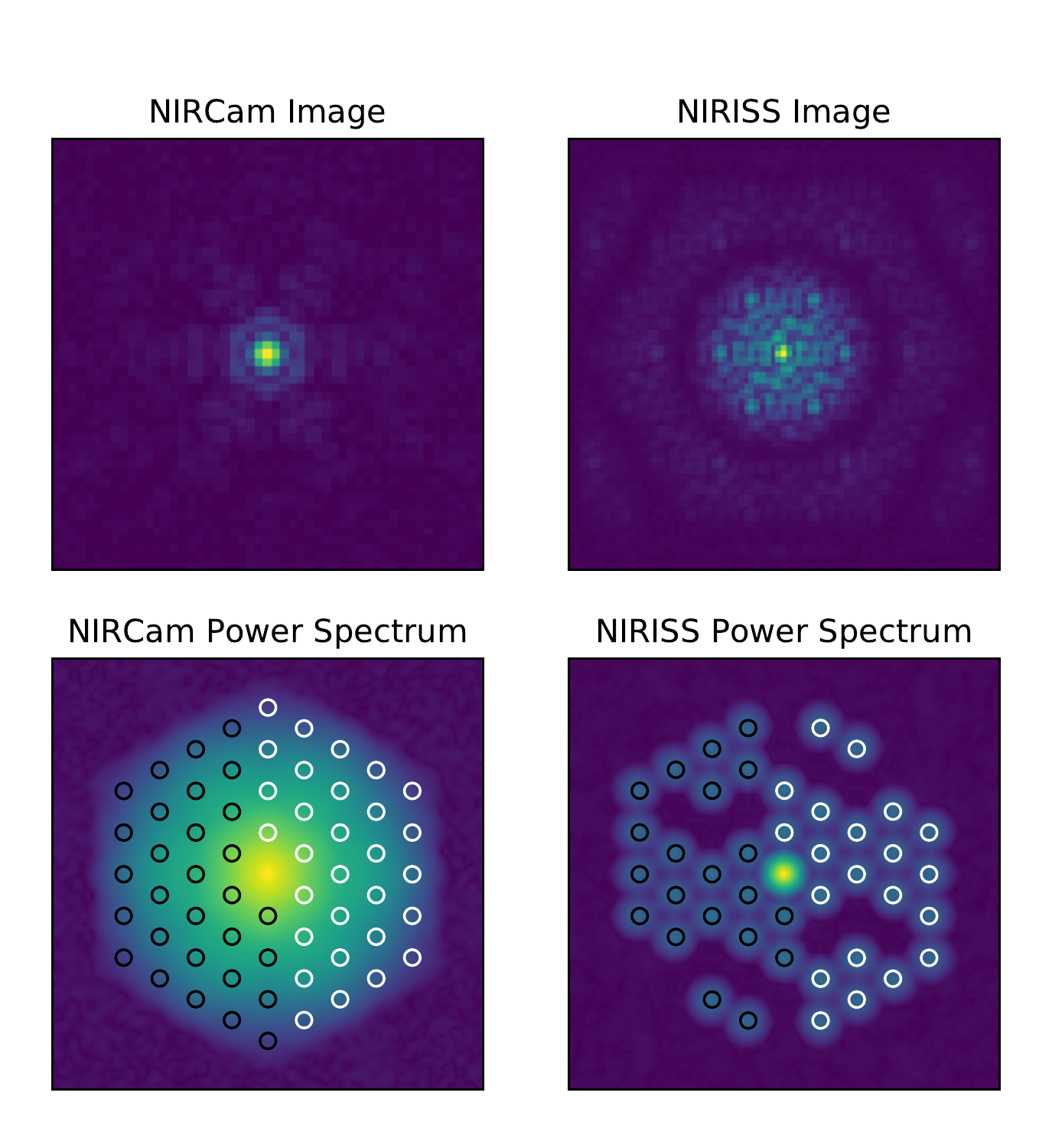}
\end{tabular}
\end{center}
\caption[Simulated NIRCam and NIRISS Point Spread Functions and Power Spectra] 
{ \label{fig:exampleframes} 
Simulated NIRCam (left) and NIRISS (right) integrations (top) and power spectra (bottom).
This dataset uses the F430M filters and residual OPD files comparable to those in Figure \ref{fig:jwstopd}. 
The scattered points show the sampling locations for calculating kernel phase; they are shown in two colors since the Fourier transform is symmetric.
}
\end{figure}

For each pointing, we generate science target and calibrator frames using different optical path difference maps to simulate PSF evolution.
We begin by randomly choosing one of ten optical path difference maps included in WebbPSF.
We then fit a hexike basis\cite{2004OptL...29.2840U} to each mirror segment, including up to 100 coefficients.
We evolve each of the hexike coefficients ($C_n$) by a factor drawn from a one-mean uniform distribution with a width ($2h$) tuned to result in an rms residual WFEs of $\sim10$ nm (see Equation \ref{eq:ev}).
This is consistent with the thermal evolution expected for \emph{JWST} over $\sim$hour long timescales\cite{2018SPIE10698E..09P}.
\begin{equation}
C_{n,seg,cal} = \mathrm{Unif}\left(1-h,1+h\right) C_{n,seg,targ}.
\label{eq:ev}
\end{equation}
For both NIRCam and NIRISS, this corresponds to $h = 0.2$.
Figure \ref{fig:jwstopd} shows an example initial, final, and residual OPD map for a single simulation for both instruments, and Figure \ref{fig:exampleframes} shows example images and power spectra.
Figures \ref{fig:jwsthistsbright} and \ref{fig:jwsthistsfaint} show histograms of the raw and calibrated NIRCam filled-aperture kernel phases, and NIRISS NRM kernel phases for target stars with apparent M$_s$ band magnitudes of 6.3, and 11.3, respectively.

\subsection{Keck OSIRIS}

\begin{figure} [h!]
\begin{center}
\begin{tabular}{c} 
\includegraphics[width=0.6\textwidth]{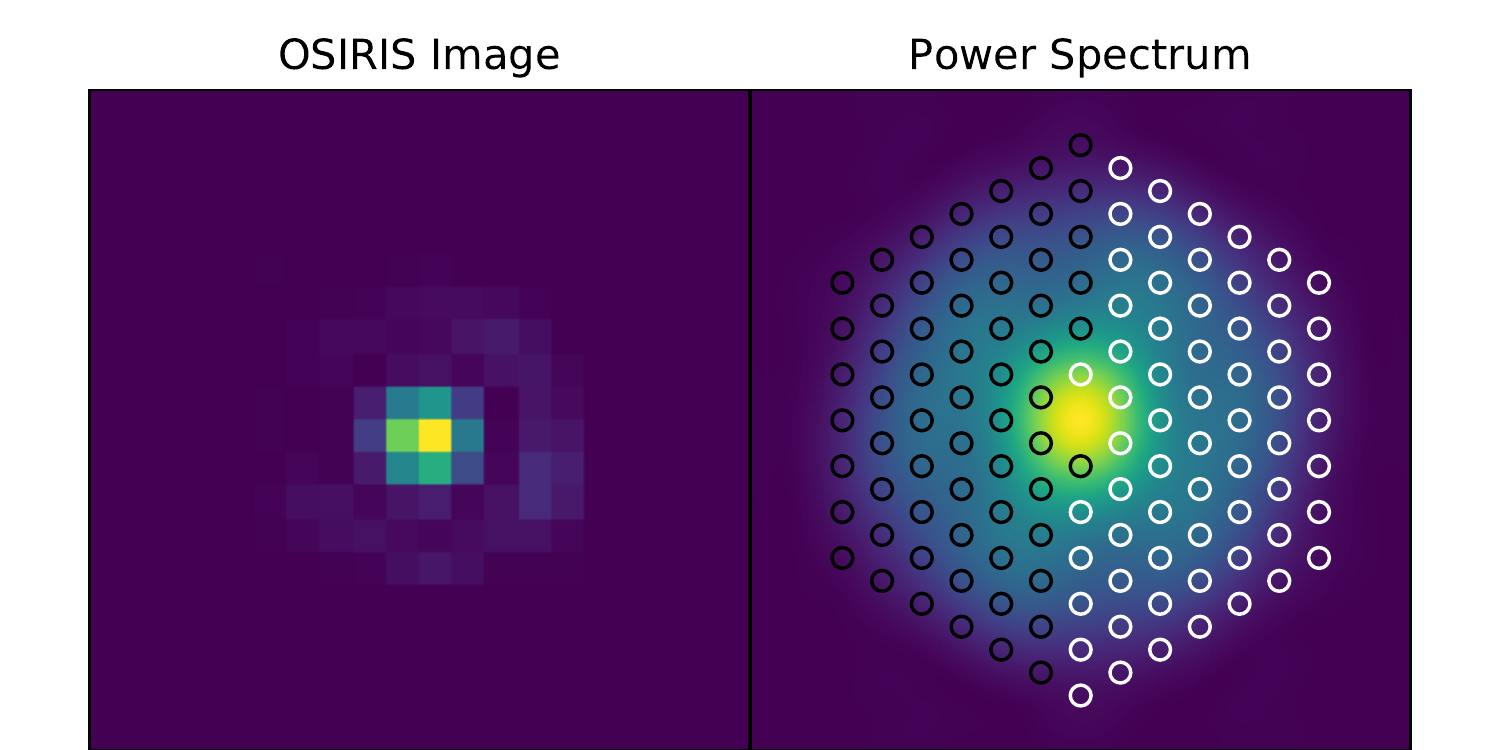}
\end{tabular}
\end{center}
\caption[Simulated OSIRIS Image and Power Spectrum] 
{ \label{fig:osirispsfs} 
Example simulated OSIRIS image (left) and power spectrum (right). 
}
\end{figure}

OSIRIS is a K band integral field spectrograph on Keck.
We simulate OSIRIS kernel phase observations in the Kn3 bandpass, since it covers the accretion-tracing emission line Br-$\gamma$, which would be of interest for observing protoplanet candidates.
This mode contains 433 0.25-nm wavelength bins between 2.121 and 2.229 $\mu$m. 
With one subaperture per mirror segment it yields 45 kernel phases.
We assume 6 hours of available observing time and account for readout, dither, and slew overheads.
We optimize our observing strategy in an identical manner to NIRC2, designing visits that either reach the contrast limit or a length of 3 hours including overheads.
We keep the peak counts per frame under 30,000 (the OSIRIS saturation limit), and use a readout time of 0.829 seconds per frame, a dither overhead of 6 seconds, and a target acquisition overhead of 90 seconds. 
We use an OPD evolution prescription identical to NIRC2, and account for detector, sky, and Poisson noise in the same way as for NIRC2.
Table \ref{tab:osirisnoise} lists the relevant detector and noise parameters for OSIRIS, and Table \ref{tab:osirisobs} lists the number of frames and integration times for each target brightness.
Figure \ref{fig:osirispsfs} shows an example image and power spectrum, and Figure \ref{fig:osirishists} shows kernel phase histograms for targets with K$_s$ band brightnesses of 6 and 11 magnitudes.

\begin{deluxetable}{l c}
\tablecaption{OSIRIS Noise and Detector Parameters\label{tab:osirisnoise}}
\tablecolumns{2}
\tablewidth{0pt}
\tablehead{}
\startdata
$\lambda_c$\tablenotemark{a} $~$ $\mu$m& 2.175\\
$\lambda/D$\tablenotemark{b} $~$ (pix) &  2.24 \\
S\tablenotemark{c} $~$ (mag/arcsec$^2$) & 11.2\\
ZP\tablenotemark{d} $~$ (mag) & 25.1 \\
DC\tablenotemark{e} $~$ (e$\mathrm{^-}$ / pix / s)\hspace{110pt} & 0.035 \\
RN\tablenotemark{f} $~$ (e$\mathrm{^-}$) & 11.0 \\
WD\tablenotemark{g} $~$ (DN) & 33,000 \\
g\tablenotemark{h} $~$ (e$\mathrm{^-}$/DN) & 4.35
\enddata
\tablenotetext{a}{Central wavelength in microns}
\tablenotetext{b}{Detector PSF sampling in pixels}
\tablenotetext{c}{Sky background in magnitudes per square arcsecond}
\tablenotetext{d}{Zero point in magnitudes for Strehl = 1.0}
\tablenotetext{e}{Dark current in electrons per pixel per second}
\tablenotetext{f}{Read noise in electrons per pixel}
\tablenotetext{g}{Well depth in counts}
\tablenotetext{h}{Gain in electrons per count}
\end{deluxetable}

\subsection{\emph{JWST} NIRSpec}

We use the Pandeia engine and WebbPSF software to generate simulated data for NIRSpec IFU observations.
We simulate observations for stars with target magnitudes between $\sim 6$ and 13. 
We use the G395H grating and the F290LP filter, which yield $\gtrsim4000$ wavelength bins from 2.87 to 5.27 microns. 
We follow the same methods for OPD evolution and calibration as for NIRCam / NIRISS, evolving the hexike coefficients to achieve 10 nm rms residual OPD ($h=0.11$ in Equation \ref{eq:ev}).
We use the NRSRAPID readout mode.

Since the NIRSpec spaxels are 0.1 arcseconds across they do not Nyquist sample the PSF.
We thus generate a nine-point small grid dither for each target observation, with the goal of reconstructing a Nyquist sampled image.
To ensure adequate signal to noise for each dither position, we allocate 3 hours for each target and calibrator observation and assume a single observing roll angle.
As we did for NIRCam and NIRISS, we find the maximum number of groups ($n_g$) that can be included in an integration before saturation.
We then split the observations up into nine sets (one for each dither) of $n_{int}$ integrations, each of which has an associated frame-time overhead. 
Table \ref{tab:nirspecdet} lists the relevant detector and noise parameters, and \ref{tab:nirspecobs} lists the observing parameters for each target star magnitude.

\begin{deluxetable}{lc}
\tablecaption{NIRSpec Noise and Detector Parameters\label{tab:nirspecdet}}
\tablecolumns{2}
\tablewidth{0pt}
\tablehead{}
\startdata
$\lambda_c$\tablenotemark{a} $~$ $\mu$m& 4.07\\
$\lambda/D$\tablenotemark{b} $~$ (pix) & 1.29 \\
DC\tablenotemark{c} $~$ (e$\mathrm{^-}$ / pix / s)\hspace{80pt} & 0.0092 \\
RN\tablenotemark{d} $~$ (e$\mathrm{^-}$) & 6 \\
WD\tablenotemark{e} $~$ (DN) & 60,000 \\
g\tablenotemark{f} $~$ (e$\mathrm{^-}$/DN) & $\sim$ 1
\enddata
\tablenotetext{a}{Central wavelength in microns}
\tablenotetext{b}{Detector PSF sampling in pixels}
\tablenotetext{c}{Dark current in electrons per pixel per second}
\tablenotetext{d}{Read noise in electrons per pixel}
\tablenotetext{e}{Well depth in counts}
\tablenotetext{f}{Gain in electrons per count}
\end{deluxetable}

\begin{figure} [h!]
\begin{center}
\begin{tabular}{c}  
\includegraphics[width=0.8\textwidth]{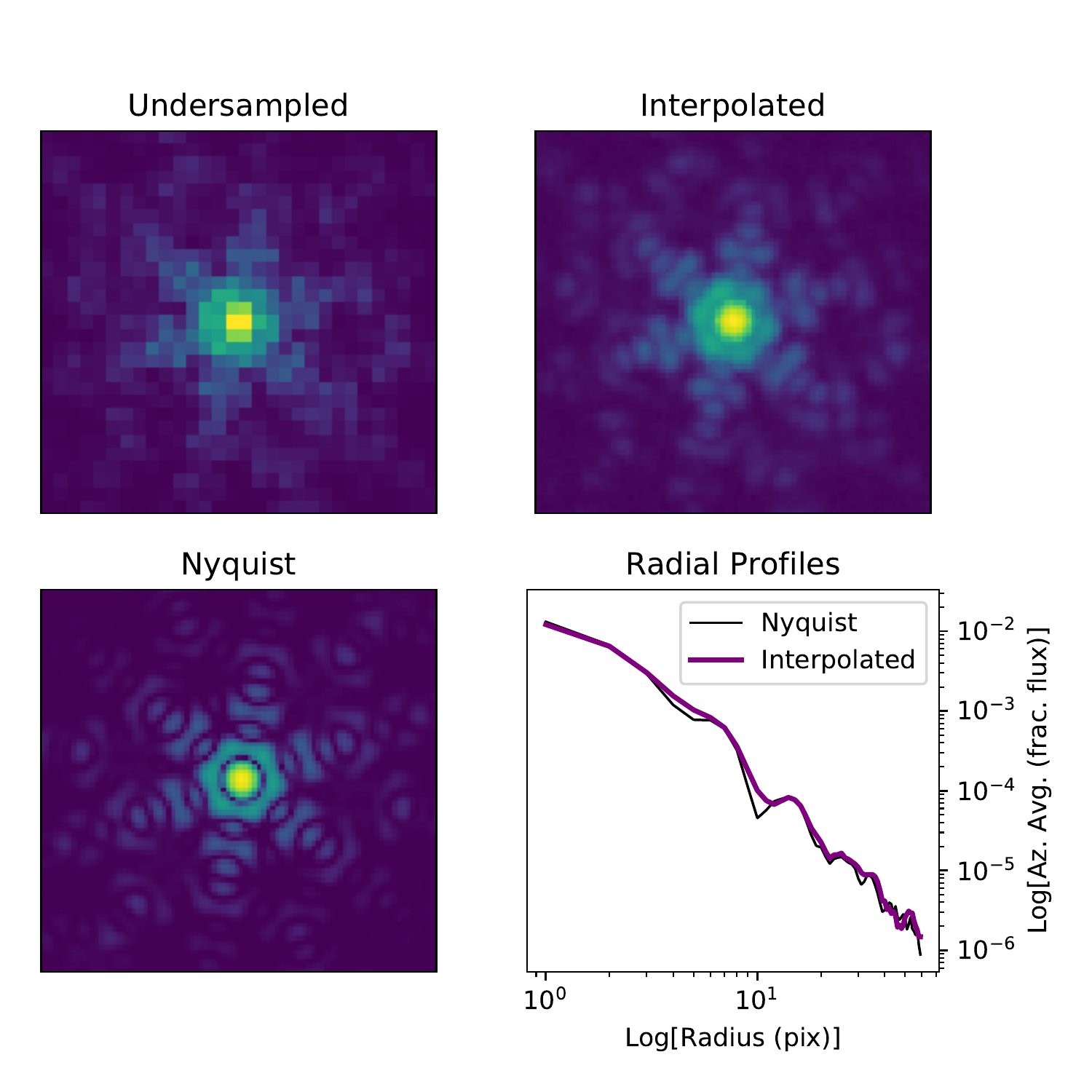}
\end{tabular}
\end{center}
\caption[Simulated NIRSpec Point Spread Functions] 
{ \label{fig:nirspecim} 
Undersampled (top left), and interpolated (top right) simulated NIRSpec images, with a simulated Nyquist sampled PSF (bottom left) for comparison. The bottom right panel shows the radial profiles for the Nyquist sampled (black, thin line), and the interpolated (purple, thick line) PSF. While similar large-scale changes are seen in both, the interpolated image does not have the same small-scale variations as the Nyquist sampled PSF. 
}
\end{figure}

\begin{figure} [h!]
\begin{center}
\begin{tabular}{c} 
\includegraphics[width=0.8\textwidth]{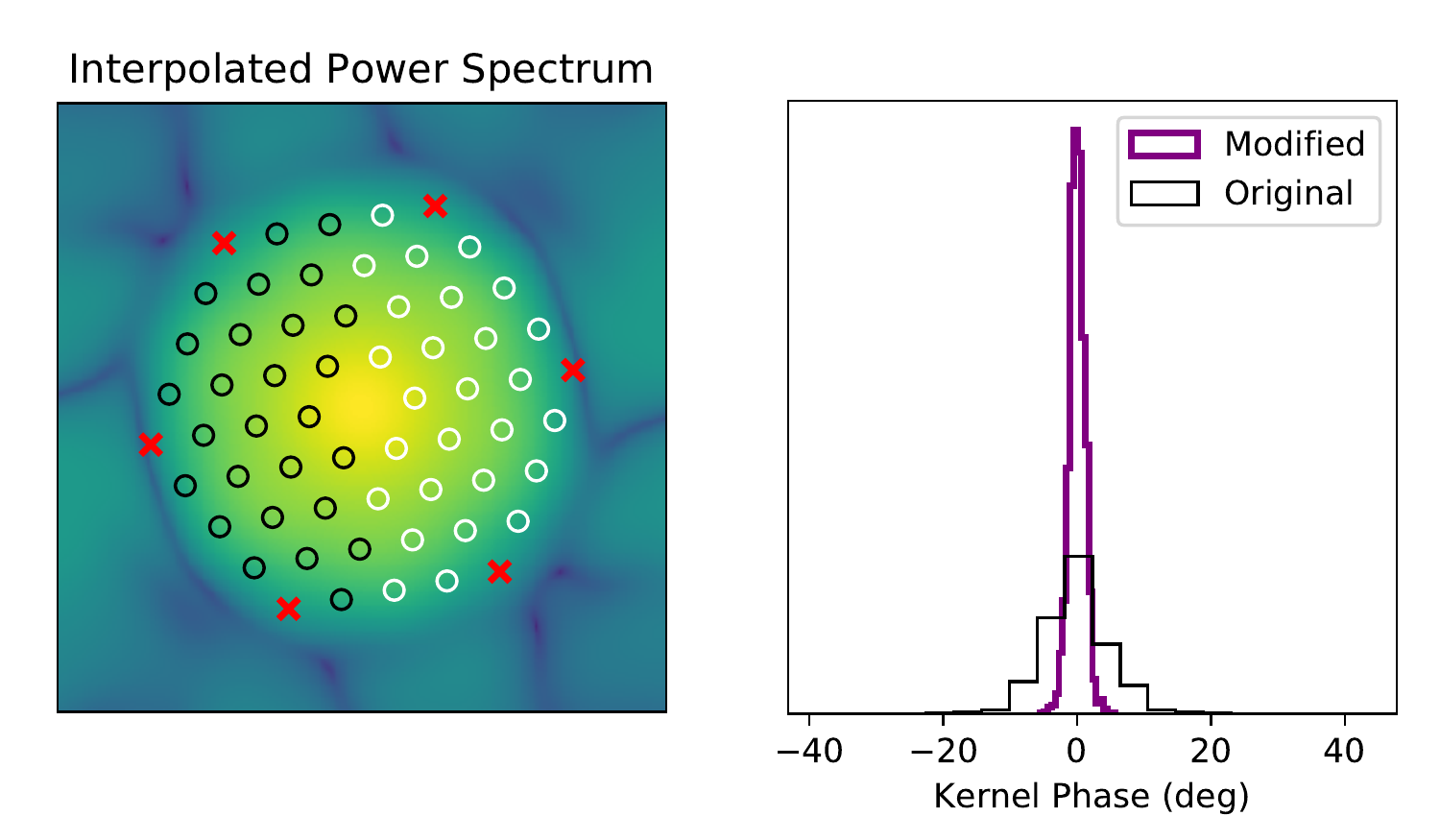}
\end{tabular}
\end{center}
\caption[Simulated NIRSpec Power Spectrum and Kernel Phases] 
{ \label{fig:nirspecpspec} 
Left: Simulated NIRSpec power spectrum of an interpolated Nyquist sampled image, displayed on a log scale to highlight the features at high spatial frequencies. The hollow circles show the (u,v) sampling points we use to calculate kernel phases. Their histogram is shown in purple in the right panel. The red x's show the (u,v) points that would have been used for a perfect Nyquist sampled observation, but that we omit from the kernel phase model to reduce the observed scatter. The black histogram shows the kernel phases calculated for a point source when these (u,v) points are included in the kernel phase model. The standard deviations of the purple and black histograms are 1.1$^\circ$ and 5.2 $^\circ$, respectively.
}
\end{figure}

We use the first nine available pointings in the NIRSpec \textbf{\emph{SMALL}} \textbf{\emph{CYCLING}} pattern to create a simulated small grid dither. 
We assign the flux at each pixel in each undersampled image to the angular coordinates located at the center of that pixel.
We then use a linear interpolation to assign fluxes to the angular coordinates at the centers of the pixels in the Nyquist sampled image (which have sizes of $\sim0.03"$) .
Figure \ref{fig:nirspecim} shows example undersampled and interpolated images, compared to a simulated Nyquist sampled PSF.
This interpolation does not account for the spatial extent of each pixel; using a more sophisticated interpolation scheme may improve the results.
We also note that up to 60 pointings can be used in the \textbf{\emph{CYCLING}} pattern, and having more pointings may improve the reconstructed PSF quality.
Since the PSF reconstruction is imperfect and some signal is lost on the highest spatial frequencies, we do not include the longest baselines in the kernel phase projection model.
This makes the number of observable kernel phases 18.
This modified projection degrades the kernel phase effective resolution, but decreases the noise significantly (see Figure \ref{fig:nirspecpspec}).

\section{RESULTS: CONTRAST CURVES AND PLANET DETECTION LIMITS}\label{sec:results}

\subsection{NIRC2}

\begin{figure} [h!]
\begin{center}
\begin{tabular}{c}  
\includegraphics[width=0.85\textwidth]{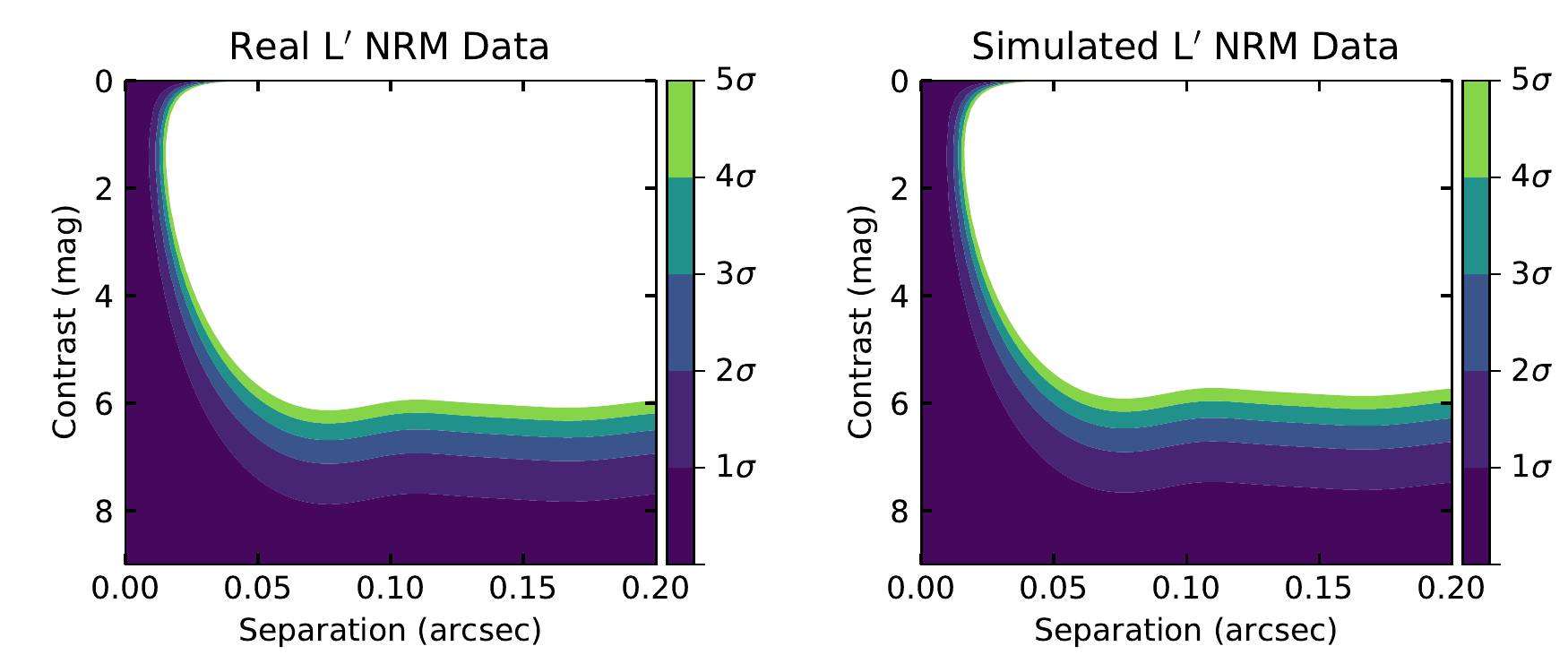}
\end{tabular}
\end{center}
\caption[NIRC2 Observed and Simulated L$'$ NRM Contrast Curves] 
{ \label{fig:ccLpcomp} 
Contrast curves for a single target observation made from real L$'$ NIRC2 NRM observations (left), and simulated observations (right). The contrast curves are nearly identical, with 5$\sigma$ contrast of $>5.5$ magnitudes.}
\end{figure}

As a check, we compare the contrast curves for real Keck NRM observations of PSF calibrators to those for simulated observations with the same target star brightness and exposure parameters. 
The L$'$ target star magnitude is 5.8, and the observation consists of a single cube of 20 20-second images.
The results are shown in Figure \ref{fig:ccLpcomp}; the contrast curves are nearly identical.

\begin{figure} [h!]
\begin{center}
\begin{tabular}{c} 
\includegraphics[width=\textwidth]{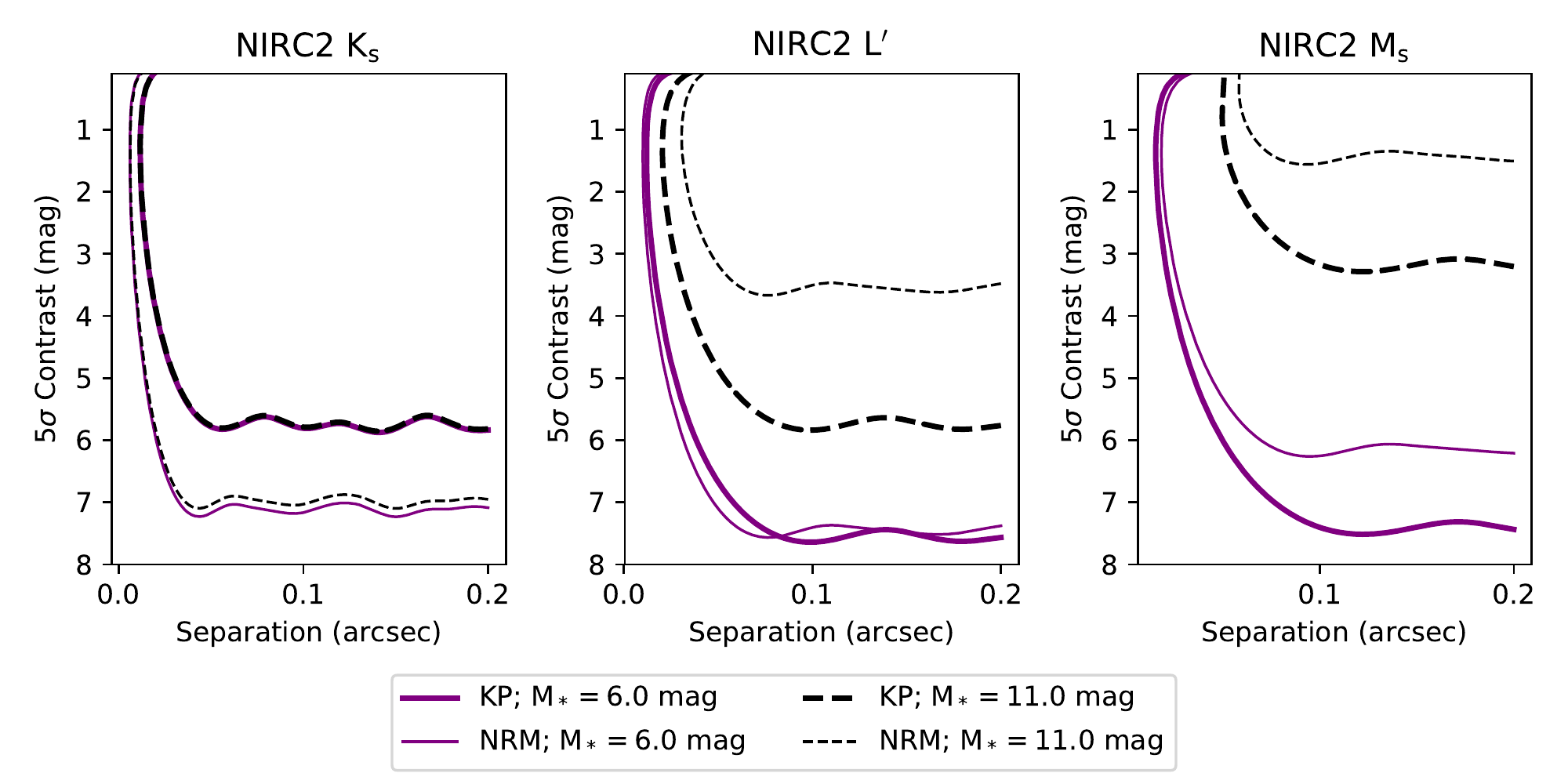}
\end{tabular}
\end{center}
\caption[NIRC2 NRM and Kernel Phase Contrast Curves] 
{ \label{fig:NIRC2cccomp} 
Kernel phase (thick lines) and NRM (thin lines) for $\mathrm{K_s}$ (left), $\mathrm{L'}$ (center), and $\mathrm{M_s}$ (right) bands. Solid purple lines show observations of 6th apparent magnitude stars, while black dashed lines show 11th apparent magnitude stars. 
}
\end{figure}

Figure \ref{fig:NIRC2cccomp} compares the simulated NIRC2 contrast curves for NRM and kernel phase at two representative target brightnesses: apparent magnitudes of 6 and 11.
The 6th magnitude case represents the contrast limited regime for K$\mathrm{_s}$ and L$'$; comparing the NRM and kernel phase contrast curves here shows that kernel phase performance degrades at low Strehl. 
Kernel phase can achieve comparable contrast to NRM in the bright limit at high ($\gtrsim0.85$) Strehl.
This is evident in the observed kernel phase scatter (see Figure \ref{fig:n2histsbright}) - the calibrated scatter is much lower for NRM observations at $\mathrm{K_s}$ band, but they are comparable at $\mathrm{L'}$.
When NRM and kernel phase are both in the contrast limit NRM provides slightly higher contrast very close to the core of the PSF (i.e. within $\sim 80$ mas or $\lambda/D$ for the L$'$ $\mathrm{M_*} = 6.0$ contrast curves). 

The L$'$ contrast curves for the $\mathrm{M_*} = 11$ star, and all of the M$\mathrm{_s}$ contrast curves demonstrate kernel phase and NRM's different behaviors with the same sky brightness.
The NRM PSF is spread over more pixels than the filled-aperture PSF.
Thus, contrast degrades more quickly with an increasing sky brightness for NRM observations.
This is also apparent in the kernel phase histograms (see Figure \ref{fig:n2histsfaint}); the raw scatter in the L$'$ and $\mathrm{M_s}$ observations is larger for NRM than for kernel phase. 
Furthermore, calibration decreases the L$'$ kernel phase scatter more significantly for filled aperture observations than for NRM observations.
This shows that random noise dominates the NRM observations at lower sky background levels compared to filled-aperture observations.

\begin{figure} [h!]
\begin{center}
\begin{tabular}{c}  
\includegraphics[width=\textwidth]{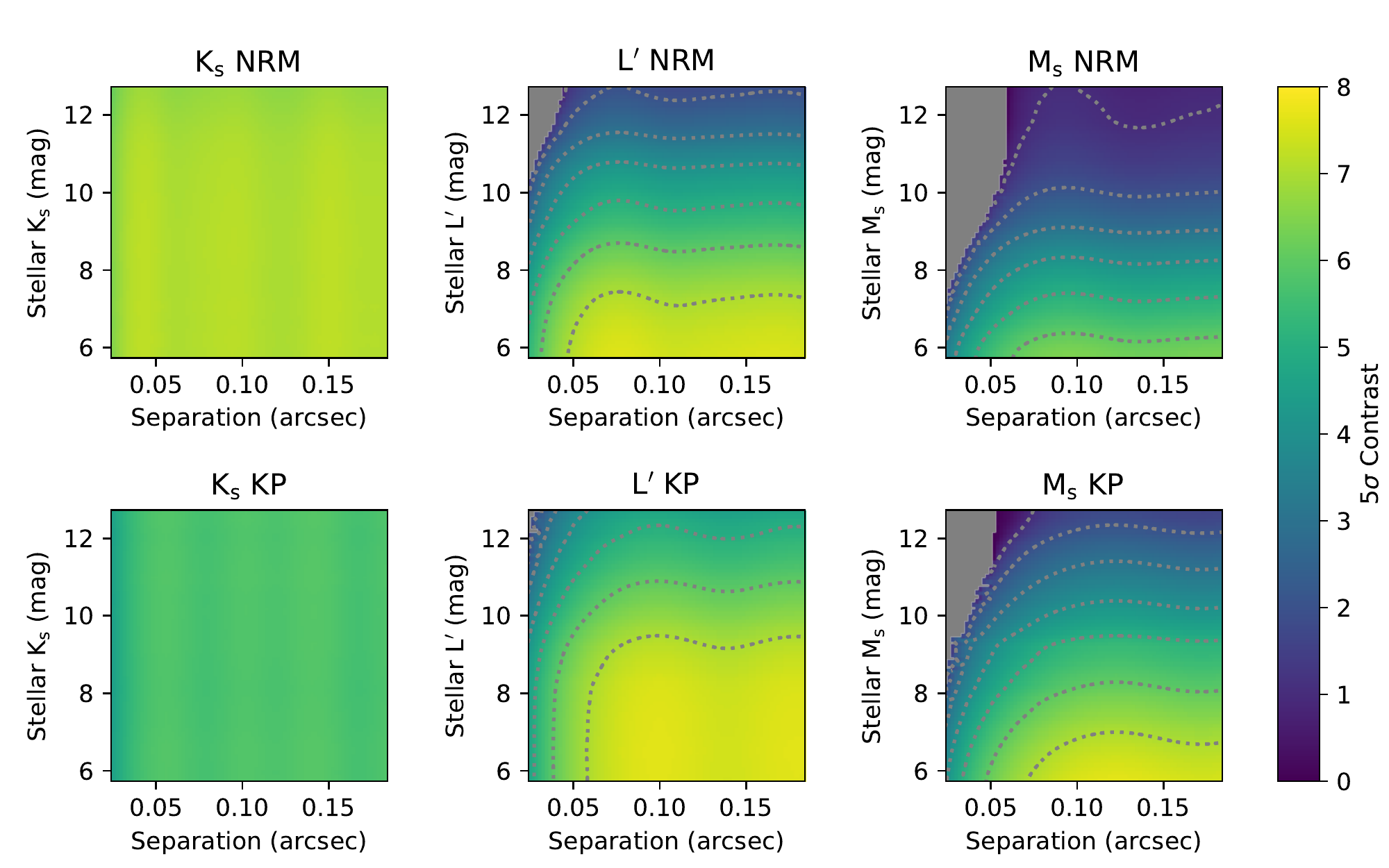}
\end{tabular}
\end{center}
\caption[NIRC2 NRM and Kernel Phase Contrast] 
{ \label{fig:NIRC2ccs} 
The colorscale shows the 5$\sigma$ single companion contrast for NRM (top) and filled-aperture kernel phase observations (bottom), as a function of target star apparent magnitude and angular separation. Left, center, and right panels show $\mathrm{K_s}$, $\mathrm{L'}$, and $\mathrm{M_s}$ bands, respectively.
Grey shaded regions are regions of the parameter space where no companion model was significant at the 5$\sigma$ level, compared to the null (single point source) model.
Grey dotted lines contour 1 magnitude increments in contrast.
}
\end{figure}

The contrast as a function of stellar apparent magnitude supports these points.
Figure \ref{fig:NIRC2ccs} shows this: at $\mathrm{K_s}$ the achievable contrast is constant with stellar brightness, and higher for NRM than for kernel phase. 
At L$'$, the kernel phase observations stay close to the contrast limit until an apparent magnitude of $\mathrm{L'}\sim9$, while NRM degrades more quickly. 
Both techniques degrade quickly at M$\mathrm{_s}$, although kernel phase provides higher achievable contrast.

\begin{figure} [h!]
\begin{center}
\begin{tabular}{c}  
\includegraphics[width=\textwidth]{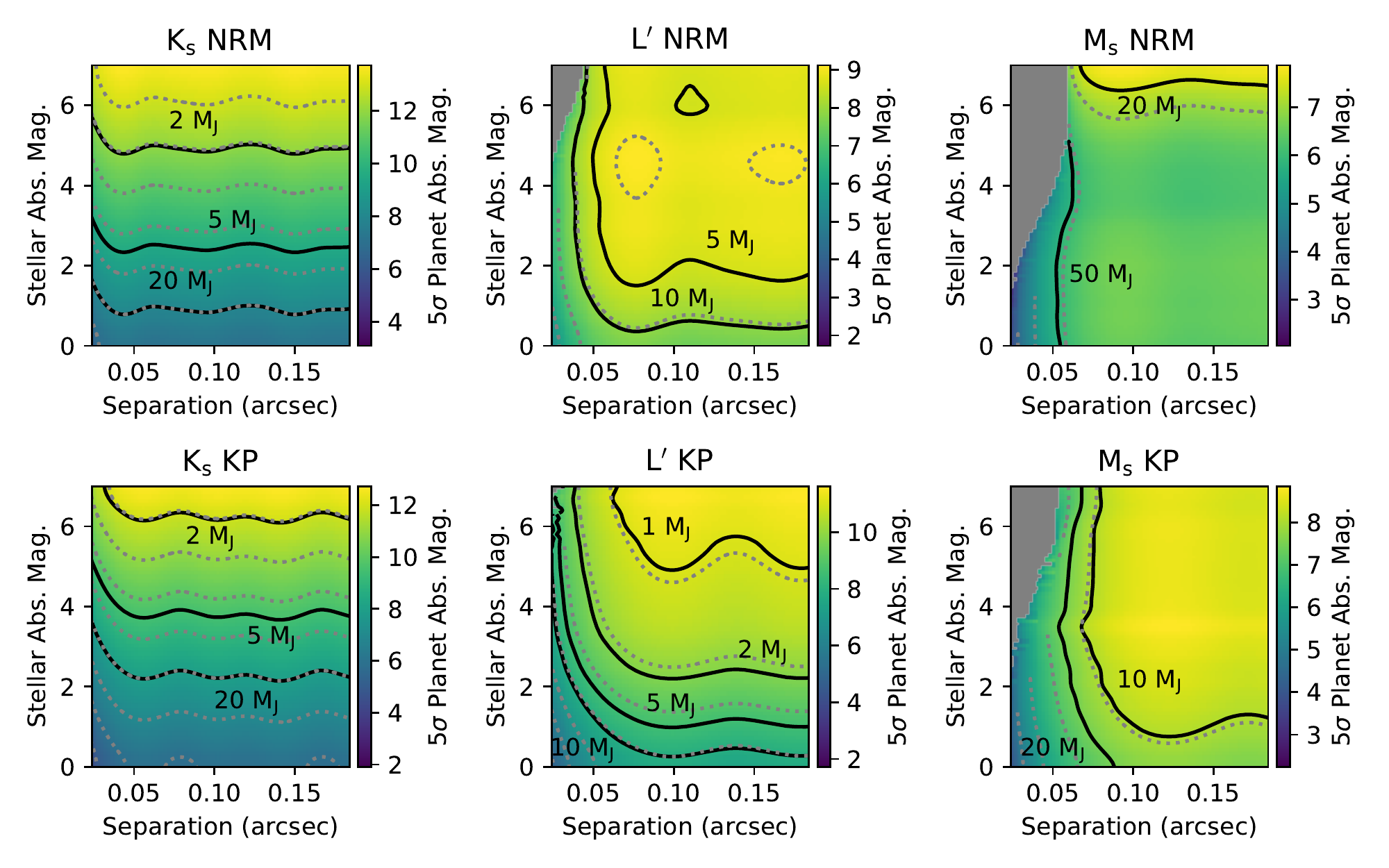}
\end{tabular}
\end{center}
\caption[NIRC2 1 Myr Planet Mass Limits] 
{ \label{fig:nirc2pcomp} 
5$\sigma$ planet mass limits using 1 Myr DUSTY models (solid lines) for all NIRC2 bandpasses and both techniques.
The color scale shows the 5$\sigma$ absolute companion magnitude, calculated using the contrasts shown in Figure \ref{fig:NIRC2ccs} and by translating the stellar apparent magnitudes to absolute magnitudes at 140 pc. 
Grey shaded regions are regions of the parameter space where no companion model was significant at the 5$\sigma$ level, compared to the null (single point source) model.
Grey dotted lines contour 1 magnitude increments in planet absolute magnitude.
The changing noise regimes (OPD error dominated to random error dominated) in the L$'$ and M$\mathrm{_s}$ NRM simulations cause the achievable contrast to fall off non-uniformly as the target brightness changes (see Figure \ref{fig:NIRC2ccs}, upper center and upper right panels). 
This causes local regions in the L$'$ and M$\mathrm{_s}$ stellar magnitude - separation parameter space to be better suited for detecting lower planet masses (see upper center and upper right panels here, colorscale and contours)
}
\end{figure}

\begin{figure} [h!]
\begin{center}
\begin{tabular}{c}  
\includegraphics[width=\textwidth]{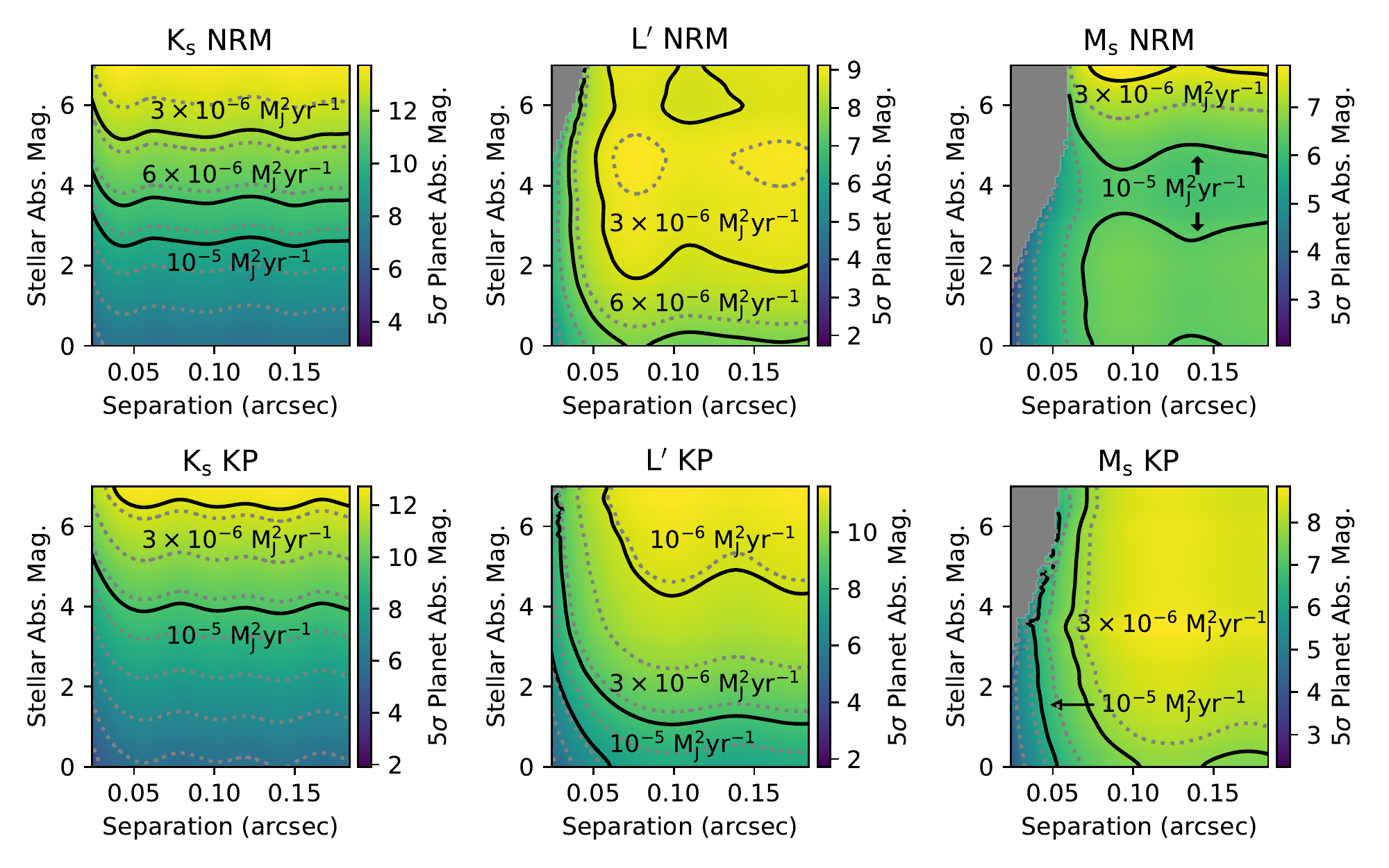}
\end{tabular}
\end{center}
\caption[NIRC2 Circumplanetary Accretion Disk Limits] 
{ \label{fig:nirc2acccomp} 
5$\sigma$ circumplanetary accretion disk (planet mass times accretion rate) limits (solid lines) for all NIRC2 bandpasses and both techniques.
The color scale shows the 5$\sigma$ absolute companion magnitude, calculated using the contrasts shown in Figure \ref{fig:NIRC2ccs} and by translating the stellar apparent magnitudes to absolute magnitudes at 140 pc. 
Grey shaded regions are regions of the parameter space where no companion model was significant at the 5$\sigma$ level, compared to the null (single point source) model.
Grey dotted lines contour 1 magnitude increments in planet absolute magnitude.
The changing noise regimes (OPD error dominated to random error dominated) in the L$'$ and M$\mathrm{_s}$ NRM simulations cause the achievable contrast to fall off non-uniformly as the target brightness changes (see Figure \ref{fig:NIRC2ccs}, upper center and upper right panels). 
This causes local regions in the L$'$ and M$\mathrm{_s}$ stellar magnitude - separation parameter space to be better suited for detecting lower planet masses times accretion rates (see upper center and upper right panels here, colorscale and contours). 
}
\end{figure}

We convert our 5$\sigma$ contrast curves to hot-start planet mass limits using DUSTY models \cite{2003A&A...402..701B}, and to planet mass times accretion rate limits using circumplanetary accretion disk models \cite{2015ApJ...799...16Z,2015ApJ...803L...4E}.
For the DUSTY models, we assume an age of $\sim1$ Myr, since the apparent magnitude range we explore corresponds to the absolute magnitudes of $\sim2$ Myr old stars with masses $\lesssim5~\mathrm{M_\odot}$\cite{2012MNRAS.427..127B} at the distance of Taurus ($\sim140$ pc\cite{2007ApJ...671.1813T}).
For the circumplanetary accretion disk models, we assume full disks with an inner disk radius of 2 Jupiter radii.\cite{2015ApJ...799...16Z}
Figures \ref{fig:nirc2pcomp} and \ref{fig:nirc2acccomp} show the planet and accretion disk limits, respectively, for a range of separations and stellar absolute magnitudes assuming a distance of 140 pc. 
Depending on the stellar magnitude, both NRM and kernel phase at $\mathrm{K_s}$ and $\mathrm{L'}$ can reach planet masses of a few to a few tens of Jupiter masses, or planet masses times accretion rates of a few times $\sim 10^{-6}$ to $\sim10^{-5}~\mathrm{M_J^2~yr^{-1}}$.

In the bright limit, L$'$ NRM and L$'$ kernel phase are both sensitive to lower planet masses and accretion rates than $\mathrm{K_s}$ NRM.
For fainter stars ($\gtrsim$ 10th apparent magnitude or $\gtrsim$5th absolute magnitude), $\mathrm{K_s}$ NRM outperforms L$'$ NRM due to the higher sky background at L$'$.
However for these stellar brightnesses, L$'$ kernel phase performs best. 
Despite brighter expected fluxes for hot-start planets at M$\mathrm{_s}$, the high sky background prevents the detection of planets less massive than $\sim 10~\mathrm{M_J}$.
We note that these mass limits look worse for older planets, and for models other than hot-start models.
For example, the ``warm-start" absolute magnitudes predicted by Spiegel and Burrows (2012)\cite{2012ApJ...745..174S} are not detectable even for 1 Myr old 10 $\mathrm{M_J}$ planets for either technique or any bandpass.

\subsection{JWST: NIRCam Kernel Phase and NIRISS Aperture Masking Interferometry}

\begin{figure} [h!]
\begin{center}
\begin{tabular}{c} 
\includegraphics[width=0.8\textwidth]{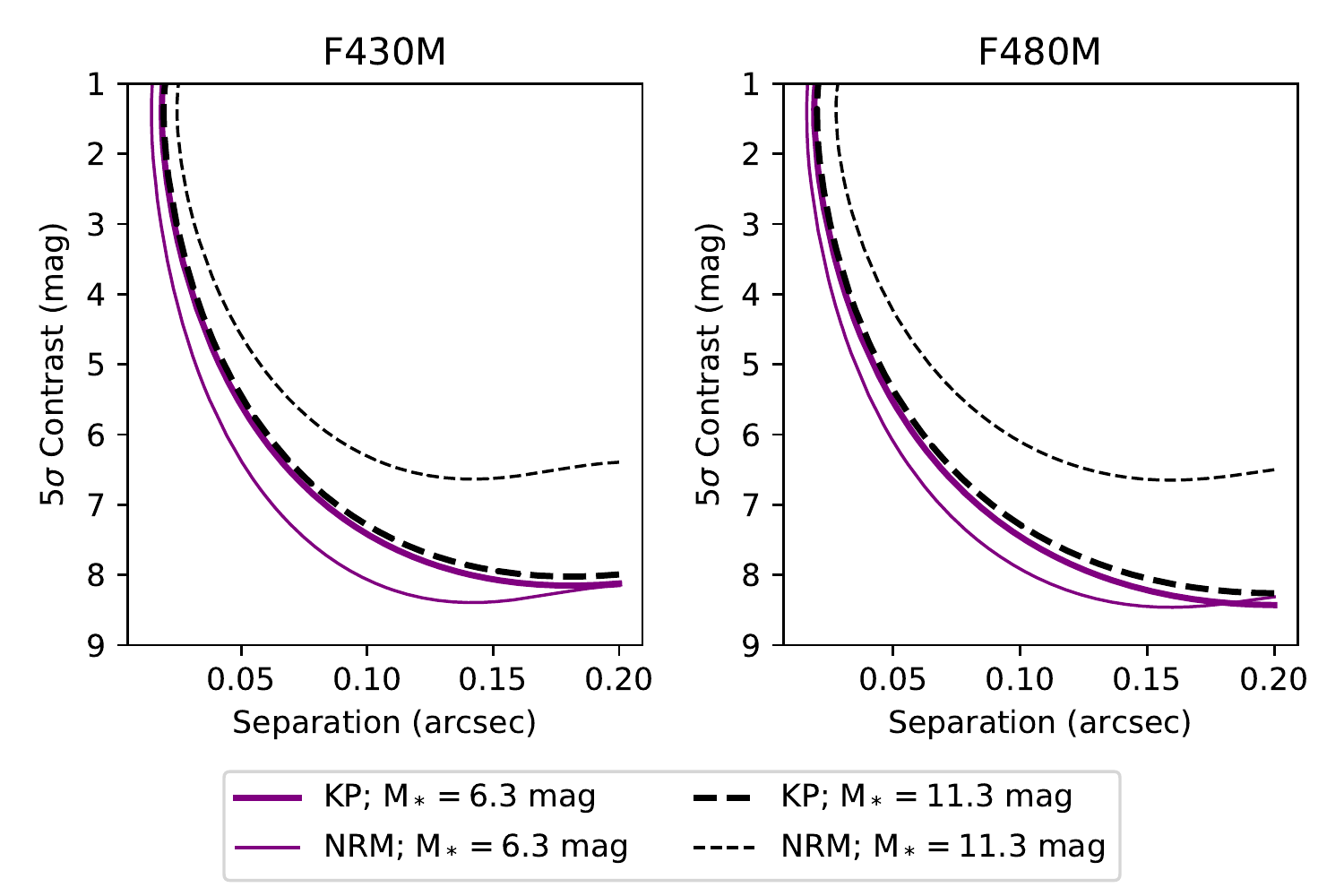}
\end{tabular}
\end{center}
\caption[NIRISS NRM and NIRCam Kernel Phase Contrast Curves] 
{ \label{fig:jwstbbcomp} 
NIRCam kernel phase (thick lines) and NIRISS NRM (thin lines) for F430M (left) and F480M (right) bands. Solid purple lines show observations of 6.3 apparent magnitude stars, while black dashed lines show 11.3 apparent magnitude stars. 
}
\end{figure}

\begin{figure} [h!]
\begin{center}
\begin{tabular}{c}  
\includegraphics[width=0.8\textwidth]{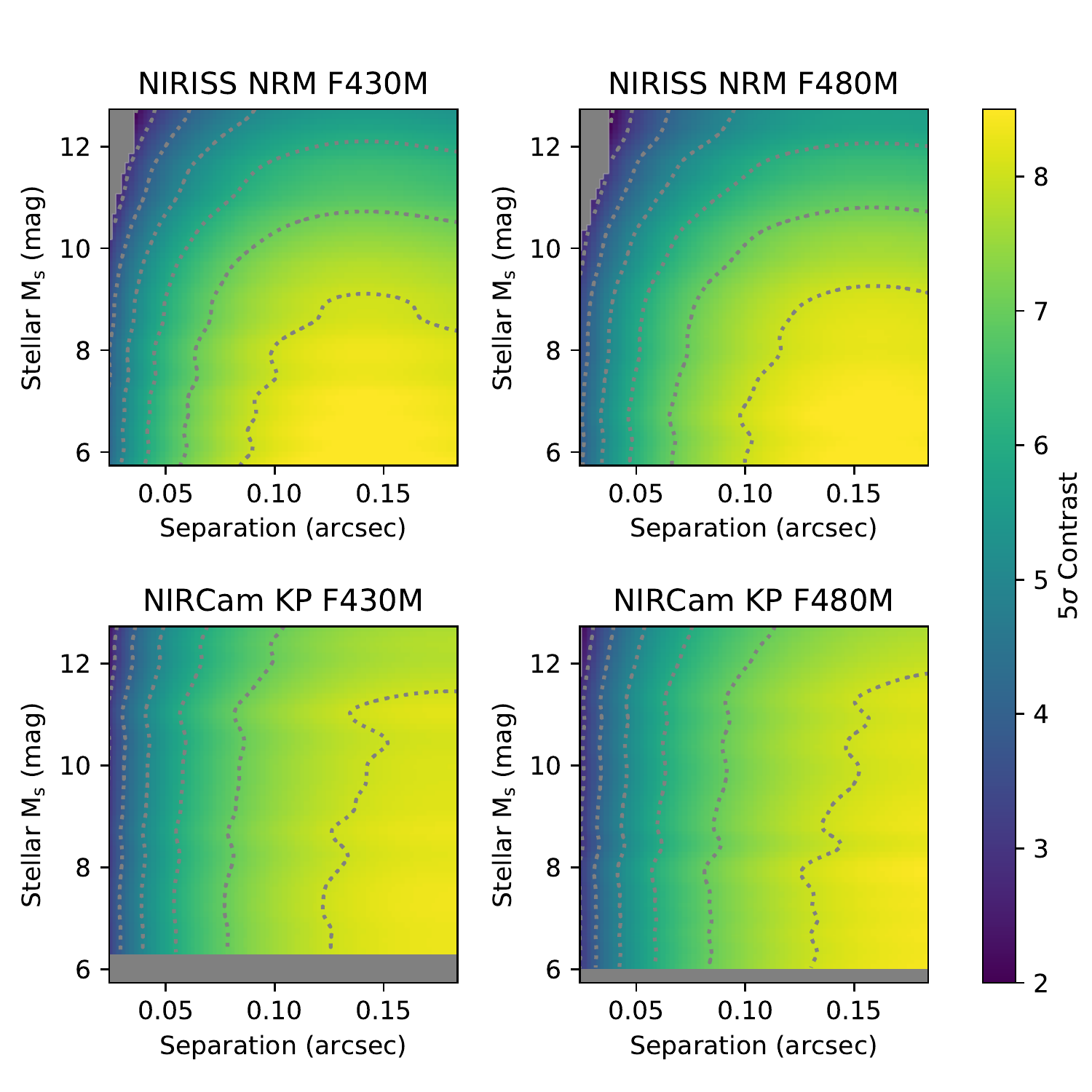}
\end{tabular}
\end{center}
\caption[NIRISS NRM and NIRCam Kernel Phase Contrast] 
{ \label{fig:jwstcontrastcomp} 
The colorscale shows the 5$\sigma$ single companion contrast for NIRISS NRM (top) and NIRCam filled-aperture kernel phase observations (bottom), as a function of target star apparent magnitude and angular separation. 
The grey dotted lines show 1 magnitude increments in companion contrast.
Left and right panels show F430M and F480M bands, respectively.
Grey shaded regions are regions of the parameter space where either no companion model was significant at the 5$\sigma$ level, compared to the null (single point source) model, or regions where the target magnitude was below the bright limit. 
Grey dotted lines contour 1 magnitude increments in contrast.
}
\end{figure}

Figure \ref{fig:jwstbbcomp} shows representative contrast curves for $\mathrm{M_*} = 6.3$ and $\mathrm{M_*} = 11.3$ apparent magnitude stars for NIRCam and NIRISS.
Again, in the bright limit ($\mathrm{M_*} = 6.3$), NRM outperforms kernel phase within $\lambda / D$ ($\sim 150$ mas), by $0.5-1.0$ magnitudes.
At larger separations, kernel phase provides comparable contrast.
The faint $\mathrm{M_*} = 11.3$ mag contrast curves show that at lower signal to noise NIRCam kernel phase can provide higher contrast than NRM on NIRISS. 
Figures \ref{fig:jwsthistsbright} and \ref{fig:jwsthistsfaint} support this.
The raw and calibrated kernel phase scatters for NIRCam and NIRISS are comparable in the bright case, but in the faint case the NIRCam kernel phases calibrate to a lower noise level than the NIRISS kernel phases.
This is apparent in Figure \ref{fig:jwstcontrastcomp}, which shows that NIRISS's contrast falls off as the stellar apparent magnitude becomes greater than $\sim 9$, but NIRCam's contrast does not. 
We also note that for apparent magnitudes less than $\sim6-6.5$, NIRCam is at its saturation limit and thus kernel phase cannot be used.

\begin{figure} [h!]
\begin{center}
\begin{tabular}{c}  
\includegraphics[width=0.9\textwidth]{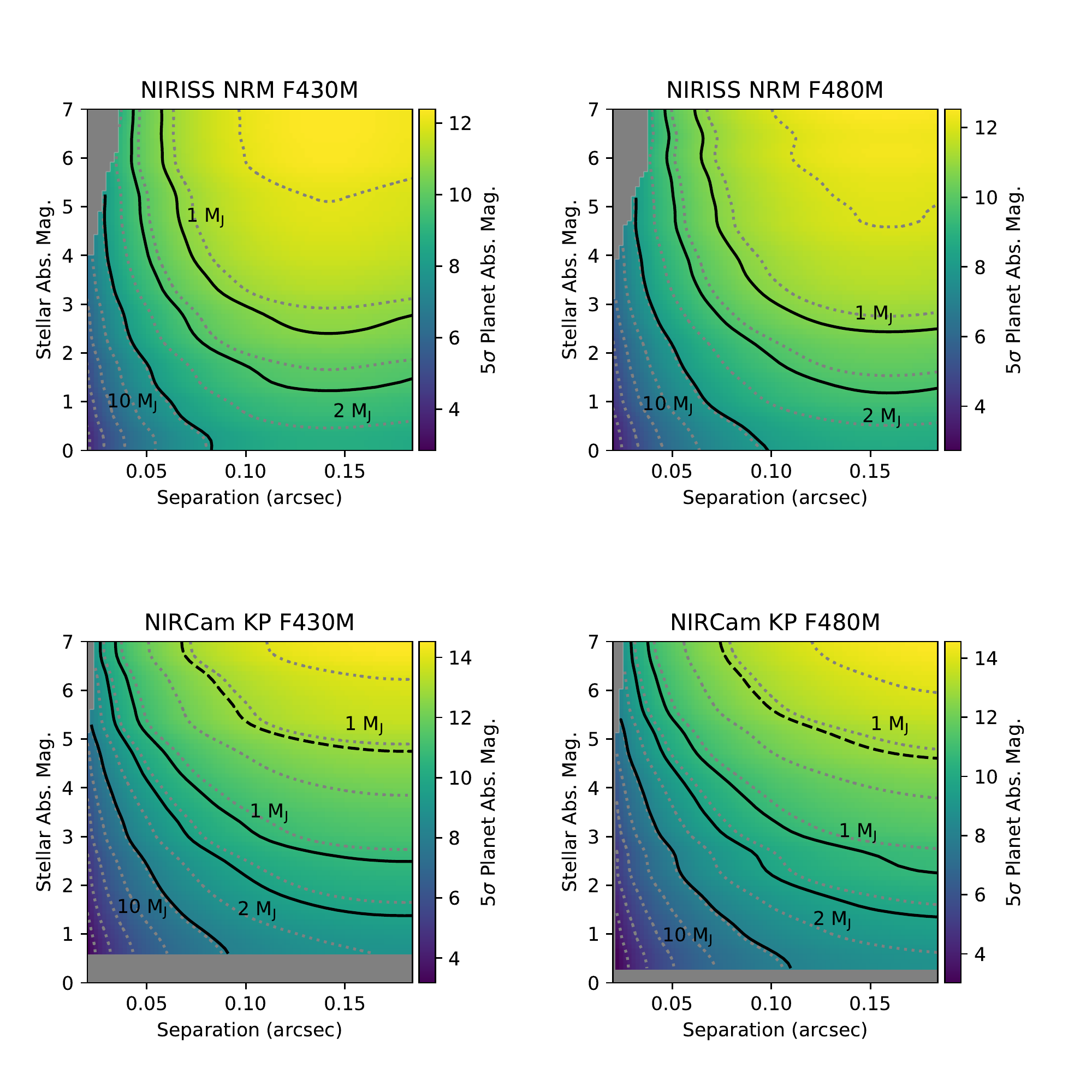}
\end{tabular}
\end{center}
\caption[NIRCam and NIRISS 1 Myr Planet Mass Limits] 
{ \label{fig:jwstpcomp} 
5$\sigma$ planet mass limits using 1 Myr DUSTY models (solid lines) and 1 Myr warm-start Spiegel and Burrows (2012) models (dashed lines) for the F430M and F480M bandpasses and both techniques / instruments.
The color scale shows the 5$\sigma$ absolute companion magnitude, calculated using the contrasts shown in Figure \ref{fig:jwstcontrastcomp} and by translating the stellar apparent magnitudes to absolute magnitudes at 140 pc. 
The grey dotted lines show 1 magnitude increments in planet absolute magnitude.
For clarity, since the 1 Myr warm-start models are tightly clustered in absolute magnitude, we only contour the 1 $\mathrm{M_J}$ model.
Warm-start planets with masses less than $10~\mathrm{M_J}$ are only detectable in the NIRCam observations for high stellar absolute magnitudes.
Grey shaded regions are regions of the parameter space where either no companion model was significant at the 5$\sigma$ level, compared to the null (single point source) model, or regions where the target magnitude was below the bright limit. 
}
\end{figure}

\begin{figure} [h!]
\begin{center}
\begin{tabular}{c}  
\includegraphics[width=0.9\textwidth]{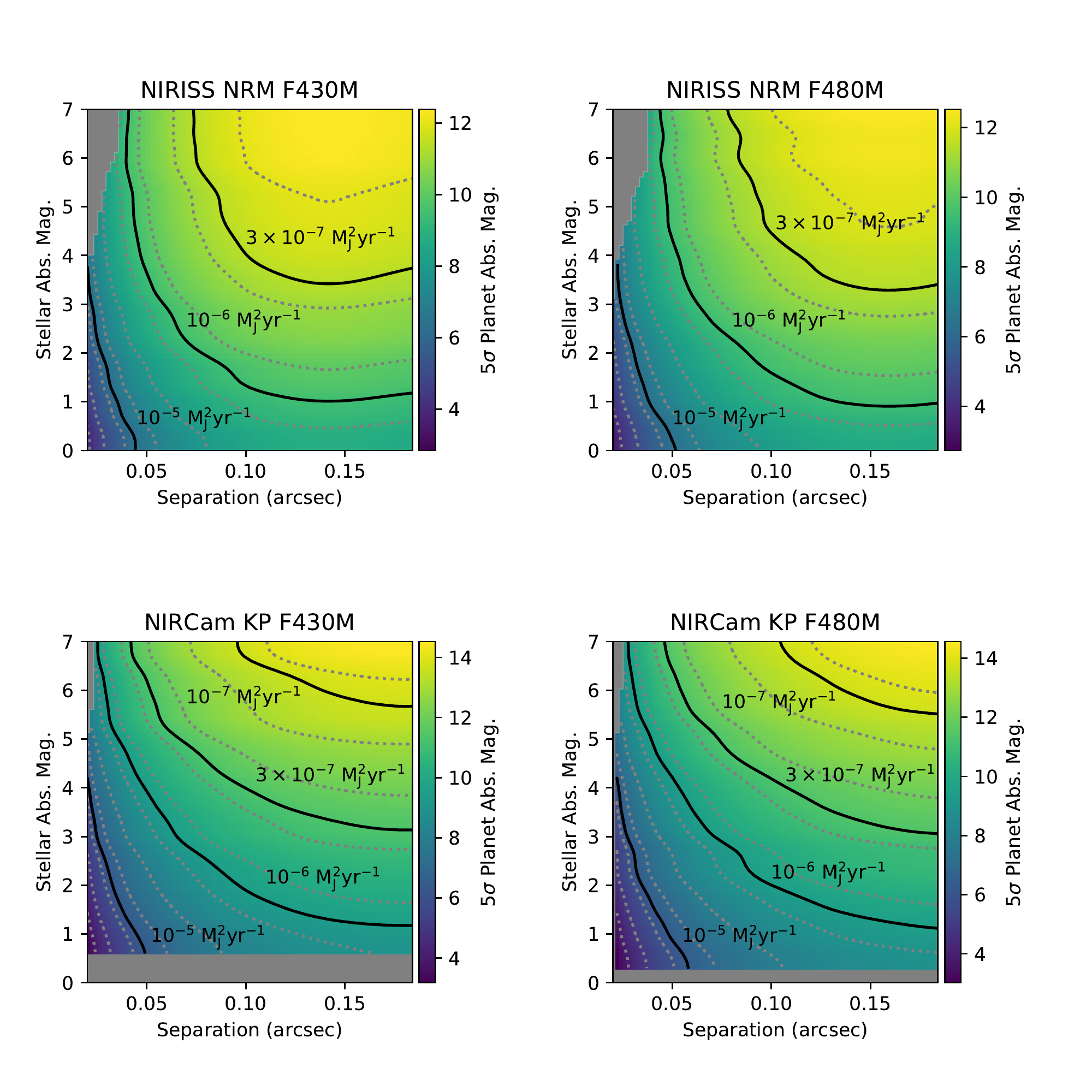}
\end{tabular}
\end{center}
\caption[NIRCam and NIRISS Circumplanetary Accretion Disk Limits] 
{ \label{fig:jwstacccomp} 
5$\sigma$ circumplanetary accretion disk (planet mass times accretion rate) limits (solid lines) for all NIRC2 bandpasses and both techniques.
The color scale shows the 5$\sigma$ absolute companion magnitude, calculated using the contrasts shown in Figure \ref{fig:jwstcontrastcomp} and by translating the stellar apparent magnitudes to absolute magnitudes at 140 pc. 
The grey dotted lines show 1 magnitude increments in planet absolute magnitude.
Grey shaded regions are regions of the parameter space where either no companion model was significant at the 5$\sigma$ level, compared to the null (single point source) model, or regions where the target magnitude was below the bright limit. 
}
\end{figure}

Figures \ref{fig:jwstpcomp} - \ref{fig:jwstacccomp} show the 5$\sigma$ planet and circumplanetary accretion disk limits for a range of stellar absolute magnitudes and angular separations. 
The absolute magnitudes were calculated assuming a distance of 140 pc. 
The greater stability of JWST leads to higher contrast than that achieved with NIRC2, making lower mass planets and lower accretion rates more accessible for brighter stars. 
Planet masses of 1$\mathrm{M_J}$ are detectable for more than half the stellar absolute magnitudes in the hot-start case, and for the high stellar absolute magnitudes in the warm-start case for NIRCam.
The range of detectable planet masses times accretion rates reaches below $10^{-7}~\mathrm{M_J^2~yr^{-1}}$.
Comparing the top and bottom panels of Figures \ref{fig:jwstpcomp} -- \ref{fig:jwstacccomp} shows that, with NIRISS NRM in the bright limit ($\mathrm{M_{abs}} \lesssim 2-3$), slightly lower mass (accretion rate) planets can be detected at smaller angular separations. 
The maximum detectable absolute planet magnitude is higher for NIRCam, as a result of the improved contrast for fainter stars; this is particularly useful for detecting planets that have formed via non-hot-start scenarios.

\subsection{Keck OSIRIS}
\begin{figure} [h!]
\begin{center}
\begin{tabular}{c} 
\includegraphics[width=0.5\textwidth]{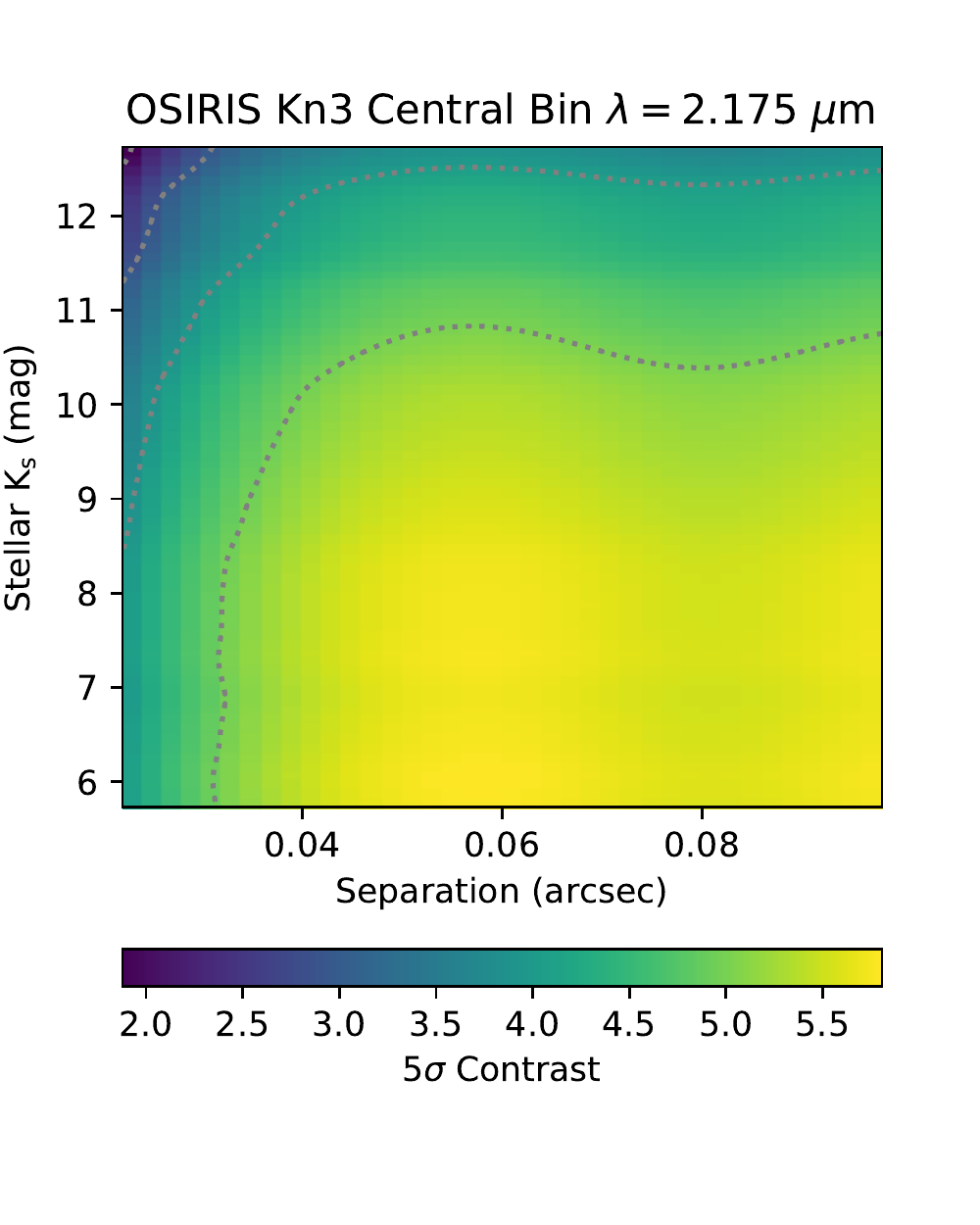}
\end{tabular}
\end{center}
\caption[OSIRIS Kernel Phase Contrast] 
{ \label{fig:osirisccs} 
The colorscale shows the 5$\sigma$ single companion contrast for OSIRIS kernel phase observations as a function of target star apparent magnitude and angular separation.
Grey dotted lines contour 1 magnitude increments in contrast.
}
\end{figure}

Figure \ref{fig:osirisccs} shows the achievable kernel phase contrast for OSIRIS as a function of $\mathrm{K_s}$ stellar apparent magnitude for the central wavelength bin in the Kn3 bandpass ($\lambda = 2.175~\mu$m).
Contrasts of $\sim5$ magnitudes are detectable at the $5\sigma$ level. 
This is lower than the achievable contrast for NIRC2 at broadband $\mathrm{K_s}$, and the calibrated kernel phase scatter is also higher for OSIRIS (see Figure \ref{fig:osirishists}) for target stars with the same brightness.
This may be due to the fact that a single OSIRIS wavelength bin has 1700 times lower throughput than NIRC2: random noise sources from the sky and detector may be a more significant noise source for OSIRIS, degrading contrast.

\begin{figure} [h!]
\begin{center}
\begin{tabular}{c}  
\includegraphics[width=0.5\textwidth]{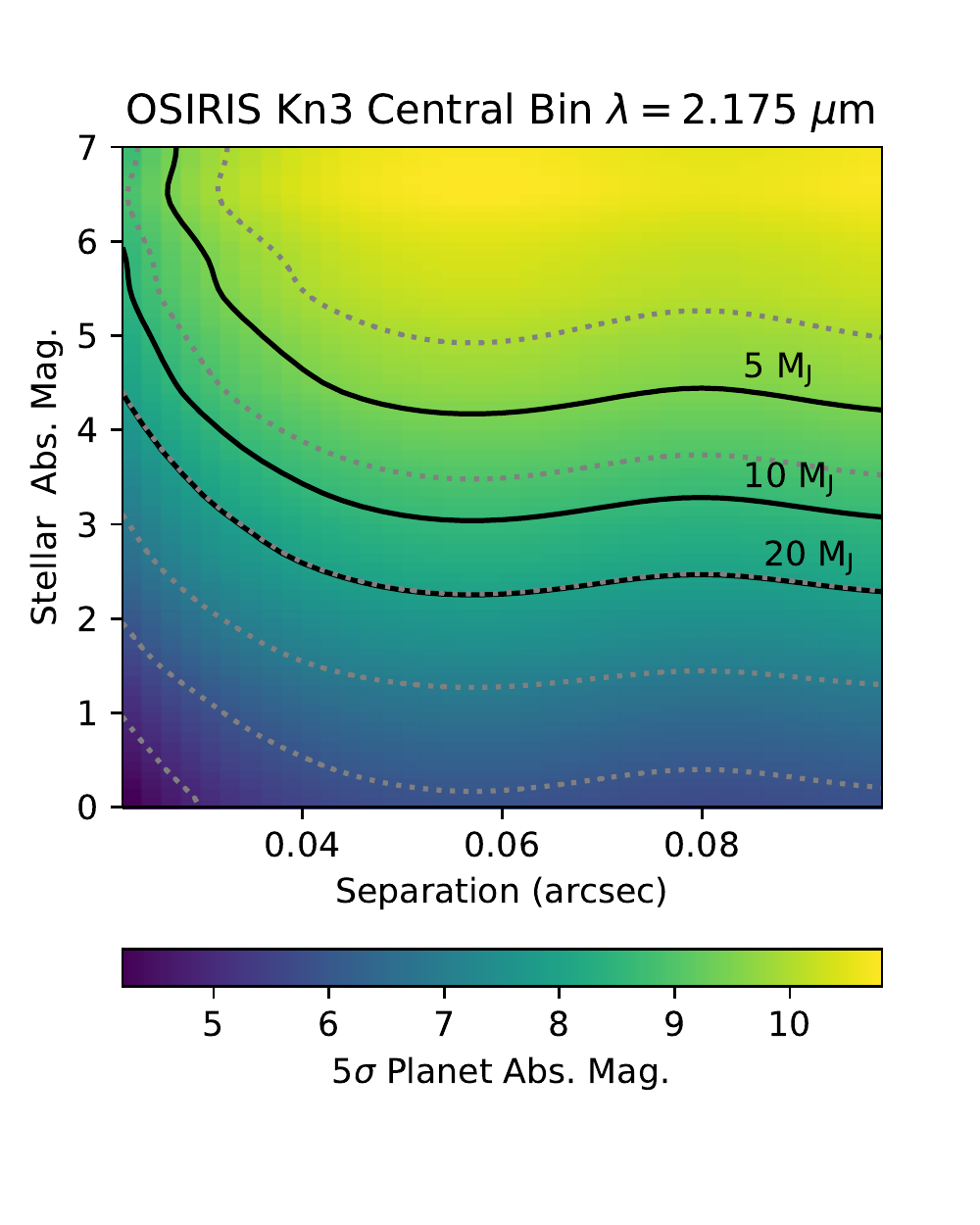}
\end{tabular}
\end{center}
\caption[OSIRIS 1 Myr Planet Mass Limits] 
{ \label{fig:osirisps} 
5$\sigma$ planet mass limits using 1 Myr DUSTY models (solid lines) for OSIRIS kernel phase.
The color scale shows the 5$\sigma$ absolute companion magnitude, calculated using the contrasts shown in Figure \ref{fig:osirisccs} and by translating the stellar apparent magnitudes to absolute magnitudes at 140 pc. 
Grey dotted lines contour 1 magnitude increments in planet absolute magnitude.
}
\end{figure}

\begin{figure} [h!]
\begin{center}
\begin{tabular}{c}  
\includegraphics[width=0.5\textwidth]{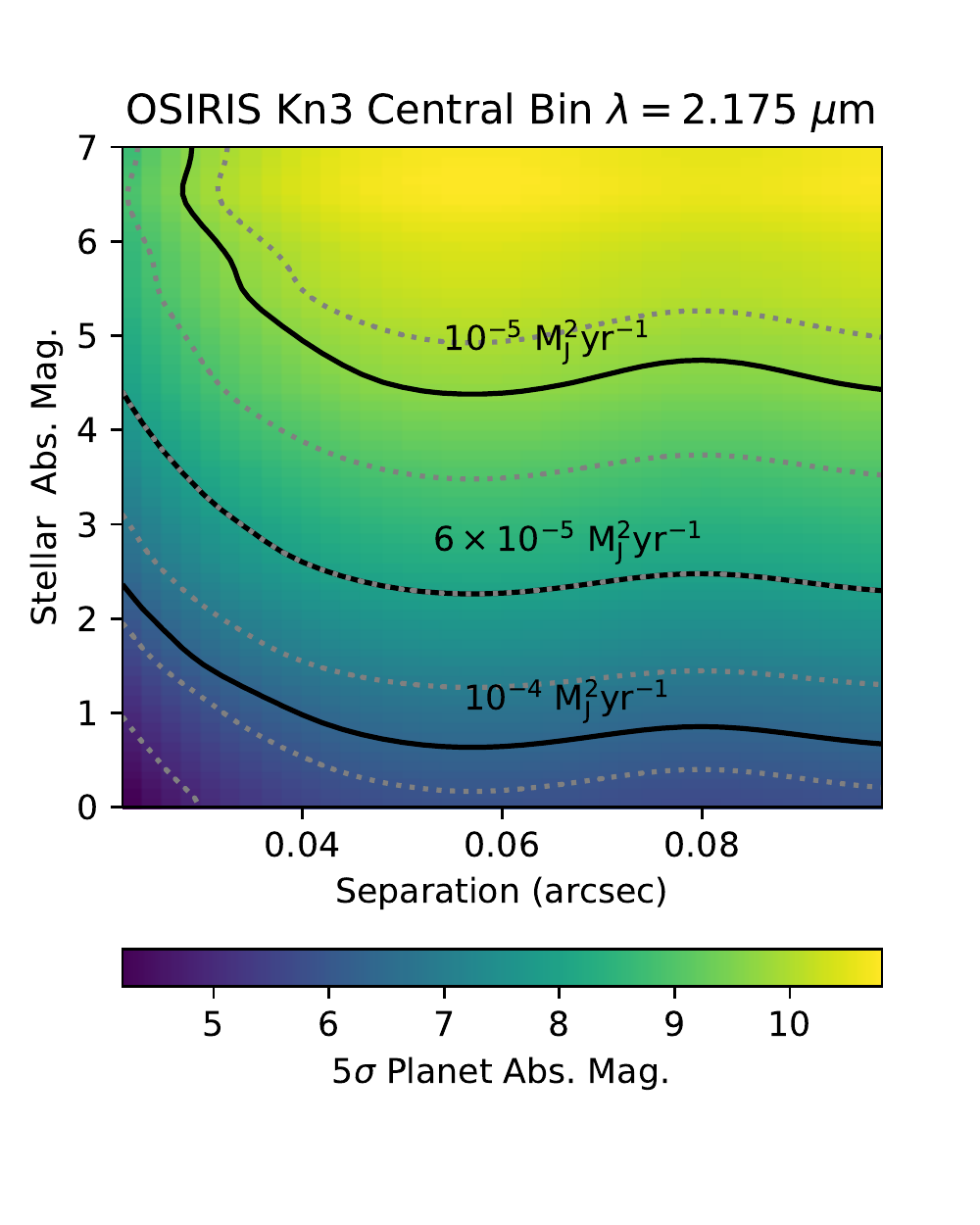}
\end{tabular}
\end{center}
\caption[OSIRIS Circumplanetary Accretion Disk Limits] 
{ \label{fig:osirisacc} 
5$\sigma$ circumplanetary accretion disk (planet mass times accretion rate) limits (solid lines) for OSIRIS.
The color scale shows the 5$\sigma$ absolute companion magnitude, calculated using the contrasts shown in Figure \ref{fig:osirisccs} and by translating the stellar apparent magnitudes to absolute magnitudes at 140 pc. 
Grey dotted lines contour 1 magnitude increments in planet absolute magnitude.
}
\end{figure}

Figures \ref{fig:osirisps} and \ref{fig:osirisacc} translate these contrast limits to planet mass limits as a function of stellar absolute magnitude.
We again assume a distance of 140 pc.
With OSIRIS, planet masses of $5-20~\mathrm{M_J}$ are detectable for stellar absolute magnitudes greater than $\sim2$.
Warm start models with masses less than $10~\mathrm{M_J}$ are not detectable.
Planet masses times accretion rates of $\sim10^{-5}~\mathrm{M_J^2~yr^{-1}}$ are detectable for the fainter stars ($\mathrm{K_s} \sim 10$; $\mathrm{M_{K_s}} \sim 5$). 
We note that these detection limits would be worse for the shortest wavelength bin ($\lambda = 2.121~\mu$m) and better for the longest wavelength bin ($\lambda = 2.229~\mu$m) due to the wavelength dependence of the Strehl ratio.

\subsection{JWST NIRSpec}
\begin{figure} [h!]
\begin{center}
\begin{tabular}{c}  
\includegraphics[width=0.6\textwidth]{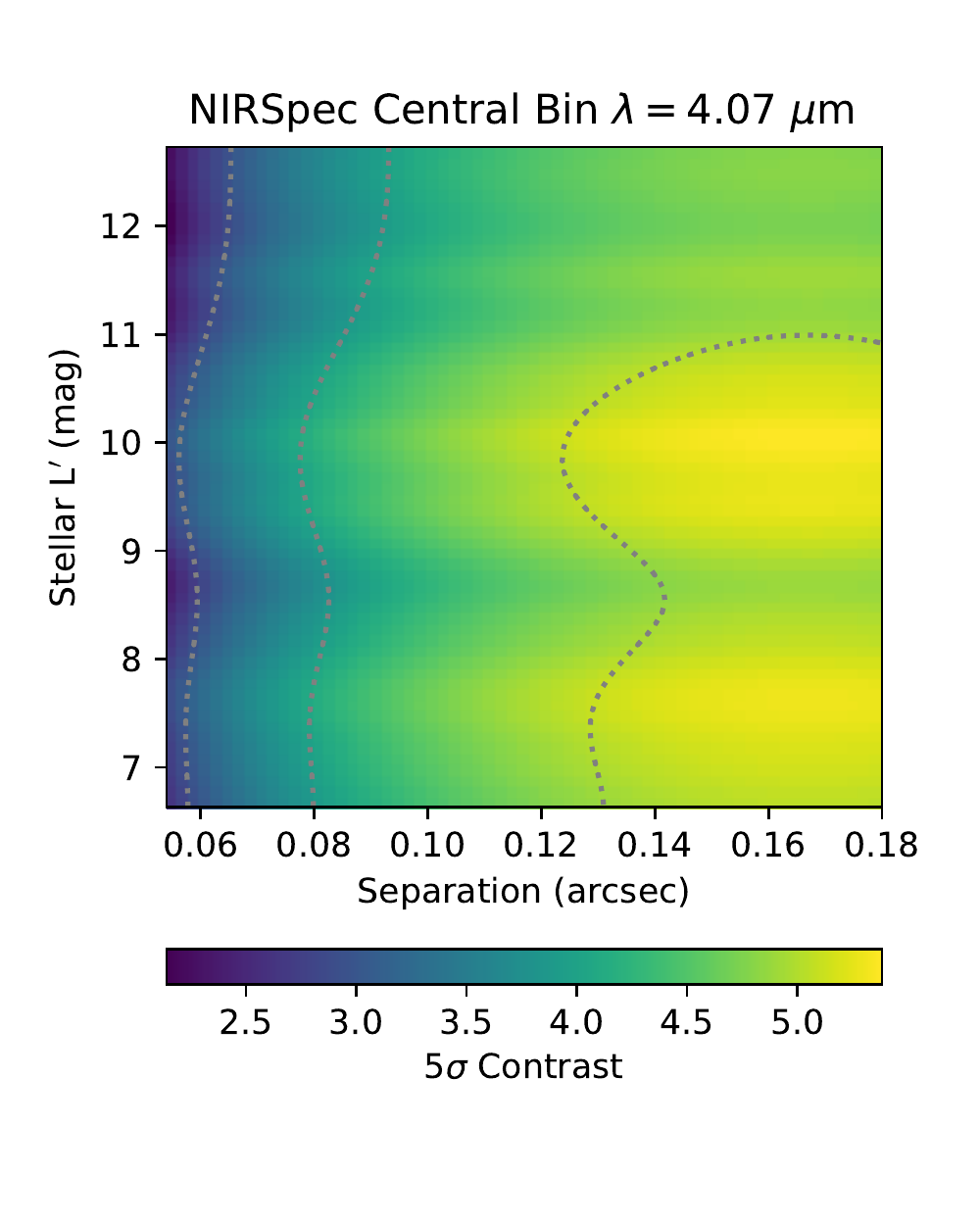}
\end{tabular}
\end{center}
\caption[NIRSpec Kernel Phase Contrast] 
{ \label{fig:nirspecccs} 
The colorscale shows the 5$\sigma$ single companion contrast for NIRSpec kernel phase observations as a function of target star apparent magnitude and angular separation.
Grey dotted lines show 1 magnitude increments in contrast.
}
\end{figure}

\begin{figure} [h!]
\begin{center}
\begin{tabular}{c}  
\includegraphics[width=0.5\textwidth]{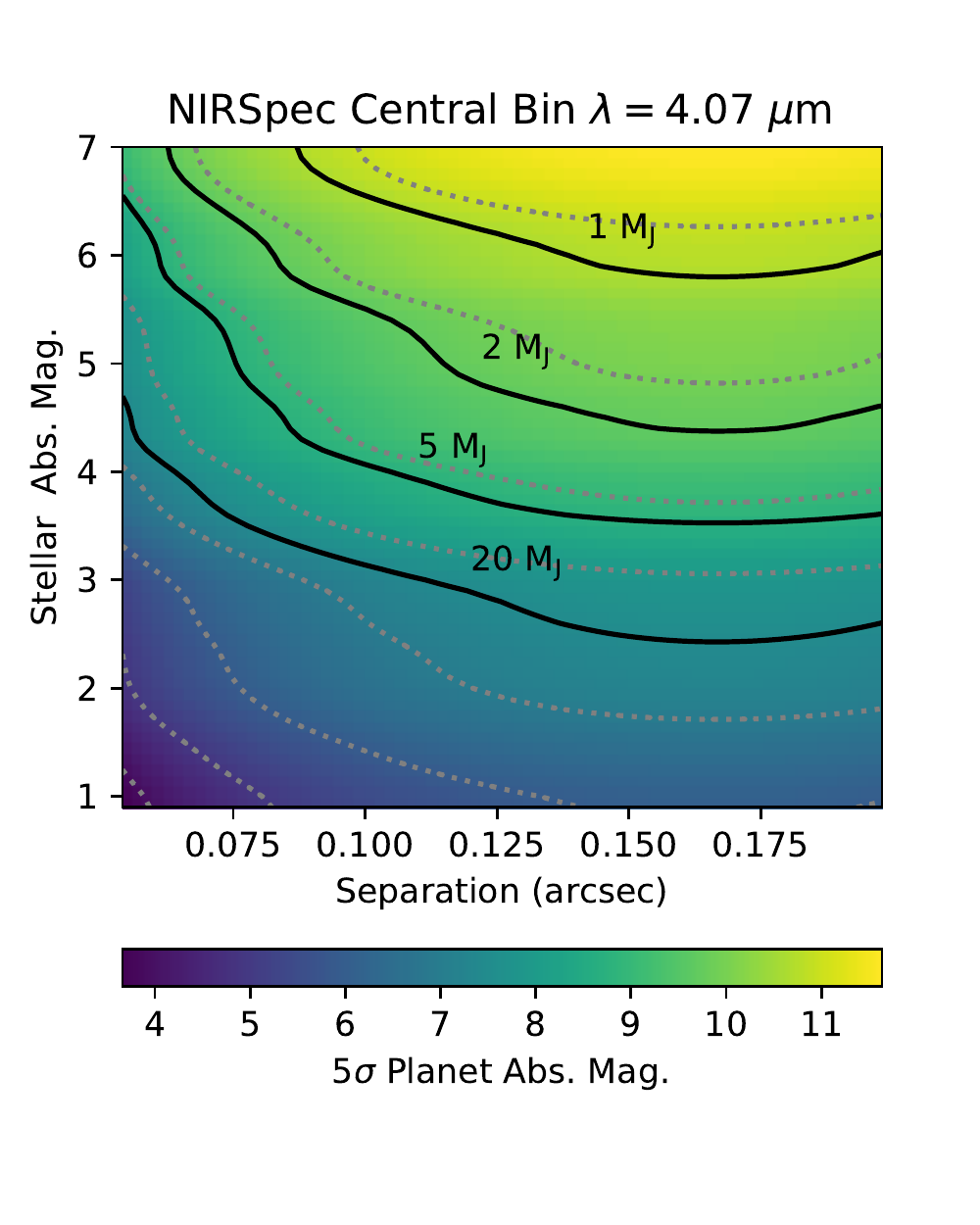}
\end{tabular}
\end{center}
\caption[NIRSpec 1 Myr Planet Mass Limits] 
{ \label{fig:nirspecps} 
5$\sigma$ planet mass limits using 1 Myr DUSTY models (solid lines) for NIRSpec kernel phase.
The color scale shows the 5$\sigma$ absolute companion magnitude, calculated using the contrasts shown in Figure \ref{fig:nirspecccs} and by translating the stellar apparent magnitudes to absolute magnitudes at 140 pc. 
Grey dotted lines show 1 magnitude increments in planet absolute magnitude.
}
\end{figure}

\begin{figure}[h!]
\begin{center}
\begin{tabular}{c}  
\includegraphics[width=0.5\textwidth]{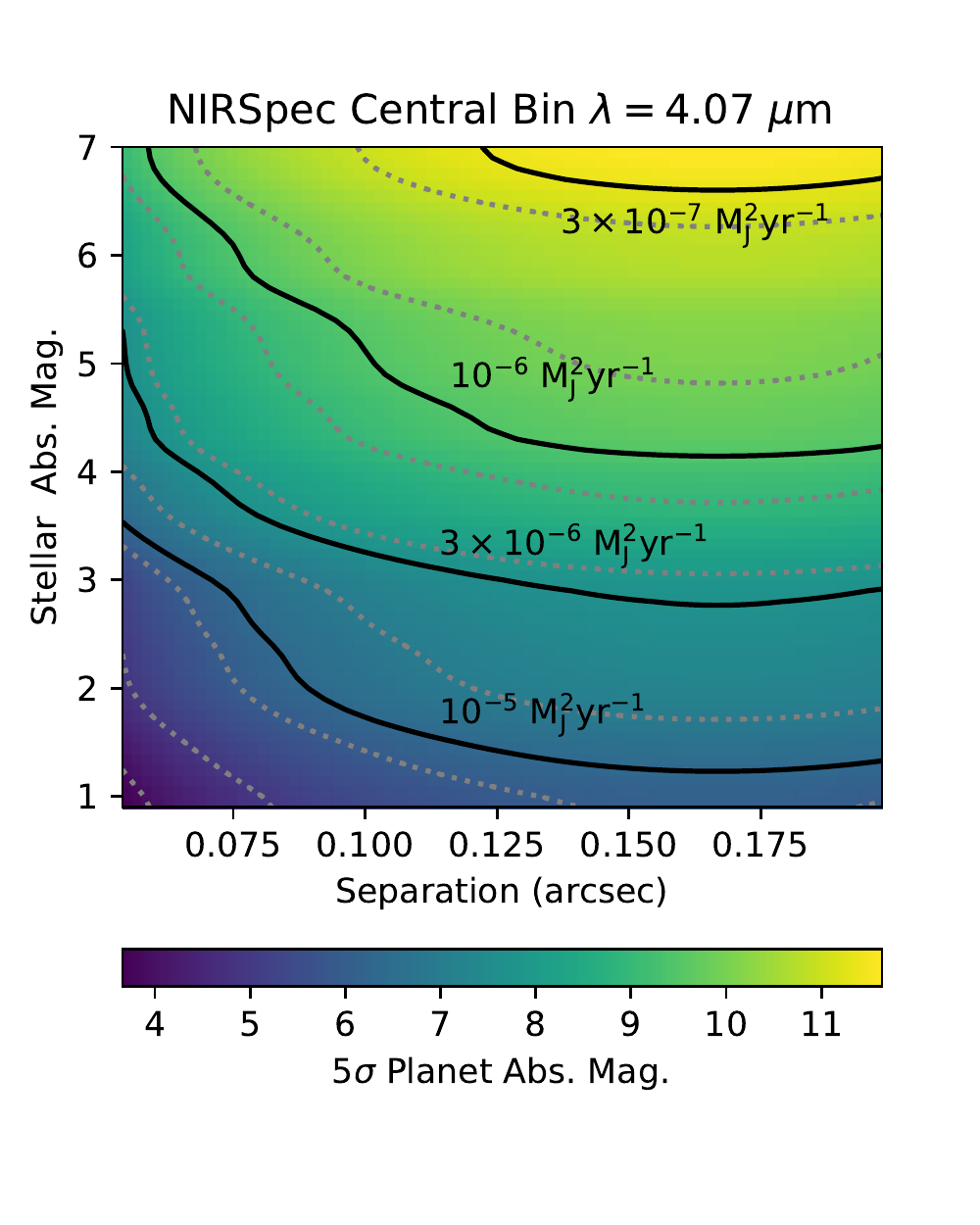}
\end{tabular}
\end{center}
\caption[NIRSpec Circumplanetary Accretion Dsk Limits] 
{ \label{fig:nirspecaccs} 
5$\sigma$ circumplanetary accretion disk (planet mass times accretion rate) limits (solid lines) for NIRSpec kernel phase.
The color scale shows the 5$\sigma$ absolute companion magnitude, calculated using the contrasts shown in Figure \ref{fig:nirspecccs} and by translating the stellar apparent magnitudes to absolute magnitudes at 140 pc. 
Grey dotted lines show 1 magnitude increments in planet absolute magnitude.
}
\end{figure}

Figure \ref{fig:nirspecccs} shows the achievable kernel phase contrast for \emph{JWST} NIRSpec as a function of $\mathrm{L'}$ stellar apparent magnitude for the central wavelength bin ($\lambda = 4.07~\mu$m), and Figure \ref{fig:nirspechists} shows example histograms of the raw and calibrated kernel phases.
Contrasts of $\sim5$ magnitudes are detectable at the $5\sigma$ level for all target star magnitudes.
Figures \ref{fig:nirspecps} and \ref{fig:nirspecaccs} translate these contrast limits to planet mass limits as a function of stellar absolute magnitude, assuming a distance of 140 pc.
With NIRSpec, planet masses of a few $\mathrm{M_J}$ are detectable for stellar absolute magnitudes greater than $\sim4-5$.
Warm start models with masses less than $10~\mathrm{M_J}$ are not detectable.
Planet masses times accretion rates of $\sim10^{-5}~\mathrm{M_J^2~yr^{-1}}$ are detectable for all target star absolute magnitudes, and lower accretion rates $\sim$ a few times $10^{-7}~\mathrm{M_J^2~yr^{-1}}$ to $\sim10^{-6}~\mathrm{M_J^2~yr^{-1}}$ are detectable for stars with absolute magnitudes greater than $\sim5$.
We note that these detection limits would be worse for the shortest wavelength bin ($\lambda = 2.87~\mu$m), partly due to lower Strehl, and partly due to the fact that the PSF is more poorly sampled in this wavelength bin.
They would be better for the longest wavelength bin ($\lambda = 5.27~\mu$m) where the Strehl is higher and the PSF is larger.

\section{DISCUSSION}\label{sec:discussion}

The contrast curves and derived planet mass / accretion rate limits for NIRC2, NIRCam, and NIRISS show that filled-aperture kernel phase is a viable alternative to non-redundant masking for high Strehl ($\sim0.8-0.9$) observations.
This corresponds to wavelengths redder than $\sim3.8$ microns for ground-based observations.
At these wavelengths, the sky brightness is a lesser problem for kernel phase observations, since the NRM point spread function is spread out over more pixels than the filled-aperture PSF.
Space based observations with a thermal background would also benefit from kernel phase for the same reason.
The simulations demonstrated this: filled-aperture kernel phase provided lower noise levels and thus better planet mass / accretion rate limits for faint stars (apparent magnitude $\gtrsim 8$ mag for NIRC2 L$'$, and apparent magnitude $\gtrsim 9-10$ for \emph{JWST} NIRCam and NIRISS).
For both Keck and \emph{JWST} in the bright limit (apparent magnitude $\sim 6$), kernel phase can achieve comparable contrast to NRM at separations outside $\lambda / D$.
However, at angular separations within $\lambda / D$, non-redundant masking provided slightly higher contrast ($\sim0.5-1$ mag) than filled-aperture kernel phase.

While NRM can provide slightly higher contrast close-in, the required integration times are much longer - a factor of $\sim10$ to achieve similar signal to noise as kernel phase.
With limited observing resources, using kernel phase for stars that are faint compared to the thermal background would save time. 
This would be useful in the context of a large survey for young planets, since the two techniques can place similar constraints on (accreting) planet populations in the contrast limit, with kernel phase outperforming NRM when the sky background becomes significant.
NRM's better performance close to the PSF core in the contrast limit suggests that an intermediate case such as a redundant mask may be the ideal observing setup for carrying out faster observations without a loss in contrast in the bright limit.

The OSIRIS and NIRSpec simulations show that kernel phase on an IFS can reach contrasts of $5-6$ magnitudes.
This is an exciting mode for characterizing young planets at smaller angular separation than that typically achieved with an IFS.
For fully-formed planets, this will lead to better atmospheric constraints than possible with narrowband imaging.\cite{2014ApJ...792...17S}
For accreting planets, it will help to distinguish between different formation scenarios: e.g. hot-start versus circumplanetary disk accretion.\cite{2015ApJ...803L...4E,2015ApJ...799...16Z,2015Natur.527..342S}
The single-bin contrast curves show that kernel phase on an integral field spectrograph is capable of simultaneous detection and characterization; individual wavelength bins can be treated independently.

The broadband simulations show an improvement in both achievable contrast and planet mass / accretion rate limits at L$'$ compared to Ks. 
This suggests that a mid-infrared integral field spectrograph would be particularly useful for planet characterization with these techniques. 
Arizona Lenslets for Exoplanet Spectroscopy (ALES)\cite{2015SPIE.9605E..1DS} on the Large Binocular Telescope is a $3-5~\mu$m integral field spectrograph with a non-redundant masking mode. 
ALES provides a spectral resolution of R$\sim$20, and the non-redundant mask has 6 holes with a maximum baseline of 8 meters in single-aperture mode, and 12 holes with a maximum baseline of 23 meters in dual-aperture mode.
ALES' redder wavelength range should lead to higher contrast (and lower planet mass / accretion rate limits) compared to shorter-wavelength ground-based integral field spectrographs.
ALES has not yet been characterized for non-redundant masking or filled-aperture kernel phase observations; this will be the subject of future work.

Current ground based observations can only reach planet masses of several Jupiter masses, and planet masses times accretion rates of a few times $\sim10^{-6}$ to $\sim10^{-5}~\mathrm{M_J^2~yr^{-1}}$.
Furthermore, the resolution limit of $8-10$ meter class telescopes means that they can only probe spatial separations of $\gtrsim10$ AU in the near- to mid-infrared.  
Both of these limits will improve as the next generation of extremely large telescopes comes online; building signal to noise on faint stars will take less time, and the factor of $\sim3$ boost in resolution will probe spatial scales of a few AU.
The 23-meter baseline dual-aperture Large Binocular Telescope can achieve similar resolution now and has an NRM mode, which has been demonstrated to work even without operational co-phasing\cite{2017ApJ...844...22S}.
NRM and kernel phase on LBT could produce ELT-like planet detections before the ELTs are operational, and allow us to develop these tools for use on the next generation of telescopes. 

Both NRM and filled-aperture kernel phase on \emph{James Webb} will expand the planet detection parameter space beyond that of ground-based observations.
While \emph{JWST} will not have higher resolution than current ground-based facilities, its greater stability will lead to lower kernel phase scatter and provide higher contrast.
Furthermore, it will not have the same limitations on target star brightness as an AO-corrected telescope.
These factors combined mean that \emph{JWST} will detect and characterize lower mass / accretion rate planets than we can observe from the ground.

\section{CONCLUSIONS}\label{sec:conclusions}

We presented contrast curves for non-redundant masking and filled-aperture kernel phase on several broadband imagers and integral field spectrographs.
The observations were simulated to carefully control noise sources, and when possible we used real observations to anchor our OPD prescription.
The simulated contrast curves show that for high Strehl, kernel phase performs comparably to or better than NRM outside of $\lambda / D$.
The compactness of the kernel phase PSF makes it particularly well suited for low SNR observations where random noise sources dominate.
In the contrast limited regime, masking outperforms filled-aperture kernel phase by $0.5-1$ magnitudes within the diffraction limit.
This slightly lower contrast suggests that in the bright limit, redundant masks may be a good compromise to reach high contrast with shorter exposure times than NRM.

Both NRM and kernel phase are capable of detecting giant, recently-formed planets and accretion signatures from forming planets.
Filled-aperture kernel phase, applied on an integral field spectrograph, will be capable of simultaneous detection and characterization, and will reach smaller separations than traditional IFS high contrast imaging.
Both techniques, applied on the next generation of adaptive optics systems and space- and ground-based observing facilities, will expand the planet detection parameter space in volume, semi-major axis, and contrast.
This will greatly inform our understanding of planet formation and evolution.

\subsection*{Disclosures}
The authors have no relevant financial interests in the manuscript and no other potential conflicts of interest to disclose.

\acknowledgments 
Steph Sallum is supported by an NSF Astronomy and Astrophysics Postdoctoral Fellowship under award AST-1701489.
The authors would like to acknowledge Zack Briesemeister and Jordan Stone for thoughtful conversations.  
An earlier version of this paper has been submitted as an SPIE conference proceeding for Astronomical Telescopes and Instrumentation: Optical and Infrared Interferometry and Imaging VI.

\bibliography{references2}   
\bibliographystyle{spiejour}  

\counterwithin{figure}{section}
\counterwithin{table}{section}
\appendix

\section{Kernel Phase Projection and Weighting}\label{app:kpweights}
\subsection{Projection}
We use the ``Martinache" projection\cite{2010ApJ...724..464M} to calculate kernel phases; this forms orthonormal combinations of Fourier phases that eliminate instrumental phase. 
The projection is based on representing Fourier phases as linear combinations of pupil plane phases:
\begin{equation}
\Phi = \mathrm{R^{-1}} \cdot \mathrm{A} \cdot \phi + \Phi_0,
\label{eq:kpproj}
\end{equation}
where $\Phi$ represents a vector of Fourier phases with length M, R is a diagonal matrix containing the redundancy of each kernel phase, A is a matrix that describes how pupil plane phases ($\phi$) are combined to create Fourier phases, and $\Phi_0$ is a vector of Fourier phases that are intrinsic to the source. 
The kernel of $\mathrm{R^{-1} \cdot A}$, $\mathrm{K}$,\cite{2016MNRAS.463.3573P} found using singular value decomposition, projects Fourier phases into kernel phases and eliminates the instrumental phase term, $\phi$.\footnote{We note that the kernel of A can also be used as a projection after multiplying Equation \ref{eq:kpproj} through by R. We choose to apply the kernel of $\mathrm{R^{-1} \cdot A}$, which has been used in previous studies.\cite{2016MNRAS.463.3573P}}
The matrix $\mathrm{K}$ is analogous to the closure phase projection in interferometric observations, but is not restricted to values of only 0, 1, and -1.

\subsection{Weighted Averaging}
In interferometric observations, closure phases are calculated from bispectra, the product of three complex visibilities on baselines forming a triangle \cite{1986Natur.320..595B}.
When closure phases are calculated for a cube of images, the bispectra are averaged over the cube, and the phase of the average bispectrum is taken as the average closure phase \cite{2015poi..book.....B}.
Averaging bispectra rather than phases has been demonstrated to perform better in noisy conditions\cite{1988A&A...198..375W}.
We thus generalize this form of vector averaging, which upweights higher signal to noise observations by including amplitude information in the average, from closure phase to kernel phase. 

For each frame, for the $i^{th}$ kernel phase, we calculate a complex quantity by taking a weighted product of complex visibilities:
\begin{equation}
\prod_{j=1}^{M} V_j^{K_{i,j}},
\end{equation}
where $K_{i,j}$ is value in the $i^{th}$ row and $j^{th}$ column of the kernel phase projection matrix, $j$ is the index that describes the M Fourier phases, and $V_j$ is the $j^{th}$ complex visibility, which has both amplitude and phase.
The average kernel phase over a cube of $n_{im}$ images is then:
\begin{equation}
\mathrm{Arg}\left(\frac{1}{n_{im}} \sum_{k=1}^{n_{im}} \prod_{j=1}^{M} V_j^{K_{i,j}}\right).
\end{equation}
This ensures that the Fourier phases are combined using the weights contained in the matrix K, but that amplitude information is used to weight the kernel phases by signal-to-noise when averaging them over multiple frames.
We checked that this weighting scheme performs similarly or better than averaging kernel phases themselves for individual frames, depending on the noise regime. 

\section{Simulated Observation Planning}\label{app:obsplanning}

\begin{ThreePartTable}
\begin{TableNotes}
\footnotesize
\item [a] Target apparent magnitude
\item [b] Total number of coadds per visit
\item [c] Total number of frames per visit
\item [d] Total number of dithers per visit
\item [e] Integration time per coadd
\item [f] Exposure time per visit
\item [g] Total observing time (including overheads) per visit
\item [h] Number of science target visits in one half night
\item [i] Observing efficiency (exposure time divided by total observing time)
\end{TableNotes}
\begin{longtable}{lcccccccc}
\caption{NIRC2 Ks NRM Observation Details\label{tab:ksnrmobs}}\\
\toprule
\colhead{Ks (mag)\tablenotemark{a}} & \colhead{$n_{coadds}$\tablenotemark{b}} &
		 \colhead{$n_{frames}$\tablenotemark{c}} & \colhead{$n_{dit}$\tablenotemark{d}} &
		 \colhead{$t_{coadd}$\tablenotemark{e}} & \colhead{$t_{exp}$\tablenotemark{f}} &
		 \colhead{$t_{tot}$\tablenotemark{g}} & \colhead{$n_{vis}$\tablenotemark{h}}&
		 \colhead{$eff$\tablenotemark{i}} \\
\midrule
\endfirsthead
\multicolumn{9}{c}
{\tablename\ \thetable\ -- \textit{Continued from previous page}} \\
\toprule
\colhead{Ks (mag)\tablenotemark{a}} & \colhead{$n_{coadds}$\tablenotemark{b}} &
		 \colhead{$n_{frames}$\tablenotemark{c}} & \colhead{$n_{dit}$\tablenotemark{d}} &
		 \colhead{$t_{coadd}$\tablenotemark{e}} & \colhead{$t_{exp}$\tablenotemark{f}} &
		 \colhead{$t_{tot}$\tablenotemark{g}} & \colhead{$n_{vis}$\tablenotemark{h}}&
		 \colhead{$eff$\tablenotemark{i}} \\
\midrule
\endhead
\bottomrule
\multicolumn{9}{r}{\textit{Continued on next page}} \\
\endfoot
\bottomrule
\insertTableNotes
\endlastfoot

5.7 & 3877 & 9 & 2 & 0.0453 & 175 & 491 & 22 & 0.36\\
5.8 & 3662 & 10 & 2 & 0.0502 & 184 & 491 & 22 & 0.37\\
5.9 & 3487 & 10 & 2 & 0.0553 & 193 & 491 & 22 & 0.39\\
6.0 & 3329 & 10 & 2 & 0.0603 & 201 & 491 & 22 & 0.41\\
6.1 & 3033 & 11 & 2 & 0.0703 & 213 & 491 & 22 & 0.43\\
6.2 & 2912 & 11 & 2 & 0.0753 & 219 & 491 & 22 & 0.45\\
6.3 & 2783 & 12 & 2 & 0.0803 & 224 & 491 & 22 & 0.46\\
6.4 & 2584 & 12 & 2 & 0.0904 & 234 & 491 & 22 & 0.48\\
\end{longtable}
\end{ThreePartTable}

\begin{ThreePartTable}
\begin{TableNotes}
\footnotesize
\item [a] Target apparent magnitude
\item [b] Total number of coadds per visit
\item [c] Total number of frames per visit
\item [d] Total number of dithers per visit
\item [e] Integration time per coadd
\item [f] Exposure time per visit
\item [g] Total observing time (including overheads) per visit
\item [h] Number of science target visits in one half night
\item [i] Observing efficiency (exposure time divided by total observing time)
\end{TableNotes}
\begin{longtable}{lcccccccc}
\caption{NIRC2 Kp NRM Observation Details\label{tab:kskpobs}}\\
\toprule
\colhead{Ks (mag)\tablenotemark{a}} & \colhead{$n_{coadds}$\tablenotemark{b}} &
		 \colhead{$n_{frames}$\tablenotemark{c}} & \colhead{$n_{dit}$\tablenotemark{d}} &
		 \colhead{$t_{coadd}$\tablenotemark{e}} & \colhead{$t_{exp}$\tablenotemark{f}} &
		 \colhead{$t_{tot}$\tablenotemark{g}} & \colhead{$n_{vis}$\tablenotemark{h}}&
		 \colhead{$eff$\tablenotemark{i}} \\
\midrule
\endfirsthead
\multicolumn{9}{c}
{\tablename\ \thetable\ -- \textit{Continued from previous page}} \\
\toprule
\colhead{Ks (mag)\tablenotemark{a}} & \colhead{$n_{coadds}$\tablenotemark{b}} &
		 \colhead{$n_{frames}$\tablenotemark{c}} & \colhead{$n_{dit}$\tablenotemark{d}} &
		 \colhead{$t_{coadd}$\tablenotemark{e}} & \colhead{$t_{exp}$\tablenotemark{f}} &
		 \colhead{$t_{tot}$\tablenotemark{g}} & \colhead{$n_{vis}$\tablenotemark{h}}&
		 \colhead{$eff$\tablenotemark{i}} \\
\midrule
\endhead
\bottomrule
\multicolumn{9}{r}{\textit{Continued on next page}} \\
\endfoot
\bottomrule
\insertTableNotes
\endlastfoot

5.7 & 6975 & 2 & 2 & 0.0051 & 36 & 491 & 22 & 0.07\\
5.8 & 6975 & 2 & 2 & 0.0051 & 36 & 491 & 22 & 0.07\\
5.9 & 6975 & 2 & 2 & 0.0051 & 36 & 491 & 22 & 0.07\\
6.0 & 6321 & 4 & 2 & 0.0101 & 64 & 491 & 22 & 0.13\\
6.1 & 6321 & 4 & 2 & 0.0101 & 64 & 491 & 22 & 0.13\\
6.2 & 6321 & 4 & 2 & 0.0101 & 64 & 491 & 22 & 0.13\\
6.3 & 6321 & 4 & 2 & 0.0101 & 64 & 491 & 22 & 0.13\\
6.4 & 6321 & 4 & 2 & 0.0101 & 64 & 491 & 22 & 0.13\\
\end{longtable}
\end{ThreePartTable}

\begin{ThreePartTable}
\begin{TableNotes}
\footnotesize
\item [a] Target apparent magnitude
\item [b] Total number of coadds per visit
\item [c] Total number of frames per visit
\item [d] Total number of dithers per visit
\item [e] Integration time per coadd
\item [f] Exposure time per visit
\item [g] Total observing time (including overheads) per visit
\item [h] Number of science target visits in one half night
\item [i] Observing efficiency (exposure time divided by total observing time)
\end{TableNotes}
\begin{longtable}{lcccccccc}
\caption{NIRC2 L$'$ NRM Observation Details\label{tab:lpnrmobs}}\\
\toprule
\colhead{L$'$ (mag)\tablenotemark{a}} & \colhead{$n_{coadds}$\tablenotemark{b}} &
		 \colhead{$n_{frames}$\tablenotemark{c}} & \colhead{$n_{dit}$\tablenotemark{d}} &
		 \colhead{$t_{coadd}$\tablenotemark{e}} & \colhead{$t_{exp}$\tablenotemark{f}} &
		 \colhead{$t_{tot}$\tablenotemark{g}} & \colhead{$n_{vis}$\tablenotemark{h}}&
		 \colhead{$eff$\tablenotemark{i}} \\
\midrule
\endfirsthead
\multicolumn{9}{c}
{\tablename\ \thetable\ -- \textit{Continued from previous page}} \\
\toprule
\colhead{L$'$ (mag)\tablenotemark{a}} & \colhead{$n_{coadds}$\tablenotemark{b}} &
		 \colhead{$n_{frames}$\tablenotemark{c}} & \colhead{$n_{dit}$\tablenotemark{d}} &
		 \colhead{$t_{coadd}$\tablenotemark{e}} & \colhead{$t_{exp}$\tablenotemark{f}} &
		 \colhead{$t_{tot}$\tablenotemark{g}} & \colhead{$n_{vis}$\tablenotemark{h}}&
		 \colhead{$eff$\tablenotemark{i}}\\
\midrule
\endhead
\bottomrule
\multicolumn{9}{r}{\textit{Continued on next page}} \\
\endfoot
\bottomrule
\insertTableNotes
\endlastfoot
5.7 & 1702 & 14 & 2 & 0.1606 & 273 & 491 & 22 & 0.56\\
5.8 & 1662 & 14 & 2 & 0.1656 & 275 & 491 & 21 & 0.56\\
5.9 & 1588 & 14 & 2 & 0.1757 & 279 & 491 & 22 & 0.57\\
6.0 & 1513 & 14 & 2 & 0.1869 & 283 & 491 & 22 & 0.58\\
6.1 & 1480 & 15 & 2 & 0.1907 & 282 & 491 & 22 & 0.57\\
6.2 & 1421 & 15 & 2 & 0.2007 & 285 & 491 & 22 & 0.58\\
6.3 & 1393 & 15 & 2 & 0.2057 & 287 & 491 & 22 & 0.58\\
6.4 & 1340 & 15 & 2 & 0.2158 & 289 & 491 & 22 & 0.59\\
\end{longtable}
\end{ThreePartTable}

\begin{ThreePartTable}
\begin{TableNotes}
\footnotesize
\item [a] Target apparent magnitude
\item [b] Total number of coadds per visit
\item [c] Total number of frames per visit
\item [d] Total number of dithers per visit
\item [e] Integration time per coadd
\item [f] Exposure time per visit
\item [g] Total observing time (including overheads) per visit
\item [h] Number of science target visits in one half night
\item [i] Observing efficiency (exposure time divided by total observing time)
\end{TableNotes}
\begin{longtable}{lcccccccc}
\caption{NIRC2 L$'$ KP Observation Details\label{tab:lpkpobs}}\\
\toprule
\colhead{L$'$ (mag)\tablenotemark{a}} & \colhead{$n_{coadds}$\tablenotemark{b}} &
		 \colhead{$n_{frames}$\tablenotemark{c}} & \colhead{$n_{dit}$\tablenotemark{d}} &
		 \colhead{$t_{coadd}$\tablenotemark{e}} & \colhead{$t_{exp}$\tablenotemark{f}} &
		 \colhead{$t_{tot}$\tablenotemark{g}} & \colhead{$n_{vis}$\tablenotemark{h}}&
		 \colhead{$eff$\tablenotemark{i}} \\
\midrule
\endfirsthead
\multicolumn{9}{c}
{\tablename\ \thetable\ -- \textit{Continued from previous page}} \\
\toprule
\colhead{L$'$ (mag)\tablenotemark{a}} & \colhead{$n_{coadds}$\tablenotemark{b}} &
		 \colhead{$n_{frames}$\tablenotemark{c}} & \colhead{$n_{dit}$\tablenotemark{d}} &
		 \colhead{$t_{coadd}$\tablenotemark{e}} & \colhead{$t_{exp}$\tablenotemark{f}} &
		 \colhead{$t_{tot}$\tablenotemark{g}} & \colhead{$n_{vis}$\tablenotemark{h}}&
		 \colhead{$eff$\tablenotemark{i}}\\
\midrule
\endhead
\bottomrule
\multicolumn{9}{r}{\textit{Continued on next page}} \\
\endfoot
\bottomrule
\insertTableNotes
\endlastfoot
5.7 & 4659 & 7 & 2 & 0.0302 & 141 & 491 & 22 & 0.29\\
5.8 & 4359 & 8 & 2 & 0.0352 & 154 & 491 & 22 & 0.31\\
5.9 & 4093 & 9 & 2 & 0.0402 & 165 & 491 & 22 & 0.34\\
6.0 & 4093 & 9 & 2 & 0.0402 & 165 & 491 & 22 & 0.34\\
6.1 & 3877 & 9 & 2 & 0.0453 & 175 & 491 & 22 & 0.36\\
6.2 & 3662 & 10 & 2 & 0.0502 & 184 & 491 & 22 & 0.37\\
6.3 & 3662 & 10 & 2 & 0.0502 & 184 & 491 & 22 & 0.37\\
6.4 & 3487 & 10 & 2 & 0.0553 & 193 & 491 & 22 & 0.39\\
\end{longtable}
\end{ThreePartTable}

\begin{ThreePartTable}
\begin{TableNotes}
\footnotesize
\item [a] Target apparent magnitude
\item [b] Total number of coadds per visit
\item [c] Total number of frames per visit
\item [d] Total number of dithers per visit
\item [e] Integration time per coadd
\item [f] Exposure time per visit
\item [g] Total observing time (including overheads) per visit
\item [h] Number of science target visits in one half night
\item [i] Observing efficiency (exposure time divided by total observing time)
\end{TableNotes}
\begin{longtable}{lcccccccc}
\caption{NIRC2 Ms NRM Observation Details\label{tab:msnrmobs}}\\
\toprule
\colhead{Ms (mag)\tablenotemark{a}} & \colhead{$n_{coadds}$\tablenotemark{b}} &
		 \colhead{$n_{frames}$\tablenotemark{c}} & \colhead{$n_{dit}$\tablenotemark{d}} &
		 \colhead{$t_{coadd}$\tablenotemark{e}} & \colhead{$t_{exp}$\tablenotemark{f}} &
		 \colhead{$t_{tot}$\tablenotemark{g}} & \colhead{$n_{vis}$\tablenotemark{h}}&
		 \colhead{$eff$\tablenotemark{i}} \\
\midrule
\endfirsthead
\multicolumn{9}{c}
{\tablename\ \thetable\ -- \textit{Continued from previous page}} \\
\toprule
\colhead{Ms (mag)\tablenotemark{a}} & \colhead{$n_{coadds}$\tablenotemark{b}} &
		 \colhead{$n_{frames}$\tablenotemark{c}} & \colhead{$n_{dit}$\tablenotemark{d}} &
		 \colhead{$t_{coadd}$\tablenotemark{e}} & \colhead{$t_{exp}$\tablenotemark{f}} &
		 \colhead{$t_{tot}$\tablenotemark{g}} & \colhead{$n_{vis}$\tablenotemark{h}}&
		 \colhead{$eff$\tablenotemark{i}}\\
\midrule
\endhead
\bottomrule
\multicolumn{9}{r}{\textit{Continued on next page}} \\
\endfoot
\bottomrule
\insertTableNotes
\endlastfoot
5.7 & 5330 & 48 & 5 & 0.1973 & 1052 & 1543 & 7 & 0.68\\
5.8 & 6407 & 58 & 6 & 0.1915 & 1227 & 1800 & 6 & 0.68\\
5.9 & 7702 & 70 & 7 & 0.1935 & 1490 & 2160 & 5 & 0.69\\
6.0 & 9260 & 84 & 9 & 0.2063 & 1910 & 2700 & 4 & 0.71\\
6.1 & 11131 & 101 & 11 & 0.2396 & 2667 & 3600 & 3 & 0.74\\
6.2 & 13382 & 121 & 13 & 0.1868 & 2499 & 3600 & 3 & 0.69\\
6.3 & 16087 & 145 & 15 & 0.2548 & 4100 & 5400 & 2 & 0.76\\
6.4 & 19340 & 175 & 18 & 0.1993 & 3854 & 5400 & 2 & 0.71\\
\end{longtable}
\end{ThreePartTable}

\begin{ThreePartTable}
\begin{TableNotes}
\footnotesize
\item [a] Target apparent magnitude
\item [b] Total number of coadds per visit
\item [c] Total number of frames per visit
\item [d] Total number of dithers per visit
\item [e] Integration time per coadd
\item [f] Exposure time per visit
\item [g] Total observing time (including overheads) per visit
\item [h] Number of science target visits in one half night
\item [i] Observing efficiency (exposure time divided by total observing time)
\end{TableNotes}
\begin{longtable}{lcccccccc}
\caption{NIRC2 Ms KP Observation Details\label{tab:mskpobs}}\\
\toprule
\colhead{Ms (mag)\tablenotemark{a}} & \colhead{$n_{coadds}$\tablenotemark{b}} &
		 \colhead{$n_{frames}$\tablenotemark{c}} & \colhead{$n_{dit}$\tablenotemark{d}} &
		 \colhead{$t_{coadd}$\tablenotemark{e}} & \colhead{$t_{exp}$\tablenotemark{f}} &
		 \colhead{$t_{tot}$\tablenotemark{g}} & \colhead{$n_{vis}$\tablenotemark{h}}&
		 \colhead{$eff$\tablenotemark{i}} \\
\midrule
\endfirsthead
\multicolumn{9}{c}
{\tablename\ \thetable\ -- \textit{Continued from previous page}} \\
\toprule
\colhead{Ms (mag)\tablenotemark{a}} & \colhead{$n_{coadds}$\tablenotemark{b}} &
		 \colhead{$n_{frames}$\tablenotemark{c}} & \colhead{$n_{dit}$\tablenotemark{d}} &
		 \colhead{$t_{coadd}$\tablenotemark{e}} & \colhead{$t_{exp}$\tablenotemark{f}} &
		 \colhead{$t_{tot}$\tablenotemark{g}} & \colhead{$n_{vis}$\tablenotemark{h}}&
		 \colhead{$eff$\tablenotemark{i}}\\
\midrule
\endhead
\bottomrule
\multicolumn{9}{r}{\textit{Continued on next page}} \\
\endfoot
\bottomrule
\insertTableNotes
\endlastfoot
5.7 & 2055 & 13 & 2 & 0.1255 & 258 & 491 & 22 & 0.53\\
5.8 & 1998 & 13 & 2 & 0.1305 & 261 & 491 & 22 & 0.53\\
5.9 & 1932 & 14 & 2 & 0.1355 & 262 & 491 & 22 & 0.53\\
6.0 & 1881 & 14 & 2 & 0.1405 & 264 & 491 & 22 & 0.54\\
6.1 & 1881 & 14 & 2 & 0.1405 & 264 & 491 & 22 & 0.54\\
6.2 & 1833 & 14 & 2 & 0.1455 & 267 & 491 & 22 & 0.54\\
6.3 & 1833 & 14 & 2 & 0.1455 & 267 & 491 & 22 & 0.54\\
6.4 & 1787 & 14 & 2 & 0.1506 & 269 & 491 & 22 & 0.55\\
\end{longtable}
\end{ThreePartTable}

\begin{ThreePartTable}
\begin{TableNotes}
\footnotesize
\item [a] Target apparent magnitude
\item [b] Readout mode
\item [c] Total number of groups per integration
\item [d] Total number of $n_g$ length integrations in 90 minutes
\item [e] Additional groups (in one final shorter integration to finish a 90 minute observing block)
\item [f] Total integration time in 90 minutes
\item [g] Observing efficiency (rounded to 0.01)
\end{TableNotes}
\begin{longtable}{lcccccc}
\caption{NIRCam F430M Observation Details\label{tab:nircamf430}}\\
\toprule
\colhead{Ms (mag)\tablenotemark{a}} & \colhead{MODE\tablenotemark{b}} &
		 \colhead{$n_{g}$\tablenotemark{c}} & \colhead{$n_{int}$\tablenotemark{d}} &
		 \colhead{$n_{g,rem}$\tablenotemark{e}} & \colhead{$t_{tot}$ (s)\tablenotemark{f}} &
		 \colhead{$eff$\tablenotemark{g}} \\
\midrule
\endfirsthead
\multicolumn{7}{c}
{\tablename\ \thetable\ -- \textit{Continued from previous page}} \\
\toprule
\colhead{Ms (mag)\tablenotemark{a}} & \colhead{MODE\tablenotemark{b}} &
		 \colhead{$n_{g}$\tablenotemark{c}} & \colhead{$n_{int}$\tablenotemark{d}} &
		 \colhead{$n_{g,rem}$\tablenotemark{e}} & \colhead{$t_{tot}$ (s)\tablenotemark{f}} &
		 \colhead{$eff$\tablenotemark{g}}\\
\midrule
\endhead
\bottomrule
\multicolumn{7}{r}{\textit{Continued on next page}} \\
\endfoot
\bottomrule
\insertTableNotes
\endlastfoot
6.3 & rapid & 2 & 54655 & \NA & 2700 & 0.5\\
6.4 & rapid & 2 & 54655 & \NA & 2700 & 0.5\\
6.5 & rapid & 2 & 54655 & \NA & 2700 & 0.5\\
6.6 & rapid & 2 & 54655 & \NA & 2700 & 0.5\\
6.7 & rapid & 3 & 36436 & \NA & 3600 & 0.67\\
6.8 & rapid & 3 & 36436 & \NA & 3600 & 0.67\\
6.9 & rapid & 3 & 36436 & \NA & 3600 & 0.67\\
7.0 & rapid & 4 & 27327 & \NA & 4050 & 0.75\\
\end{longtable}
\end{ThreePartTable}

\begin{ThreePartTable}
\begin{TableNotes}
\footnotesize
\item [a] Target apparent magnitude
\item [b] Readout mode
\item [c] Total number of groups per integration
\item [d] Total number of $n_g$ length integrations in 90 minutes
\item [e] Additional groups (in one final shorter integration to finish a 90 minute observing block)
\item [f] Total integration time in 90 minutes
\item [g] Observing efficiency (rounded to 0.01)
\end{TableNotes}
\begin{longtable}{lcccccccc}
\caption{NIRCam F480M Observation Details\label{tab:nircamf480}}\\
\toprule
\colhead{Ms (mag)\tablenotemark{a}} & \colhead{MODE\tablenotemark{b}} &
		 \colhead{$n_{g}$\tablenotemark{c}} & \colhead{$n_{int}$\tablenotemark{d}} &
		 \colhead{$n_{g,rem}$\tablenotemark{e}} & \colhead{$t_{tot}$ (s)\tablenotemark{f}} &
		 \colhead{$eff$\tablenotemark{g}} \\
\midrule
\endfirsthead
\multicolumn{7}{c}
{\tablename\ \thetable\ -- \textit{Continued from previous page}} \\
\toprule
\colhead{Ms (mag)\tablenotemark{a}} & \colhead{MODE\tablenotemark{b}} &
		 \colhead{$n_{g}$\tablenotemark{c}} & \colhead{$n_{int}$\tablenotemark{d}} &
		 \colhead{$n_{g,rem}$\tablenotemark{e}} & \colhead{$t_{tot}$ (s)\tablenotemark{f}} &
		 \colhead{$eff$\tablenotemark{g}}\\
\midrule
\endhead
\bottomrule
\multicolumn{7}{r}{\textit{Continued on next page}} \\
\endfoot
\bottomrule
\insertTableNotes
\endlastfoot
6.0 & rapid & 2 & 54655 & \NA & 2700 & 0.5\\
6.1 & rapid & 2 & 54655 & \NA & 2700 & 0.5\\
6.2 & rapid & 2 & 54655 & \NA & 2700 & 0.5\\
6.3 & rapid & 2 & 54655 & \NA & 2700 & 0.5\\
6.4 & rapid & 3 & 36436 & \NA & 3600 & 0.67\\
6.5 & rapid & 3 & 36436 & \NA & 3600 & 0.67\\
6.6 & rapid & 3 & 36436 & \NA & 3600 & 0.67\\
6.7 & rapid & 4 & 27327 & \NA & 4050 & 0.75\\
\end{longtable}
\end{ThreePartTable}

\begin{ThreePartTable}
\begin{TableNotes}
\footnotesize
\item [a] Target apparent magnitude
\item [b] Total number of groups per integration
\item [c] Total number of $n_g$ length integrations in 90 minutes
\item [d] Additional groups (in one final shorter integration to finish a 90 minute observing block)
\item [e] Total integration time in 90 minutes
\item [f] Observing efficiency (rounded to 0.01)
\end{TableNotes}
\begin{longtable}{lccccc}
\caption{NIRISS F430M Observation Details\label{tab:nirissf430}}\\
\toprule
\colhead{Ms (mag)\tablenotemark{a}} &
		 \colhead{$n_{g}$\tablenotemark{b}} & \colhead{$n_{int}$\tablenotemark{b}} &
		 \colhead{$n_{g,rem}$\tablenotemark{d}} & \colhead{$t_{tot}$ (s)\tablenotemark{e}} &
		 \colhead{$eff$\tablenotemark{f}} \\
\midrule
\endfirsthead
\multicolumn{6}{c}
{\tablename\ \thetable\ -- \textit{Continued from previous page}} \\
\toprule
\colhead{Ms (mag)\tablenotemark{a}} &
		 \colhead{$n_{g}$\tablenotemark{b}} & \colhead{$n_{int}$\tablenotemark{b}} &
		 \colhead{$n_{g,rem}$\tablenotemark{d}} & \colhead{$t_{tot}$ (s)\tablenotemark{e}} &
		 \colhead{$eff$\tablenotemark{f}}\\
\midrule
\endhead
\bottomrule
\multicolumn{6}{r}{\textit{Continued on next page}} \\
\endfoot
\bottomrule
\insertTableNotes
\endlastfoot
5.7 & 35 & 2070 & 31 & 5246 & 0.97\\
5.8 & 38 & 1907 & 15 & 5258 & 0.97\\
5.9 & 42 & 1725 & 31 & 5271 & 0.98\\
6.0 & 46 & 1575 & 31 & 5282 & 0.98\\
6.1 & 50 & 1449 & 31 & 5292 & 0.98\\
6.2 & 55 & 1317 & 46 & 5302 & 0.98\\
6.3 & 61 & 1188 & 13 & 5311 & 0.98\\
6.4 & 67 & 1081 & 54 & 5319 & 0.99\\
\end{longtable}
\end{ThreePartTable}

\begin{ThreePartTable}
\begin{TableNotes}
\footnotesize
\item [a] Target apparent magnitude
\item [b] Total number of groups per integration
\item [c] Total number of $n_g$ length integrations in 90 minutes
\item [d] Additional groups (in one final shorter integration to finish a 90 minute observing block)
\item [e] Total integration time in 90 minutes
\item [f] Observing efficiency (rounded to 0.01)
\end{TableNotes}
\begin{longtable}{lccccc}
\caption{NIRISS F480M Observation Details\label{tab:nirissf480}}\\
\toprule
\colhead{Ms (mag)\tablenotemark{a}} &
		 \colhead{$n_{g}$\tablenotemark{b}} & \colhead{$n_{int}$\tablenotemark{b}} &
		 \colhead{$n_{g,rem}$\tablenotemark{d}} & \colhead{$t_{tot}$ (s)\tablenotemark{e}} &
		 \colhead{$eff$\tablenotemark{f}} \\
\midrule
\endfirsthead
\multicolumn{6}{c}
{\tablename\ \thetable\ -- \textit{Continued from previous page}} \\
\toprule
\colhead{Ms (mag)\tablenotemark{a}} &
		 \colhead{$n_{g}$\tablenotemark{b}} & \colhead{$n_{int}$\tablenotemark{b}} &
		 \colhead{$n_{g,rem}$\tablenotemark{d}} & \colhead{$t_{tot}$ (s)\tablenotemark{e}} &
		 \colhead{$eff$\tablenotemark{f}}\\
\midrule
\endhead
\bottomrule
\multicolumn{6}{r}{\textit{Continued on next page}} \\
\endfoot
\bottomrule
\insertTableNotes
\endlastfoot
5.7 & 39 & 1858 & 19 & 5261 & 0.97\\
5.8 & 42 & 1725 & 31 & 5271 & 0.98\\
5.9 & 46 & 1575 & 31 & 5282 & 0.98\\
6.0 & 51 & 1421 & 10 & 5294 & 0.98\\
6.1 & 56 & 1294 & 17 & 5303 & 0.98\\
6.2 & 61 & 1188 & 13 & 5311 & 0.98\\
6.3 & 67 & 1081 & 54 & 5319 & 0.99\\
6.4 & 74 & 979 & 35 & 5327 & 0.99\\
\end{longtable}
\end{ThreePartTable}

\begin{ThreePartTable}
\begin{TableNotes}
\footnotesize
\item [a] Target apparent magnitude
\item [b] Total number of frames per visit
\item [c] Total number of dithers per visit
\item [d] Integration time per frame
\item [e] Exposure time per visit
\item [f] Total observing time (including overheads) per visit
\item [g] Number of science target visits in one half night
\item [h] Observing efficiency (exposure time divided by total observing time)
\end{TableNotes}
\begin{longtable}{lccccccc}
\caption{OSIRIS Kn3 KP Observation Details\label{tab:osirisobs}}\\
\toprule
\colhead{Ks (mag)\tablenotemark{a}} & 
		 \colhead{$n_{frames}$\tablenotemark{b}} & \colhead{$n_{dit}$\tablenotemark{c}} &
		 \colhead{$t_{frame}$\tablenotemark{d}} & \colhead{$t_{exp}$\tablenotemark{e}} &
		 \colhead{$t_{tot}$\tablenotemark{f}} & \colhead{$n_{vis}$\tablenotemark{g}}&
		 \colhead{$eff$\tablenotemark{h}}\\
\midrule
\endfirsthead
\multicolumn{8}{c}
{\tablename\ \thetable\ -- \textit{Continued from previous page}} \\
\toprule
\colhead{Ks (mag)\tablenotemark{a}} & 
		 \colhead{$n_{frames}$\tablenotemark{b}} & \colhead{$n_{dit}$\tablenotemark{c}} &
		 \colhead{$t_{frame}$\tablenotemark{d}} & \colhead{$t_{exp}$\tablenotemark{e}} &
		 \colhead{$t_{tot}$\tablenotemark{f}} & \colhead{$n_{vis}$\tablenotemark{g}}&
		 \colhead{$eff$\tablenotemark{h}}\\
\midrule
\endhead
\bottomrule
\multicolumn{8}{r}{\textit{Continued on next page}} \\
\endfoot
\bottomrule
\insertTableNotes
\endlastfoot
5.7 & 57 & 2 & 5.99 & 342 & 491 & 22 & 0.7\\
5.8 & 53 & 2 & 6.51 & 345 & 491 & 22 & 0.7\\
5.9 & 48 & 2 & 7.27 & 349 & 491 & 22 & 0.71\\
6.0 & 44 & 2 & 8.01 & 352 & 491 & 22 & 0.72\\
6.1 & 41 & 2 & 8.66 & 355 & 491 & 22 & 0.72\\
6.2 & 38 & 2 & 9.41 & 357 & 491 & 22 & 0.73\\
6.3 & 34 & 2 & 10.61 & 361 & 491 & 22 & 0.73\\
6.4 & 32 & 2 & 11.32 & 362 & 491 & 22 & 0.74\\
\end{longtable}
\end{ThreePartTable}

\begin{ThreePartTable}
\begin{TableNotes}
\footnotesize
\item [a] Target apparent magnitude
\item [b] Total number of groups per integration
\item [c] Total number of $n_g$ length integrations per small grid dither position
\item [d] Additional groups (in one final shorter integration to finish one observing block in a single small grid dither position)
\item [e] Total integration time per dither position per target
\item [f] Observing efficiency (rounded to 0.01)
\end{TableNotes}
\begin{longtable}{lccccc}
\caption{NIRSpec Observation Details\label{tab:nirspecobs}}\\
\toprule
\colhead{L$'$ (mag)\tablenotemark{a}} &
		 \colhead{$n_{g}$\tablenotemark{b}} & \colhead{$n_{int}$\tablenotemark{b}} &
		 \colhead{$n_{g,rem}$\tablenotemark{d}} & \colhead{$t_{tot}$ (s)\tablenotemark{e}} &
		 \colhead{$eff$\tablenotemark{f}} \\
\midrule
\endfirsthead
\multicolumn{6}{c}
{\tablename\ \thetable\ -- \textit{Continued from previous page}} \\
\toprule
\colhead{L$'$ (mag)\tablenotemark{a}} &
		 \colhead{$n_{g}$\tablenotemark{b}} & \colhead{$n_{int}$\tablenotemark{b}} &
		 \colhead{$n_{g,rem}$\tablenotemark{d}} & \colhead{$t_{tot}$ (s)\tablenotemark{e}} &
		 \colhead{$eff$\tablenotemark{f}}\\
\midrule
\endhead
\bottomrule
\multicolumn{6}{r}{\textit{Continued on next page}} \\
\endfoot
\bottomrule
\insertTableNotes
\endlastfoot
6.6 & 2 & 55 & \NA & 591 & 0.5\\
6.7 & 2 & 55 & \NA & 591 & 0.5\\
6.8 & 2 & 55 & \NA & 591 & 0.5\\
6.9 & 2 & 55 & \NA & 591 & 0.5\\
7.0 & 2 & 55 & \NA & 591 & 0.5\\
7.1 & 3 & 37 & \NA & 784 & 0.66\\
7.2 & 3 & 37 & \NA & 784 & 0.66\\
7.3 & 3 & 37 & \NA & 784 & 0.66\\
\end{longtable}
\end{ThreePartTable}
\pagebreak

\section{Kernel Phase Histograms}\label{app:hists}

\begin{figure}[h!]
\begin{center}
\begin{tabular}{c}  
\includegraphics[width=\textwidth]{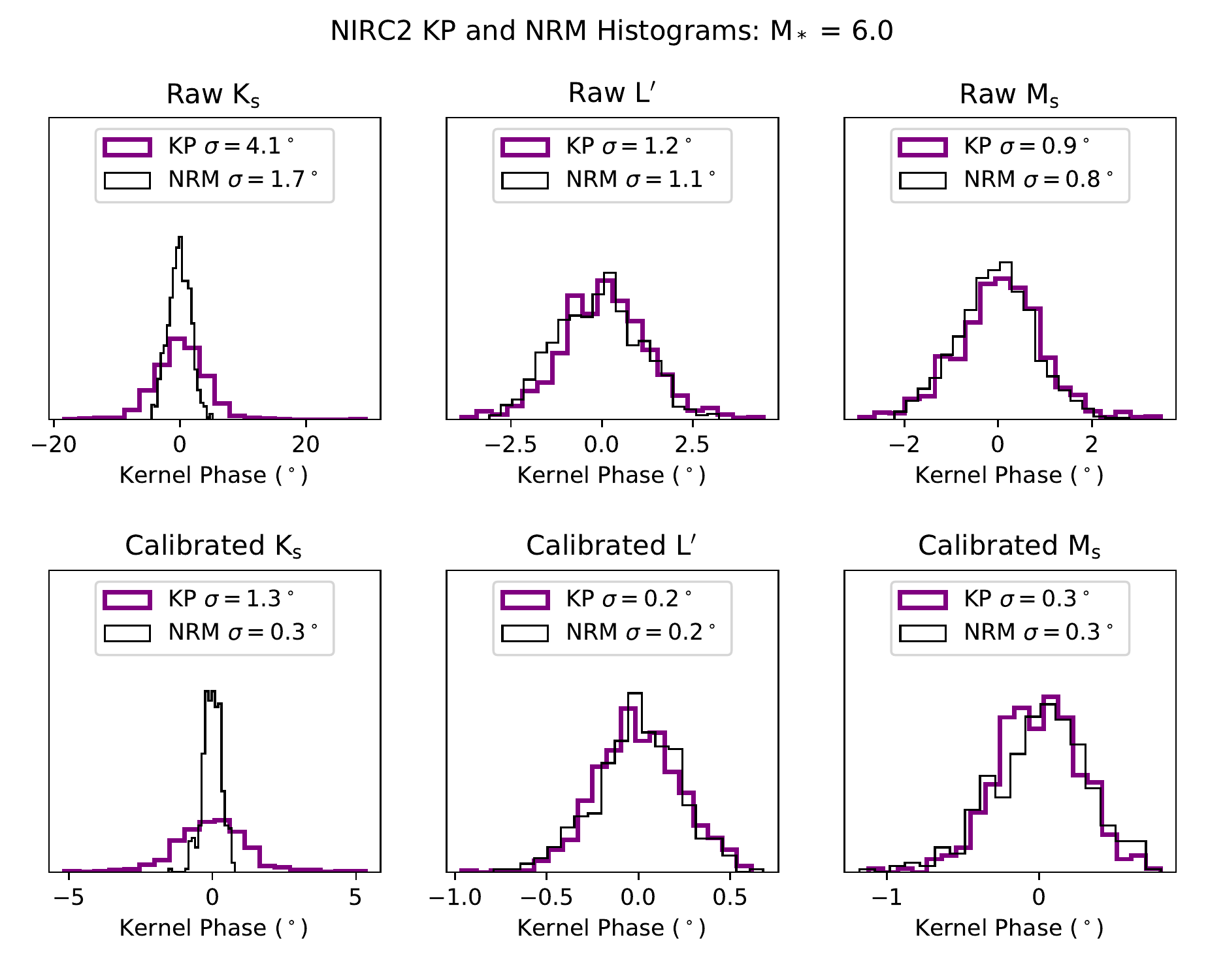}
\end{tabular}
\end{center}
\caption[NIRC2 Kernel Phase Histograms: Bright Case] 
{ \label{fig:n2histsbright}
Histograms of raw (top) and calibrated (bottom) kernel phases for K$\mathrm{_s}$ (left), L$'$ (middle), and M$\mathrm{_s}$ band (center) NIRC2 observations for a bright (6th apparent magnitude) target star.
Thick purple lines show filled-aperture kernel phases, while thin black lines show NRM kernel phases.
}
\end{figure}
\pagebreak

\begin{figure} [h!]
\begin{center}
\begin{tabular}{c}  
\includegraphics[width=\textwidth]{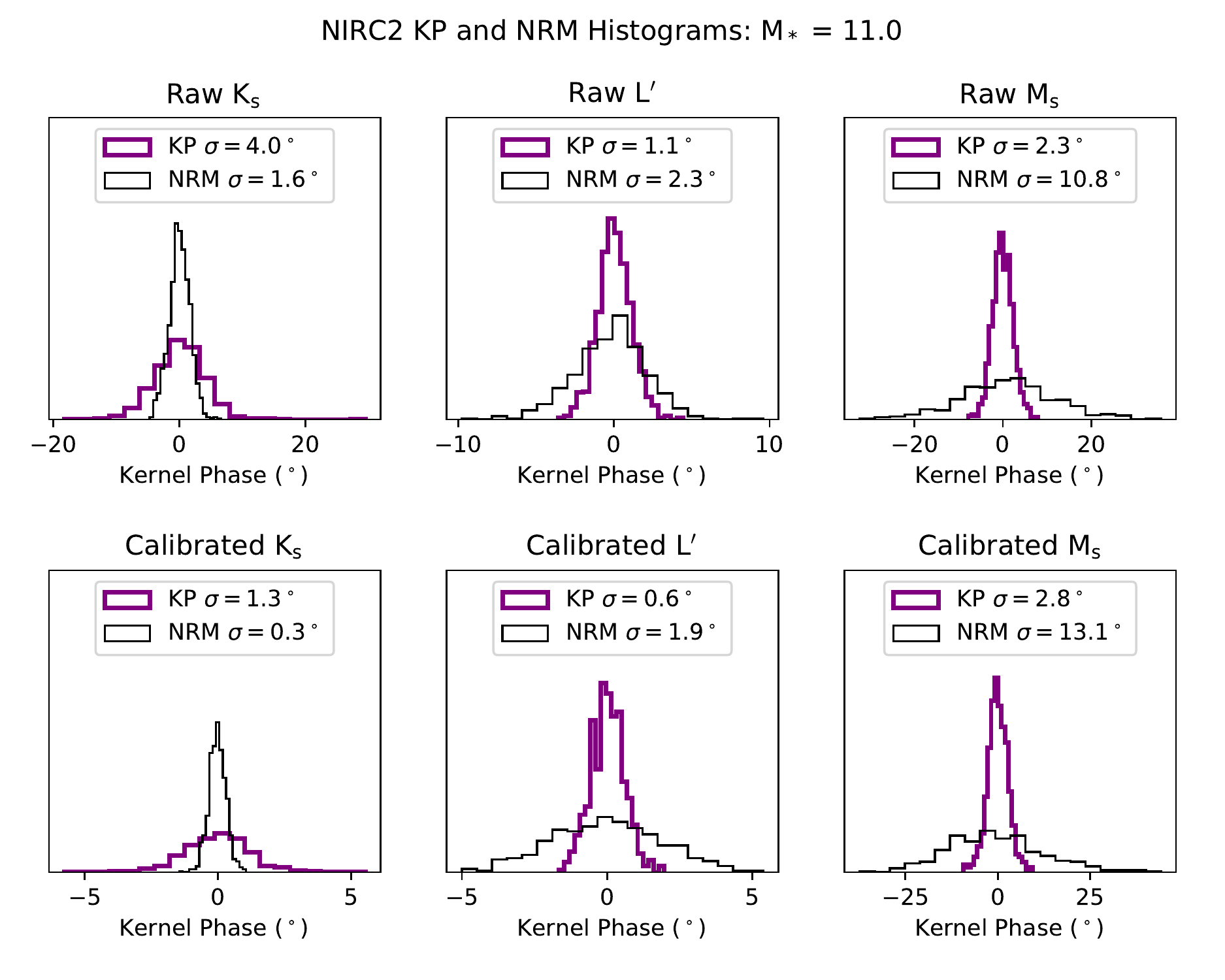}
\end{tabular}
\end{center}
\caption[NIRC2 Kernel Phase Histograms: Faint Case] 
{ \label{fig:n2histsfaint}
Histograms of raw (top) and calibrated (bottom) kernel phases for K$\mathrm{_s}$ (left), L$'$ (middle), and M$\mathrm{_s}$ band (center) NIRC2 observations for a faint (11th apparent magnitude) target star.
Thick purple lines show filled-aperture kernel phases, while thin black lines show NRM kernel phases.
}
\end{figure}
\pagebreak

\begin{figure} [h!]
\begin{center}
\begin{tabular}{c} 
\includegraphics[width=0.7\textwidth]{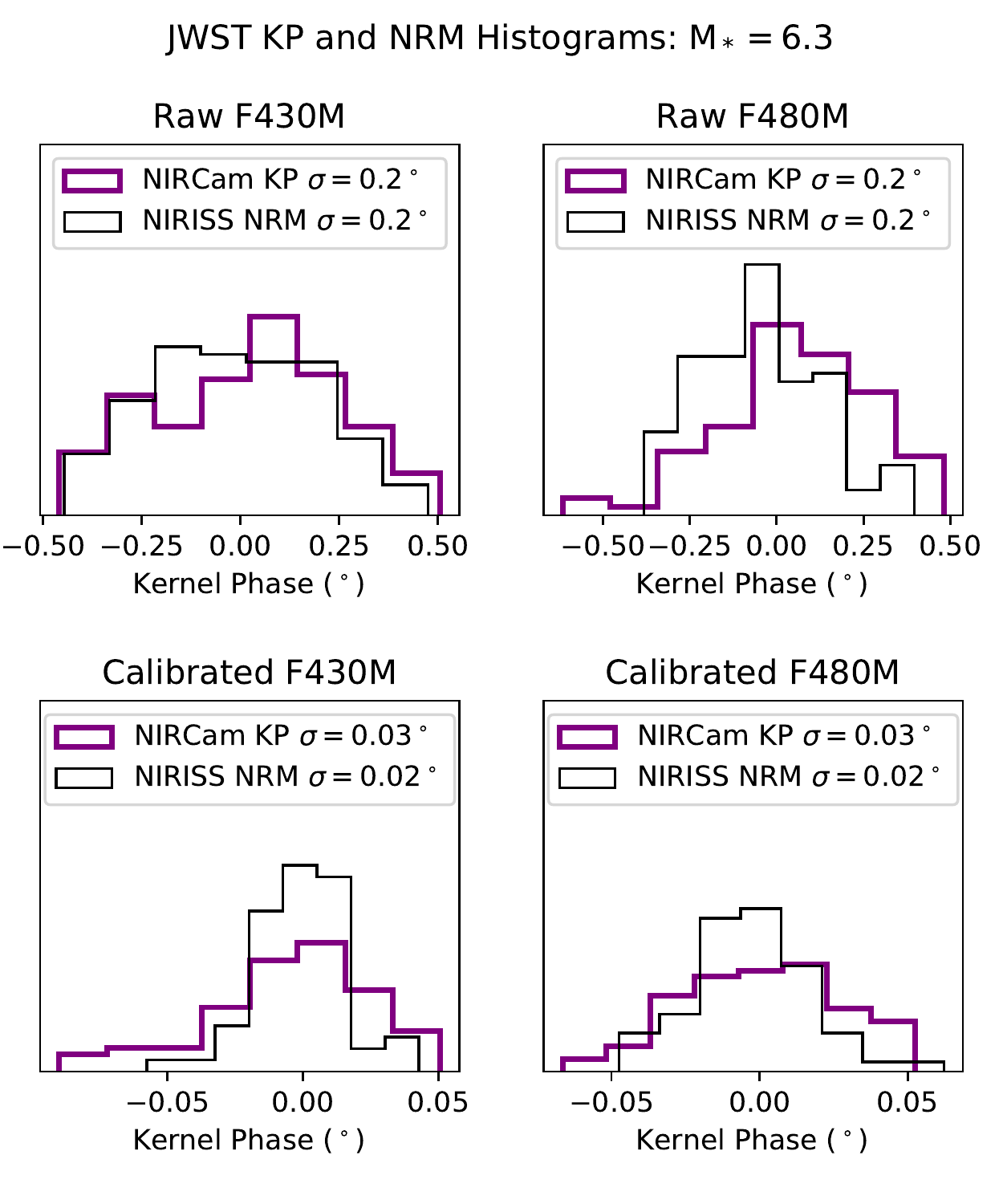}
\end{tabular}
\end{center}
\caption[NIRCam and NIRISS Kernel Phase Histograms: Bright Case] 
{ \label{fig:jwsthistsbright} 
Histograms of raw (top) and calibrated (bottom) kernel phases for F430M (left) and F480M band (right) observations for a bright (6th apparent magnitude) target star.
Thick purple lines show NIRCam filled-aperture kernel phases, while thin black lines show NIRISS NRM kernel phases.
}
\end{figure}
\pagebreak

\begin{figure} [h!]
\begin{center}
\begin{tabular}{c} 
\includegraphics[width=0.7\textwidth]{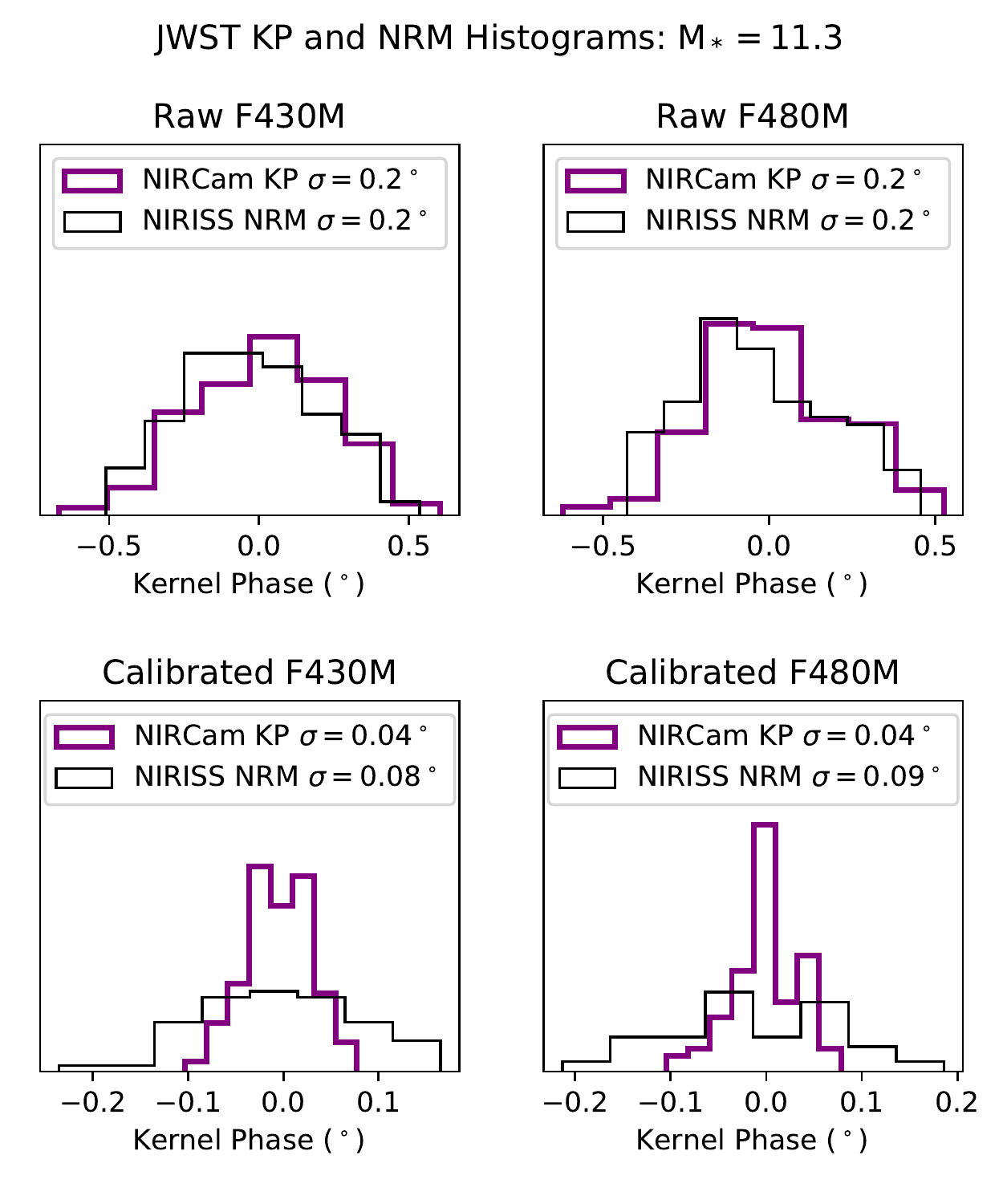}
\end{tabular}
\end{center}
\caption[NIRCam and NIRISS Kernel Phase Histograms: Faint Case] 
{ \label{fig:jwsthistsfaint} 
Histograms of raw (top) and calibrated (bottom) kernel phases for F430M (left) and F480M band (right) observations for a faint (11th apparent magnitude) target star.
Thick purple lines show NIRCam filled-aperture kernel phases, while thin black lines show NIRISS NRM kernel phases.
}
\end{figure}
\pagebreak

\begin{figure} [h!]
\begin{center}
\begin{tabular}{c} 
\includegraphics[width=0.7\textwidth]{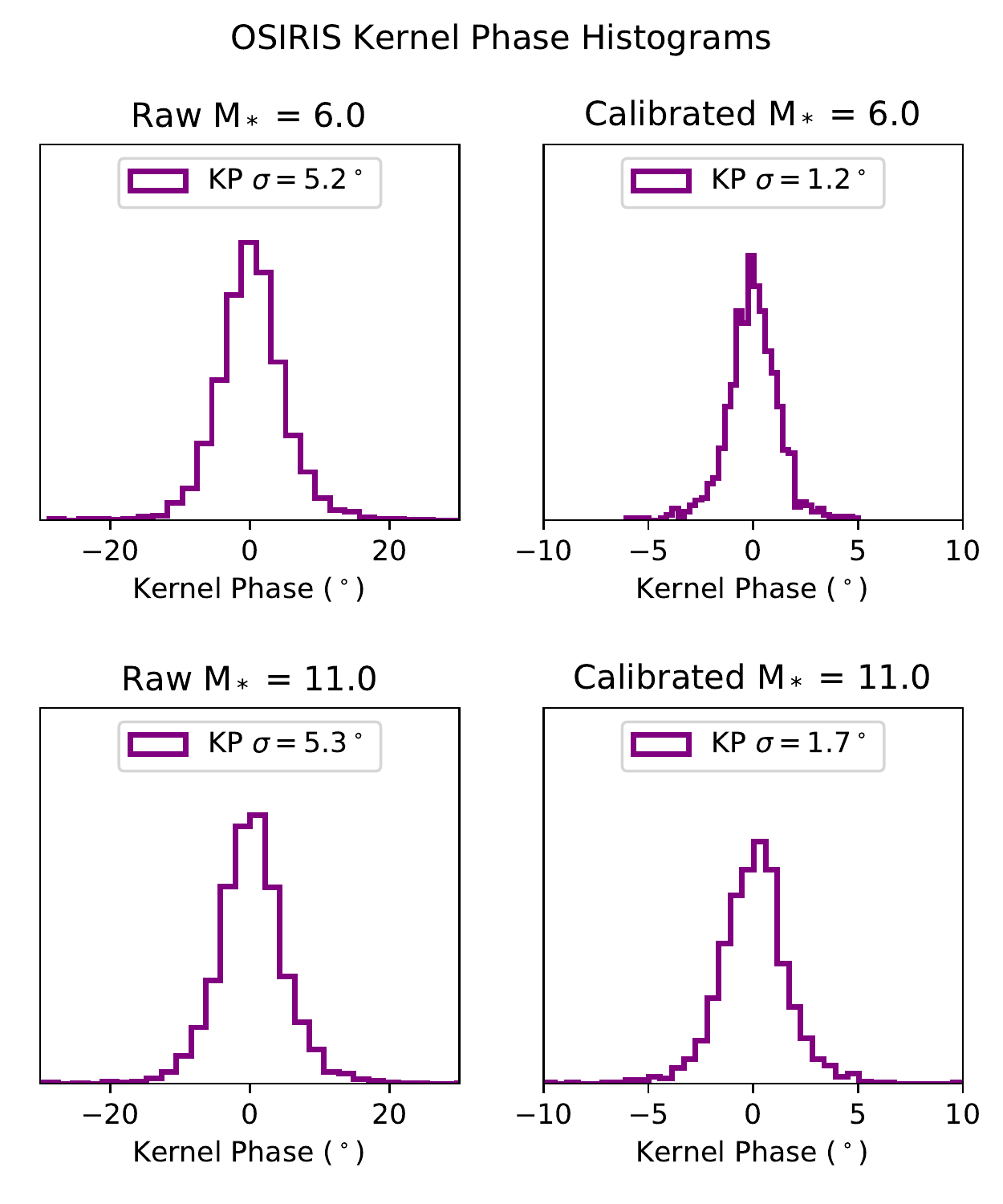}
\end{tabular}
\end{center}
\caption[OSIRIS Kernel Phase Histograms] 
{ \label{fig:osirishists} 
Histograms of raw (left) and calibrated (right) kernel phases for simulated OSIRIS observations for bright (6th apparent magnitude; top), and faint (11th apparent magnitude; bottom) target stars.
}
\end{figure}
\pagebreak

\begin{figure} [h!]
\begin{center}
\begin{tabular}{c} 
\includegraphics[width=0.7\textwidth]{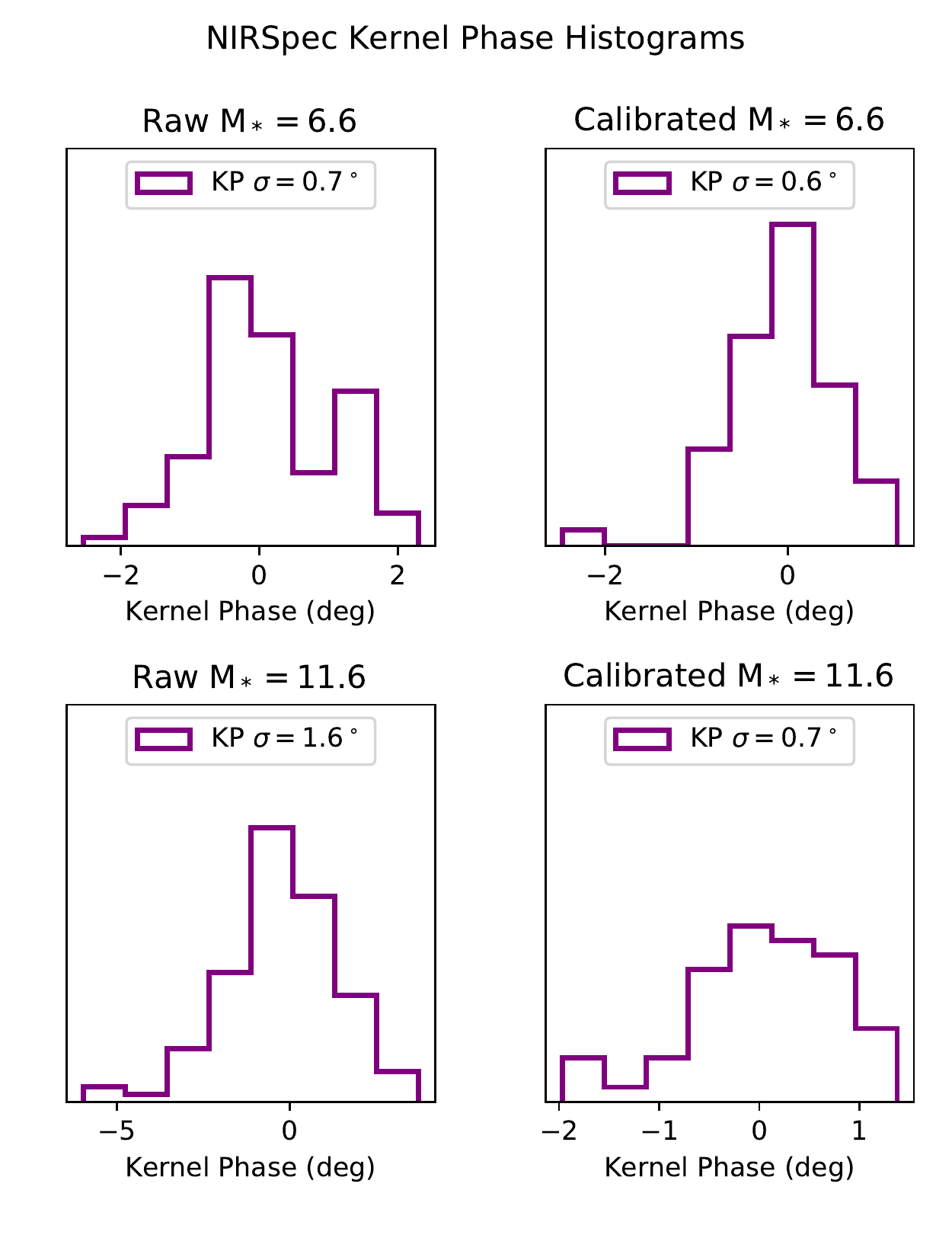}
\end{tabular}
\end{center}
\caption[NIRSpec Kernel Phase Histograms] 
{ \label{fig:nirspechists} 
Histograms of raw (left) and calibrated (right) kernel phases for simulated NIRSpec observations for bright (6th apparent magnitude; top), and faint (11th apparent magnitude; bottom) target stars.
}
\end{figure}

\pagebreak

\section{Scaling Contrast Curves for Other Target Stars and Observing Parameters}\label{app:scaling}
Since we provided contrasts for a small range of target star brightnesses, here we discuss how to adapt them to other target stars and observing scenarios.
If the amount of observing time per visit and number of visits is kept fixed, and the target star magnitude is decreased below an absolute magnitude of 6, for all observing modes presented here one would expect a similar contrast to the $\mathrm{M_* = 6}$ case.
This is because the $\mathrm{M_* = 6}$ observations represent the contrast limit, where OPD and calibration errors dominate. 
However, for all target star brightnesses, increasing the number of (assumed independent) visits would increase the achievable contrast in the bright limit by a factor of $\sqrt{\mathrm{N_{visits,new}} / \mathrm{N_{visits,old}}}$.

For the stars that are not contrast limited - e.g. for observing stars that are fainter than $\mathrm{M_* = 11-12}$ - if the observing time and number of visits were kept fixed, the contrast would decrease as the target star magnitude increased.
Since the kernel phase signal to noise is proportional to $\sqrt{\mathrm{N_{photons, target}}}$, the contrast would scale as $\sqrt{\mathrm{flux_{target, new}}}/\sqrt{\mathrm{flux_{target, old}}}$.
However the kernel phase scatter is also proportional to the square root of the number of sky background photons $\sqrt{\mathrm{N_{photons, sky}}}$; for high enough target star magnitudes eventually the background noise would dominate the kernel phases completely. 
This is already apparent in the upper regions of the $\mathrm{M_s}$ NRM and kernel phase panels in Figure \ref{fig:NIRC2ccs}.
If one were to scale the observing time per visit for stars not in the contrast limit, the contrast would increase as $\sqrt{\mathrm{t_{visit, new}/t_{visit,old}}}$, until the contrast limit is reached.

\end{document}